\documentclass[11pt]{article}

\usepackage[english]{babel}
\usepackage{amsmath}
\usepackage{amssymb}
 \usepackage{amstext,amsthm,bbm}
\usepackage{graphicx}
\usepackage{color}

\usepackage{enumerate}
\usepackage{fullpage}

\newtheorem{theorem}{Theorem}[section]
\newtheorem*{theorem*}{Theorem}
\newtheorem{corollary}[theorem]{Corollary}
\newtheorem{proposition}[theorem]{Proposition}

\newtheorem{lemma}[theorem]{Lemma}

\theoremstyle{definition}

\newtheorem{remark}[theorem]{Remark}

\numberwithin{equation}{section}
\newcommand{\R}{\mathbb R}

\newcommand{\C}{\mathbb C}

\newcommand{\eps}{\varepsilon}

\renewcommand{\det}{{\,\rm det}\:}

\newcommand{\erfc}{{\,\rm erfc}\:}
\renewcommand{\Re}{{\mathrm{Re}}}
\renewcommand{\Im}{{\mathrm{Im}}}
\renewcommand{\exp}[1]{\mathrm{exp} \left\{#1\right\}}

\begin{document}

\title{ Kac-Rice inspired approach to non-Hermitian random matrices}

\author{\Large Yan V Fyodorov \\[0.5ex] {\small Department of Mathematics, King's College London\footnote{Permanent address. e-mail: yan.fyodorov@kcl.ac.uk}}\\
{\small Strand, London, WC2R 2LS, UK}\\ and \\{\small Faculty of Physics, Bielefeld University, Postfach 100131, 33501 Bielefeld, Germany\footnote{Visiting Professor affiliation.}}}
	
\date{September 2026}
%\date{\today}

\maketitle

%%%%%%%%%%%%%%%%%%%%%%%%%%%%%%%%%%%%%%%%%%%%%%%%%%%%
\begin{abstract}
We suggest a method of analyzing the one-point joint eigenvalue--eigenvector
intensity ${\cal P}_N(z,{\bf v})$ of an eigenvalue $z$ and the associated
right eigenvector ${\bf v}$ (normalized by ${\bf v}^*{\bf v}=1$) for
non-Hermitian random matrices of a given size $N\times N$. At finite $N$
this intensity has total mass $N$. The approach  is essentially based on the Kac-Rice counting formula applied to the associated characteristic polynomial
combined with a certain integral identity for the Dirac delta function of such a polynomial. To illustrate utility of the general method we derive  ${\cal P}_N(z,{\bf v})$
 in the two particular cases:
(i)  one-parameter family of matrices interpolating between complex Ginibre and real Ginibre ensembles and (ii) a complex Ginibre matrix additively perturbed by a general fixed matrix. In particular, in the former case we analyze the formation of an excess of eigenvalues in the vicinity of the real axis on approaching the real Ginibre limit,
which eventually gives rise to the existence of a new scaling regime of "weak non-reality" as $N\to \infty$. Performing the same construction on the real unit sphere we further obtain, at finite $N$, the density of purely real eigenvalues in the real Ginibre limit. In the second case we further analyze non-Hermitian Rosenzweig-Porter model which recently attracted considerable interest in physics literature. In addition, we provide new insights into eigenvalue and eigenvector distribution for a general rank one perturbation of complex Ginibre matrices of finite size $N$, and in the structure of an outlier emerging as $N\gg 1$. Finally we discuss a generalization of the proposed method which is expected to be suitable for analysis of JPD involving both left- and right eigenvectors.

\end{abstract}

\section{Introduction}
Let $\mathbf{x}$ be a $N-$ component column vector, real or complex.  We will use $\mathbf{x}^T=(x_1,\ldots, x_N)$ to denote the corresponding transposed row vector ( and similar notation for matrices), and $\mathbf{x}^*=(\overline{x}_1,\ldots, \overline{x}_N)$ for the Hermitian conjugate, with bar standing for complex conjugation. The inner product of two such vectors will be denoted  as $\mathbf{x}_1^*\mathbf{x}_2=\sum_{i=1}^N \overline{x}_{1i}x_{2i}$.

Let  $X$ be a $N\times N$ matrix which will be in general assumed to be non-selfadjoint and not normal: $X^*\ne X, \, X^{*}X\ne XX^{*}$. We will further assume that all $N$ eigenvalues $z_a, \, a=1,\ldots,N$ of this matrix, which are in general complex numbers, are distinct.% have multiplicity one.

 Then the matrix is diagonalizable by a similarity transformation: $X=S Z S^{-1}$ where $Z=\mbox{diag}\left(z_1,\ldots,z_N\right)$ and $S$ is in general non-unitary: $S^*\ne S^{-1}$. The associated {\it right} eigenvectors defined by  $X\,\mathbf{v}_{a}=z_a \mathbf{v}_{a}$ are columns of the matrix $S$, whereas their {\it left} counterparts
 satisfying  $\mathbf{u}_{a}^* X=z_a \mathbf{u}^*_{a}$ form the rows of $S^{-1}$, and generically $\mathbf{v}_{a}$ and $\mathbf{u}_{a}$ are not related, unless the matrix is complex symmetric:
 $X^T=X$. Non-unitarity of $S$ implies that the two eigenvector families are not, in general, orthonormal: one cannot in general have $\mathbf{v}^*_{b}\mathbf{v}_{a}= \delta_{ab}$ for all $a,b$, and similarly for the left eigenvectors. Individual pairs may of course happen to be orthogonal.

  If the matrix $X$ is random, it makes sense to be interested in its statistical characteristics. In what follows we will use the notation $\left \langle \ldots \right\rangle_{X}$ to denote the expectation with respect to any specified ensemble of matrices $X$.
   In this way, for example the brackets $\left \langle \ldots \right\rangle_{GinUE}$ will represent the expectation taken with respect to the probability measure on $G$ known as the complex Ginibre ensemble, denoted here as $GinUE$ to reflect its invariance with respect to complex unitary conjugation. Its counterpart with real entries known as the real Ginibre ensemble will be denoted correspondingly with $GinOE$, to reflect invariance with respect to conjugation by real orthogonal matrices.

Throughout the paper Dirac $\delta$-symbols are used in the standard
distributional sense. We fix this convention precisely here. Whenever we
write a Dirac delta of a nontrivial function or map, we mean the corresponding
pull-back distribution: informally, the Dirac distribution at the origin is
transported back to the zero set of the map, with the appropriate Jacobian
factor. More precisely, under the regularity or transversality assumptions
stated below, it is understood through its action on test functions and, in
the eigenvalue--eigenvector identities central to the paper, through the weak
limit, in the sense of distributions, of the Gaussian regularizations
\begin{equation}\label{Gaudelta}
\delta_{g,\epsilon}^{(k)}(F)
=
\frac{1}{(2\pi\epsilon)^{k/2}}\exp{-\frac{|F|^2}{2\epsilon}},
\qquad \epsilon\downarrow0,
\end{equation}
where $k$ is the real dimension of the target space of $F$ and $|F|$ denotes
its Euclidean norm.
For a complex vector ${\bf u}\in\mathbb C^N$ we identify $\mathbb C^N$ with $\mathbb R^{2N}$ and write
\begin{equation}\label{Gaureg_AppA}
\delta_{g,\epsilon}^{(2N)}({\bf u})=\frac{1}{(2\pi\epsilon)^N}
\exp{-\frac{{\bf u}^*{\bf u}}{2\epsilon}}.
\end{equation}
All identities below involving pull-backs of Dirac distributions are identities of distributions, or equivalently of Radon measures, against the indicated class of test functions.  Appendix~A proves the fundamental sphere--delta identity directly by
heat-kernel regularization and complements this proof with two alternative
derivations, one based on Fourier/Gaussian regularization and the other on
induction. Appendix~B gives an independent geometric proof by coarea and transversality; in the regular-value setting it also identifies the Gaussian weak limit with the standard coarea pull-back.  

For $z=x+iy\in\mathbb C$ we set $d^2z:=dx\,dy$ and
$\delta^{(2)}(z):=\delta(x)\delta(y)$. For a matrix $X$ with simple
eigenvalues $z_1,\ldots,z_N$ the empirical eigenvalue measure is
\[
\rho_X(d^2z)=\sum_{a=1}^N\delta_{z_a}(d^2z),
\qquad\text{equivalently}\qquad
\rho_X(z)=\sum_{a=1}^N\delta^{(2)}(z-z_a),
\]
where the second, customary physics notation is used whenever no confusion
with an ordinary function can arise.

We further write
\[
\mathbf S_N^c=\{{\bf v}\in\mathbb C^N:{\bf v}^*{\bf v}=1\},\qquad
d\mu_N({\bf v})
=
\delta({\bf v}^*{\bf v}-1)d^{2N}{\bf v}
=
\frac12\,dS({\bf v}),
\]
so that $\int d\mu_N=\pi^N/\Gamma(N)$, and denote by
\[
d\mu_H({\bf v})=\frac{\Gamma(N)}{\pi^N}d\mu_N({\bf v})
\]
the normalized Haar measure on the complex unit sphere. We denote by
$\delta_H({\bf v}-{\bf v}_0)$ the Dirac distribution on $\mathbf S_N^c$
with respect to $d\mu_H({\bf v})$.

Choose for each eigenvalue any normalized right eigenvector ${\bf v}_a$,
${\bf v}_a^*{\bf v}_a=1$. Instead of fixing one of its coordinates, which
would fail whenever that coordinate vanishes, we average intrinsically over
the global phase and define the phase-averaged marked empirical measure.
In the customary physics notation, regarding $\Pi_X$ as a distributional
density with respect to $d^2z\,d\mu_H({\bf v})$, we write
\begin{equation}\label{empir_right_eigenv}
\Pi_X(z,{\bf v})
=
\sum_{a=1}^N\delta^{(2)}(z-z_a)
\int_0^{2\pi}
\delta_H\!\left({\bf v}-e^{i\phi}{\bf v}_a\right)
\,\frac{d\phi}{2\pi}.
\end{equation}
This definition is independent of the chosen representatives ${\bf v}_a$.
Its $z$-marginal is $\rho_X$ and its total mass is $N$.

For a random matrix ensemble with simple spectrum almost surely we denote by
$\mathcal I_N=\mathbb E_X\Pi_X$ the expected marked one-point intensity.
Whenever it has a density with respect to
$d^2z\,d\mu_H({\bf v})$, that density will be denoted by
$\mathcal P_N(z,{\bf v})$; therefore
\begin{equation}\label{intensity_normalization}
\int_{\mathbb C}\int_{\mathbf S_N^c}
\mathcal P_N(z,{\bf v})\,d\mu_H({\bf v})\,d^2z=N.
\end{equation}
The density of a uniformly sampled eigenpair is $N^{-1}\mathcal P_N$.
This distinction between the one-point intensity and the normalized
marked-eigenpair law will be kept explicit below. In particular, after the
macroscopic change of variables $z=\sqrt N\,w$, the Jacobian
$d^2z=N\,d^2w$ cancels the factor $N^{-1}$, so that
$\mathcal P_N(\sqrt N w,\cdot)$ naturally appears as the density of the
normalized marked law in the $w$ variable.

The number of approaches allowing one to successfully address spectral properties of non-Hermitian non-normal random matrices is relatively limited.
They essentially can be divided into two groups. The first approach relies upon either the Schur decomposition $X=U(Z+T)U^{*}$, where $U$ is unitary, $Z$ diagonal and $T$ is (upper) triangular, or its {\it incomplete} counterpart introduced by Edelman and collaborators in
\cite{EKS} and \cite{Ed:97} and adapted for addressing eigenvector properties in \cite{Fyo2018}. Although originally applied in the framework of classical Ginibre ensembles, recently that approach proved to be also very efficient in demonstrating the universality of obtained results for matrices with independent, identically distributed entries, see   \cite{Maltsev_Osman_2024, Osman2024_1,Osman2024_2,Osman2025}.

A different approach employs a trick going back to \cite{Girko1984} and
known as the {\it Girko Hermitization}: given a non-Hermitian $X$ of size $N$ one considers the Hermitian $2N\times 2N$ matrix
\begin{equation}\label{Hermitized}
Y_X(z,\overline{z})=\left(\begin{array}{cc} 0 &  \left(z{\bf 1}_N-X\right)\\  \left(z{\bf 1}_N-X\right)^{*} & 0\end{array}\right).
\end{equation}
Its utility comes from the fact that $z$ is an eigenvalue of $X$ if and only if the matrix  $Y_X(z,\overline{z})$ has a nontrivial kernel.
With the Wirtinger derivatives
$\partial_z=\frac12(\partial_x-i\partial_y)$ and
$\partial_{\bar z}=\frac12(\partial_x+i\partial_y)$, the empirical density
of complex eigenvalues of $X$ can be recovered from the fundamental
distributional Poincar\'e--Lelong identity
\begin{equation}
\frac{1}{\pi}\partial_{\bar z}\partial_z
\log{|\det{Y_X(z,\bar z)}|}
=
\sum_{i=1}^{N}\delta^{(2)}(z-z_i),
\end{equation}
whose integral form is frequently referred to as the Girko formula
(see, e.g., the lucid discussion in \cite{GKZ2011}). This or related formulas were of utmost importance for rigorously proving the famous Girko's circular law for the limiting mean density emerging as $N\to \infty$ and establishing its universality, see \cite{Bai97,TaoVu2008} and a review \cite{Bordenave-Chafai}.  Carefully working with the eigenvector of Hermitized Girko matrix Eq.(\ref{Hermitized})
corresponding to zero eigenvalue allows one to shed some light on corresponding eigenvectors of $X$, see e.g. \cite{Cipolloni_Erdos_Xu}.

In a version of this Hermitization approach one frequently defines the associated resolvent via
 \begin{equation}\label{HermitizedResolvent}
M_\epsilon(z,\overline{z})=\left(i\epsilon{\bf 1}_{2N}+Y_X(z,\overline{z})\right)^{-1}=\left(\begin{array}{cc} M^{(11)}_\epsilon(z,\overline{z}) &  M^{(12)}_\epsilon(z,\overline{z})\\  M^{(21)}_\epsilon(z,\overline{z})& M^{(22)}_\epsilon(z,\overline{z})\end{array}\right), \quad \epsilon>0.
\end{equation}
In terms of such a resolvent the expectation of the empirical density, which we denote by $p_N(z):=\mathbb E_X\rho_X(z)$, can be conveniently
  represented, e.g. as $p_{N}(z)=\lim_{\epsilon{\to 0}}\left\langle\frac{1}{\pi }\partial_{z}\mbox{Tr} M^{(12)}_\epsilon(z,\overline{z})\right\rangle_X$.
 Insights into statistical properties of eigenvectors can be also achieved by carefully working with
Eq.(\ref{HermitizedResolvent}), see also the explicit formulas below.  In general, the Hermitization resolvent approach remains one of principal available tools for eigenvector analysis, and has been recently utilized in some form in e.g. \cite{Cipolloni_Erdos_Xu}.

\subsection{An overview of previous works on eigenvectors}

Choose the left and right eigenvectors bi-orthonormally, $\mathbf{u}_a^*\mathbf{v}_b=\delta_{ab}$.  The standard overlap matrix is then $\mathcal{O}_{ab}=(\mathbf{u}_a^*\mathbf{u}_b)(\mathbf{v}_b^*\mathbf{v}_a)$.  Its diagonal entries satisfy $\mathcal{O}_{aa}=\|\mathbf{u}_a\|^2\|\mathbf{v}_a\|^2\ge1$.  The usual eigenvalue condition number is $\kappa_a=\sqrt{\mathcal{O}_{aa}}$; it measures the first-order sensitivity of $z_a$ to additive perturbations of $X$, see e.g. \cite{TrefethenBook}.
The Cauchy-Schwarz inequality
implies  $\mathcal{O}_{aa}\ge 1$.  For a given eigenpair equality means that the corresponding left and right eigenvectors are collinear; if this happens for every eigenpair of a simple-spectrum matrix, the eigenbasis is orthonormal and the matrix is normal.  For some classes of non-normal matrices of large size $N\gg 1$ one instead typically has $\mathcal{O}_{aa}\gg 1$, so that their eigenvalues can be much more sensitive to perturbations than in the normal case.

  The line of research addressing  statistics of the right - and left eigenvectors of non-normal random matrices, in particular those encapsulated in the properties of the non-orthogonality matrix $\mathcal{O}_{ab}$, originated from the influential papers by Chalker and Mehlig \cite{ChalkerMehlig1998, MehligChalker2000} who were the first to evaluate asymptotically, for large $N\gg 1$, the lowest moments of the form $\mathcal{O}(z)=\left\langle \sum_{a=1}^N \mathcal{O}_{aa}\delta^{(2)}(z-z_a)\right\rangle_{GinUE}$ as well as its off-diagonal counterpart
\begin{equation}\label{meanoverlap}
 \quad \mathcal{O}(z,\zeta)=\left\langle \sum_{a\ne b}^N
\mathcal{O}_{ab}\delta^{(2)}(z-z_a)\delta^{(2)}(\zeta-z_b)\right\rangle_{GinUE}.
\end{equation}

For the standard macroscopic normalization $N^{-1/2}G^{(C)}$, Chalker and Mehlig extracted the leading asymptotics of $\mathcal{O}(z)$ and $\mathcal{O}(z,\zeta)$.  In particular, in the bulk $|z|<1$ one has $\mathcal{O}(z)\sim \frac{N^2}{\pi}(1-|z|^2)$, while the mean eigenvalue intensity is asymptotic to $N/\pi$. This suggests that typically one should expect $\mathcal{O}_{aa}\sim N$ for eigenvalues inside the circle, which is parametrically larger than $\mathcal{O}_{aa}=1$ typical for normal matrices.
The Hermitized resolvent also encodes eigenvector-overlap observables; precise regularized formulas depend on the chosen Hermitization convention, and we refer to \cite{Janik1999} and subsequent work for such representations.

  The general interest in theoretical physics community in the non-normality in general and non-orthogonality factors in particular comes from several perspectives.
For example, in the context of dynamical systems non-normality is known to give rise to a long transient behaviour, see a general discussion in \cite{Trefethen1993,Somp2008}. In a related setting non-symmetric matrices appear very naturally via linearization around an equilibrium in a complicated nonlinear dynamical system, see e.g. \cite{May1972,FyoKhor2016,BFK2021}, and the non-orthogonality factors then control transients in a relaxation towards equilibrium \cite{Grela2017}, see e.g. discussion and further references in\cite{Fyo_etal_2025} and \cite{Troude_Sornette_2025}. Non-orthogonality was also argued to play an essential role in quantum dynamics and entanglement studies \cite{Cipolloni-KF1,Cipolloni-KF2} as well as
  in analysis of spectral outliers in non-selfadjoint matrices \cite{MetzNeri}.
 Another strong motivation comes from the field of quantum chaotic scattering, where non-selfadjoint random matrices of special type (different from the Ginibre ensembles) play a prominent role, see e.g. \cite{FyoSom1997,FyoSomRev2003} for the background information.    The corresponding non-orthogonality overlap matrix
$\mathcal{O}_{ab}$  shows up in various scattering observables, such as e.g. decay laws \cite{Savi97}, 'Petermann factors'  describing excess noise in open laser resonators \cite{Scho00}, in sensitivity of the resonance widths to small perturbations \cite{FyoSav2012} and in the shape of reflection dips \cite{FyoOsm2022}.
In recent years there was an essential progress in experimental studies of the non-orthogonality effects in that framework, see
\cite{NonorthExper,DavyGenack1,DavyGenack2}. All these applications generated a considerable effort to understand eigenvector non-orthogonality statistically, with initial emphasis on computing the Chalker-Mehlig correlators (\ref{meanoverlap}) and related objects beyond the framework of the complex Ginibre ensemble, see e.g. \cite{Janik1999,Scho00,MehligSanter2001,FyoMel2002,FyoSav2012,Burda2014,Burda2015,Belinchi2017,Burda2017,NT2018}.

In the mathematical literature a systematic rigorous research of statistical properties of eigenvectors of non-normal random matrices started much later with the work by Walters and Starr
\cite{WaltersStarr2015}. That line of research got later essential boost in works by
 Bourgade and Dubach \cite{BourgadeDubach}  and Fyodorov \cite{Fyo2018}. The former work  demonstrated a possibility to find
the law of the random variable  $\mathcal{O}_{aa}$ for the complex Ginibre ensemble, asymptotically for large $N$,  and provided valuable information
about the off-diagonal correlations between the two different eigenvectors at various scales of eigenvalue separation (the so-called 'microscopic' vs. 'mesoscopic' scales).
It also emphasized importance of eigenvector non-orthogonality factors for studying Dyson Brownian motion in the non-Hermitian context, see further studies in \cite{Grela_Warchol_2018} and \cite{Yabuoku}.
The paper \cite{Fyo2018} suggested an alternative approach to extend those results to finite-size GinUE matrices and to real eigenvalues of GinOE.
In recent years various statistical aspects of non-orthogonal eigenvectors in Ginibre (or closely related) ensembles have been the subject of vigorous activity, see e.g.
\cite{Akemann1,Akemann2,Akemann3,Banks_etal,Crowford_Rosenthal,FyoTarn2021,Dubach2021,Dubach2023,Cipolloni_etal_2024,Crumpton_etal_2025,Wurfel_etal_2024,Zhang_2024}, with more general ensembles of i.i.d. entries and universality issues addressed in \cite{BG_Zeitouni_eigvect,Luh_ORourk_1,Luh_ORourk_2,Erdos_Ji2024,Dubova_etal_2024,Osman2024_1,Osman2024_2,Cipolloni_Erdos_Xu}.

\subsection{Overview of the paper}

The main goal of the present paper is to develop an alternative approach for addressing statistics of both eigenvalues and eigenvectors of non-Hermitian
random matrices which is different from both methods outlined above, though the Hermitized matrix $Y_X(z,\overline{z})$ in fact also appears there in some form. The approach employs the framework of the so-called Kac-Rice counting formulas, see
\cite{Kac-Rice_lectures} for a detailed introduction into the subject, and the next section for an informal introduction and implementation in the present context.
In that framework we prove the following

\begin{theorem}[Phase-averaged Kac--Rice formula]\label{MainMetaTheorem}
Let $X\in\mathcal M_N(\mathbb C)$ have simple eigenvalues $z_1,\ldots,z_N$, and set
$p_X(z)=\det(X-z{\bf 1}_N)$.  For every $F\in C_c^\infty(\mathbb C\times\mathbf S_N^c)$,
\begin{equation}\label{Meta_Kac_Rice_right_eigenvec_test}
\frac1\pi\int_{\mathbb C}\int_{\mathbf S_N^c}
F(z,{\bf v})\,\delta^{(2N)}\!\left((X-z{\bf 1}_N){\bf v}\right)
|p_X'(z)|^2\,d\mu_N({\bf v})\,d^2z
=
\sum_{a=1}^N\int_0^{2\pi}F(z_a,e^{i\phi}{\bf v}_a)\,\frac{d\phi}{2\pi},
\end{equation}
where ${\bf v}_a$ is any unit right eigenvector associated with $z_a$.  The Dirac factor on the left is the distributional Gaussian limit established in Appendix~A; Appendix~B identifies the same measure geometrically by coarea.  Equivalently, the phase-averaged marked empirical measure has the distributional density
\begin{equation}\label{Meta_Kac_Rice_right_eigenvec}
\Pi_N(z,{\bf v})=\frac1\pi\delta^{(2N)}\!\left((X-z{\bf 1}_N){\bf v}\right)|p_X'(z)|^2
\end{equation}
with respect to $d^2z\,d\mu_N({\bf v})$.
\end{theorem}

The principal proof is given in Appendix~A by Gaussian regularization and the rank-one Harish--Chandra--Itzykson--Zuber integral, with coinciding nonzero singular values treated by the confluent divided-difference form.  The same appendix also gives two alternative derivations, based respectively
on Fourier/Gaussian regularization and induction. Appendix~B provides an
independent coarea proof. All three derivations require only that the relevant zeros of the determinant
be simple; the nonzero singular values of $X-z{\bf 1}_N$ may have arbitrary
multiplicities.

\begin{corollary}\label{main_corollary}
Let $X$ be a random matrix with simple spectrum almost surely.  Then Theorem~\ref{MainMetaTheorem} may be averaged directly, as an identity of finite Radon measures.  In particular, for every test function $F$ the expectation of the right-hand side of Eq.~\eqref{Meta_Kac_Rice_right_eigenvec_test} is bounded by $N\|F\|_\infty$.  Moreover, since
\[
\det Y_X(w,\bar w)=(-1)^N|p_X(w)|^2,
\qquad
|p_X'(z)|^2=(-1)^N\left.\partial_w\partial_{\bar w}\det Y_X(w,\bar w)\right|_{w=z},
\]
the expected marked intensity with respect to $d^2z\,d\mu_N({\bf v})$ can be written distributionally as
\begin{equation}\label{main_practical}
\widetilde{\mathcal P}_N(z,{\bf v})=
\frac{(-1)^N}{\pi}\,\mathbb E_X\!\left[
\delta^{(2N)}\!\left((X-z{\bf 1}_N){\bf v}\right)
\left.\partial_w\partial_{\bar w}\det Y_X(w,\bar w)\right|_{w=z}
\right].
\end{equation}
The density with respect to the normalized Haar measure is
$\mathcal P_N=(\pi^N/\Gamma(N))\widetilde{\mathcal P}_N$.
For the Gaussian ensembles treated below, the heat-kernel regularization makes the Fourier integrals absolutely convergent for every $\epsilon>0$; differentiation, expectation and integration may therefore be interchanged at fixed $\epsilon$. The passage $\epsilon\downarrow0$ is of a different nature: what the computations of Sections \ref{interpol} and 4 produce is the pointwise limit of the $\epsilon$-regularized \emph{densities}, whereas Appendix A establishes weak convergence of the regularized \emph{measures}. The two are reconciled by the following elementary statement, which we shall use without further comment.
\end{corollary}

\begin{lemma}\label{density_identification}
Let $\lambda$ be a Radon measure on a locally compact second-countable space $E$, let $\mu_{\epsilon}=f_{\epsilon}\,d\lambda$ be finite positive measures with $\mu_{\epsilon}\Rightarrow\mu$ weakly as $\epsilon\downarrow0$, and suppose that $f_{\epsilon}\to f$ locally uniformly on $E$ with $f$ continuous. Then $\mu=f\,d\lambda$.
\end{lemma}
\begin{proof}
For $F\in C_c(E)$ one has $\int Ff_{\epsilon}\,d\lambda\to\int Ff\,d\lambda$ by locally uniform convergence on the compact set $\mbox{supp}F$, and
$\int Ff_{\epsilon}\,d\lambda=\int F\,d\mu_{\epsilon}\to\int F\,d\mu$ by hypothesis. Hence $\int F\,d\mu=\int Ff\,d\lambda$ for all such $F$.
\end{proof}

In the two ensembles treated below the $\epsilon$-regularized densities converge locally uniformly on $\mathbb{C}\times{\bf S}^c_N$ --- for fixed $0\le\tau<1$ in Section \ref{interpol}, where the quadratic form $C_{\epsilon}$ of Eq.(\ref{quad_form2}) is positive definite uniformly in $\epsilon\ge0$, and for every fixed $A$ in Section 4, where the regulator can be dropped outright --- and the limits displayed there are continuous. Lemma \ref{density_identification}, applied with $E=\mathbb{C}\times{\bf S}^c_N$ and $d\lambda=d^2z\,d\mu_N({\bf v})$, therefore identifies those limits with the densities of the corresponding intensities.

Such representation would however appear rather useless, unless one suggests an efficient way to evaluate the expectation of the right-hand side over a given ensemble of random matrices $X$. Fortunately, Eq.(\ref{main_practical}) does provide such a possibility. For this purpose one may employ the Fourier integral representation for the Gaussian-regularized "heat kernel" version, see Eq.(\ref{Gaureg_AppA}).: $\delta^{(2N)}\left(\left(X-z{\bf 1}\right){\bf v}\right)=\lim_{\epsilon\to 0}\delta^{(2N)}_{g,\epsilon}\left(\left(X-z{\bf 1}\right){\bf v}\right)$, where (cf. Eq.(\ref{Gaureg_AppA}))
\begin{equation}\label{Gaudelta_vec}
  \delta^{(2N)}_{g,\epsilon}\left(\left(X-z{\bf 1}\right){\bf v}\right)
=\int_{\mathbb{C}^N}e^{-\epsilon\frac{{\bf k}^*{\bf k}}{2}+\frac{i}{2}\left[{\bf k}^*\left(X-z{\bf 1}\right){\bf v}+
{\bf v}^*\left(X^*-\overline{z}{\bf 1}\right){\bf k}\right]}\frac{d{\bf k}d{\bf k}^*}{(2\pi)^{2N}}
\end{equation}
and we denoted  $d{\bf k}d{\bf k}^*:=\prod_{i=1}^Nd \Re{k_i}d\Im{k_i}$ and ${\bf k}^*{\bf k}:=\sum_{i=1}^N\overline{k}_ik_i$ for ${\bf k}=(k_1,\ldots,k_N)^T\in \mathbb{C}^N$.
The Fourier integral representation of the $\delta-$function then can be combined with the Berezin Gaussian integral representation
 \begin{equation}\label{Berezin_int}
  \det{Y_X(w,\overline{w})}=(-1)^N\int \exp{\frac{i}{2}\left(\mathbf{\Psi}_{+}^T,\mathbf{\Psi}_{-}^T\right)
  \left(\begin{array}{cc}0 & X-w{\bf 1}\\ X^*-\overline{w}{\bf 1}& 0\end{array}\right)\left(\begin{array}{c}\mathbf{\Phi}_{+} \\ \mathbf{\Phi}_{-}\end{array}\right)}\,{\cal D}(\mathbf{\Psi},\mathbf{\Phi})\, ,
  \end{equation}
  where $\mathbf{\Psi}_{\sigma}=\left(\psi_{\sigma,1},\ldots,\psi_{\sigma,N}\right)^T$ and $\mathbf{\Phi}_{\sigma}=\left(\phi_{\sigma,1},\ldots,\phi_{\sigma,N}\right)^T$ for $\sigma=\{+,-\}$
  are vectors with anticommuting/Grassmann entries,  ${\cal D}(\mathbf{\Psi},\mathbf{\Phi})=\prod_{\sigma}\prod_{i=1}^N [2d\psi_{\sigma i} d\phi_{\sigma i}]$
  and we assume the standard Berezin rules $\int d\phi=0, \, \int \phi\, d\phi=1$ for the integration over an anticommuting variable $\phi$.
  Employing such representations the expectation over $X$ in Eq.(\ref{main_practical}) can frequently be computed,  especially  straightforwardly if entries of $X$ are independent random Gaussian variables.
   In the main part of the paper we demonstrate viability of our approach by implementing such a strategy for two different choices of the ensemble for $X$, both cases being certain nontrivial deformations of Ginibre ensembles.  In those cases we compute explicitly the associated JPD ${\cal P}_N(z,{\bf v})$ and eventually the mean eigenvalue densities, in most cases beyond currently available in the literature.

The organization of the paper is as follows.  It starts with a short overview of the Kac-Rice counting formulas which provide the main framework for arriving at the statement of Theorem \ref{MainMetaTheorem}. The Section 2 then provides a detailed summary and discussions of the main results and
 finishes with a discussion of open problems as well as with describing an extension of the method allowing one, in principle, to deal with the JPD of both right and left eigenvectors, see Eq.(\ref{left_right_KacRice}).  Section 3  contains details of evaluation of  $\mathcal{P}_N(z,{\bf v})$ of the (normalized) right eigenvector and associated eigenvalue for an ensemble of random matrices interpolating between the real Ginibre and complex Ginibre ensembles.  Our approach yields an exact and explicit formula for every $N\ge2$ (the scalar case $N=1$ is immediate), which together with the associated mean eigenvalue density (i.e. its marginal obtained after integrating over the eigenvector)  are then amenable to extracting the appropriate  scaling limits as $N\to \infty$.  That section closes with a treatment of the real-eigenvalue sector at $\tau=1$, see Section \ref{realsector}: performing the same Kac-Rice construction on the {\it real} rather than the complex unit sphere yields an exact finite-$N$ formula for the density of real eigenvalues of the real Ginibre ensemble, and along the way reproduces the partial real Schur decomposition of \cite{EKS,Ed:97} with no Jacobian computation.
In the next Section 4 we apply essentially the same method for evaluating $\mathcal{P}_N(z,{\bf v})$ on the matrices from the complex Ginibre ensemble $GinUE$ perturbed by a general fixed matrix $A$, possibly non-normal. One may further straightforwardly pass from a fixed perturbation to a random perturbation independent of the original Ginibre matrix. We choose to treat in detail a special case of random diagonal Gaussian perturbation, which appeared recently in physics literature under the name of non-Hermitian Rosenzweig-Porter model \cite{TomasiKhaimovich_NHRP22}. We concentrate on the so-called critical case of the latter model which was left unstudied in \cite{TomasiKhaimovich_NHRP22} , but is perfectly
amenable to our approach. Finally, we consider in some detail another particular case of considerable interest: a rank-one perturbation of complex Ginibre matrices.

Appendices contain technical details and proofs.

For clarity about theorem status, Theorems~\ref{Theorem1} and~\ref{Theorem2} are exact finite-$N$ identities, as are Theorem~\ref{RealMetaTheorem} and Proposition~\ref{GinOErealprop} of Section~\ref{realsector}; the single external input in Section~\ref{realsector} is the classical Gaussian expectation Eq.(\ref{Edet_classical}), which is quoted from \cite{EKS,Ed:97} and concerns no eigenvalues. Theorem~\ref{NHRP} is a fixed-$\sigma$ large-$N$ limit proved in Section~\ref{NHRPproof}. By contrast, the logarithmic scaling $\sigma^2=(\mu/2)\log N$ discussed in Remark~\ref{logNHRZ} is a physics-level prediction based on the exact finite-$N$ formulas and is not used in any theorem.

{\bf Acknowledgements}.
The author is most grateful to Gernot Akemann and Dmitry Savin for early discussions of the approach and subsequent joint work on applying methods of this paper to a special case of complex symmetric matrices, see \cite{Akemann4}. Martin Zirnbauer is gratefully acknowledged for discussions concerning an alternative regularization approach to the basic determinant identity. The author would like to thank Ivan Khaymovich and Roman Sarapin for informative comments about the content of the papers \cite{TomasiKhaimovich_NHRP22} and \cite{Sarapin25}, respectively, and is further grateful to Boris Khoruzhenko for highly useful comments and advice on the first version of this article and to Mark Crumpton for his assistance with preparing figures for the paper. 
The paper also benefited greatly from the thorough criticism of an anonymous
referee on its second version. In revising the manuscript in response to that
report, ChatGPT was used to assist in checking and restructuring parts of the
mathematical arguments, identifying points requiring additional justification,
and improving the presentation. All mathematical statements, proofs, and
conclusions were subsequently checked by the author, who bears sole
responsibility for their correctness.

The research at King's College London was supported by  EPSRC grant {\bf UKRI1015} "Non-Hermitian random matrices: theory and applications"
and at Bielefeld University by the Deutsche Forschungsgemeinschaft (DFG) grant SFB 1283/2 2021–317210226. Kind hospitality of the
Institute for Applied Mathematics at Bonn University, Germany where the first version of this article has been completed is also acknowledged with thanks.

%%%%%%%%%%%%%%%%%%%%%%%%%%%%%%%%%%%%%%%%%%%%%%%%%%%%%%%%%%%%%%%%%%%%%%%%%%%%%%%

\section{Discussion of the main results.}

\subsection{Kac-Rice counting formulas: an informal introduction}

 Given a continuously differentiable function $f(x)$ of a real argument $x\in \mathbb{R}$  one may ask about counting the number ${\cal N}_{(a,b)}(u)$ of crossings of such a function with the $u$ level in an interval $x\in[a,b]$.  The two seminal papers by Mark Kac \cite{Kac_poly_1943} and Stephen Rice \cite{Rice1944} suggested that assuming individual crossings are not degenerate ( i.e. not tangential, and occurring at distinct points) such counting can be done via the following identity:
\begin{equation}\label{Kac-Rice_counting}
{\cal N}^{(f)}_{(a,b)}(u)=\lim_{\epsilon\to 0}\frac{1}{2\epsilon}\int_a^b \mathbf{1}_{u-\epsilon,u+\epsilon}(f(x))|f'(x)|dx=\lim_{\epsilon\to 0}\int_a^b \delta_{b,\epsilon}\left(u-f(x)\right)|f'(x)|dx
\end{equation}
where $\mathbf{1}_{A}(f(x))$ stands for the indicator function for the values of $f(x)$ in a set $A$, and $\delta_{b,\epsilon}(x)$ stands for the "box version" of $\epsilon-$regularized Dirac delta-function. In fact it is easy to verify independently that the same formula holds if one uses the Gaussian regularization as in eq.(\ref{Gaudelta}). Expecting that under approriate conditions the result for counting should be independent of the particular choice of the regularization and
further assuming $f(x)$ to be random,
 one may  interpret the ensuing distributional identity
$\rho_u(x)=\delta\left(u-f(x)\right)|f'(x)|$ as the empirical density of simple roots of the random equation $f(x)=u$.
More generally, employing the standard Laplace asymptotic method for multivariable integrals it is not difficult to see that given a smooth enough vector field $\mathbf{f}: \mathbb{R}^n\to \mathbb{R}^n$ which has only isolated zeroes in $\mathbb{R}^n$, the number ${\cal N}_{D}$ of such zeroes, i.e. solutions of the system of equations $f_i({\bf x})=0, \,i=1,\ldots,n$ in $ D\subset \mathbb{R}^n$, can be counted via
\begin{equation}\label{Kac-Rice_counting_multi}
\mathcal{N}_{D}=\lim_{\epsilon\to 0} \int_D \rho_{\eps}^{(KR)}({\bf x})\,d^n{\bf x}, \quad \mbox{with}\quad  \rho_{\epsilon}^{(KR)}({\bf x})=
\delta_{g,\epsilon}
(\mathbf{f})\left|\det\left(\frac{\partial f_i}{\partial x_j}\right)\right|,
\end{equation}
where
\begin{equation}\label{heat_delta}
\delta_{g,\epsilon}(\mathbf{f})=\frac{1}{(2\pi\epsilon)^{n/2}}e^{-\frac{\sum_{i=1}^nf^2_i({\bf x})}{2\epsilon}}
\end{equation}
is the "heat kernel" Gaussian regularization of  $\delta(\mathbf{f}):=\prod_i\delta(f_i({\bf x}))$, cf. Eq.(\ref{Gaureg_AppA}). Hence $\rho^{(KR)}({\bf x})=
\prod_i\delta(f_i({\bf x}))\left|\det\left(\frac{\partial f_i}{\partial x_j}\right)\right|$ can be considered as the empirical density of the vector field zeroes.
In recent decades such distributional representations, interpreted with due care, played fundamental role in describing statistics of critical points of high-dimensional random surfaces,  with numerous applications in theory of disordered glassy systems, optimization, mathematical ecology and beyond, see \cite{Fyo_2004_landscapes,Auffinger_etal_glass,Auffinger_BenArous,FyoKhor2016,BFK2021} for the original works, and
\cite{Fyo_rev_2015, RosFyoRev} for discussion of recent developments and further references.

In particular, let $P_N(z)=\sum_{k=0}^N a_kz^k$ be a random polynomial of degree $N$ in a single complex variable $z\in \mathbb{C}$, and assume it has $N$
simple zeroes $z_1, \ldots, z_N$ with probability one. Using $n=2$ version of the Kac-Rice counting formula Eq.(\ref{Kac-Rice_counting_multi}) combined with the Cauchy-Riemann relations the empirical density $\rho_P(z):=\sum_{n=1}^N\delta^{(2)}(z-z_n)$
   of such zeroes  in the complex plane $\mathbb{C}$ can be then written as
$\rho_P(z)=\delta^{(2)}(P_N(z))|P_N'(z)|^2$. Applying such a formula to the characteristic polynomial $\det{\left(X-z\mathbf{1}\right)}$  the above relation implies that the empirical density $\rho_N(z)$ of complex eigenvalues of any non-selfadjoint matrix $X$ with non-degenerate eigenvalues can be readily represented as

\begin{equation}\label{Kac-Rice_charpol}
\rho_N(z)=\delta^{(2)}\left(\det{\left(X-z\mathbf{1}\right)}\right)\,\left|\frac{d}{dz}\det\left(X-z\mathbf{1}\right)\right|^2.
\end{equation}
In practice, however, in the RMT setting such formula looks rather useless, as it seems difficult to suggest an efficient way to evaluate the expectation of the right-hand side over a given ensemble of random matrices $X$.
One of the main messages of the present paper is that such an evaluation becomes possible because, along any holomorphic one-parameter matrix family for which the determinant has only simple zeros, the scalar Dirac factor admits a multivariate representation.  More precisely, let $\Omega\subset\mathbb C$ be open and let $B:\Omega\to\mathcal M_N(\mathbb C)$ be holomorphic.  Assume that every zero $z_0$ of $\det B(z)$ in $\Omega$ is simple.  Then, as an identity in $\mathcal D'(\Omega)$,
\begin{equation}\label{main_integral_identity}
\delta^{(2)}\!\left(\det B(z)\right)=\frac1\pi\int_{\mathbf S_N^c}
\delta^{(2N)}\!\left(B(z){\bf v}\right)d\mu_N({\bf v}).
\end{equation}
No condition on the multiplicities of the nonzero singular values of $B(z_0)$ is needed.  The principal proof in Appendix~A uses Gaussian regularization and the confluent rank-one Harish--Chandra--Itzykson--Zuber formula; Appendix~B supplies an independent coarea proof.  Notice that Eq.~\eqref{main_integral_identity} is deliberately stated as a distributional identity in the parameter $z$; no pointwise meaning is assigned to either side at a fixed singular matrix.

Applying Eq.~\eqref{main_integral_identity} to the characteristic pencil $B(z)=X-z{\bf 1}_N$ gives
\begin{equation}\label{Kac-Rice_charpol_density}
\rho_X(z)=\int_{\mathbf S_N^c}\Pi_N(z,{\bf v})\,d\mu_N({\bf v}),\qquad
\Pi_N(z,{\bf v})=\frac1\pi\delta^{(2N)}\!\left((X-z{\bf 1}_N){\bf v}\right)|p_X'(z)|^2,
\end{equation}
which is precisely the density form of Theorem~\ref{MainMetaTheorem}.  The phase averaging is intrinsic and no coordinate of ${\bf v}$ is singled out.  For a random matrix ensemble the expectation $\mathcal I_N=\mathbb E_X\Pi_X$ is the marked one-point intensity; the normalized law of a uniformly sampled eigenpair is $N^{-1}\mathcal I_N$.

   An extension of Eq.(\ref{Kac-Rice_charpol_density}) providing the Joint Empirical Density  of both right and left eigenvectors corresponding to the same eigenvalue will be briefly discussed later on in the paper, see  Section \ref{Towards}. Its further analysis and exploitation is left for the future work.

\begin{remark}  The expression for ${\cal P}_N(z,{\bf v})$ based on Eq.(\ref{Kac-Rice_charpol_density}) appeared without any derivation in a short communication by  Sommers and Iida \cite{Sommers-Iida1994} in the course of addressing statistics of eigenvectors of {\it Hermitian} matrices in a crossover between GOE and GUE, but it seems to  have never been further used beyond the original paper. The authors noted that " the formula is valid for any (not necessarily Hermitian) complex $N-$dimensional matrix", and further made some sketchy remarks towards their "lengthy derivation", which the present author found not easy to follow.  No relation or reference to Kac-Rice formulas has been  mentioned in \cite{Sommers-Iida1994}.
\end{remark}

\vspace{0.5cm}

The representation Eq.(\ref{main_practical})  opens a principal possibility to evaluate the JPD  ${\cal P}_N(z,{\bf v})$ for some random matrix ensembles
via the method exposed after the Corollary 1.2. Below we discuss the results of actual implementations of this program for two different generalizations of the classical Ginibre ensembles.

\begin{remark}
Eq.(\ref{Kac-Rice_charpol_density}) defines the JPD with respect to the measure $d^2zd\mu_{N}({\bf v})$. In the rest of the paper we however found it more convenient to present results for the JPD  with respect to the measure $d^2zd\mu_{H}({\bf v})$, where $d\mu_{H}({\bf v})=\frac{\Gamma(N)}{\pi^N}d\mu_{N}({\bf v})$ is properly normalized: $\int_{\mathbb{C}^N}d\mu_{H}({\bf v})=1$.
\end{remark}

\subsection{ Case 1 - ensemble interpolating between complex and real Ginibre}

Recall that the real Ginibre ensemble consists of $N\times N$ square matrices $G^{(R)} \in \mathcal{M}_N\left(\R\right)$ with independent
identically distributed real standard normal ${\cal N}_R(0,1)$  matrix elements $G^{(R)}_{j,k}$.
%	\begin{equation}\label{Ginelements}
%		G^{(R)}_{j,k} \sim \mathcal{N}\left(0,1\right).
%	\end{equation}
Similarly, the complex Ginibre ensemble consists of square matrices $G^{(C)}  \in \mathcal{M}_N\left(\C\right)$ with independent
identically distributed complex standard normal ${\cal N}_C(0,1)$ matrix elements $G^{(C)} _{j,k}$.
Equivalently one may write $G^{(C)}=\frac{1}{\sqrt{2}}(G^{(R)}_1+i G^{(R)}_2)$, with $G^{(R)}_{1,2}$ being two independent copies of real Ginibre matrices.
	%\begin{equation}\label{Ginelements}
	%	G^{(C)} _{j,k} = g^{\left(1\right)}_{j,k}+
	%		i\cdot g^{\left(2\right)}_{j,k},
	%		\quad
	%		\mbox{with i.i.d.}
	%		\,\,
	%		g^{\left(\cdot\right)}_{j,k}
	%		\sim \mathcal{N}\left(0,1/2\right),
	%\end{equation}

This suggests an idea of defining a one-parameter family of matrices $G_{\tau}$ (which we may call the {\it interpolating} Ginibre ensemble) according to
\begin{equation}\label{def_interpolating}
G_{\tau}=G^{(R)}_1\sqrt{\frac{1+\tau}{2}}+iG^{(R)}_2\sqrt{\frac{1-\tau}{2}},\quad \tau\in[0,1],
\end{equation}
such that in the two limiting cases $\tau=0$ and $\tau=1$ one recovers the  complex Ginibre $G^{(C)}=G_{\tau=0}$ and real Ginibre  $G^{(R)}=G_{\tau=1}$ ensemble, correspondingly.

\begin{remark}
 The book-length treatment of those two limiting cases can be found in \cite{ByunFor_book}, while the interpolating Ginibre ensemble does not seem to have been
 previously studied.
 One of the main qualitative differences between the spectral properties in the two limiting cases is that the real Ginibre matrices typically have
 of the order of $\sqrt{N}$ of purely real eigenvalues as $N\to \infty$, while no such eigenvalues appear with probability one in the complex case.
  It is clear that the latter situation takes place for any fixed $0\le \tau<1$. Still, one may expect that for $1-\tau\ll 1$ one should be able to see the effect of accumulation of eigenvalues in the vicinity of the real axis. We will see below that this is indeed the case.
   \end{remark}

Our main result for the interpolating Ginibre ensemble is the following
\begin{theorem}\label{Theorem1}
Assume $N\ge2$. For matrices from the interpolating Ginibre ensemble Eq.~(\ref{def_interpolating}) with $0\le \tau<1$, the one-point joint intensity of an eigenvalue $z$ and its phase-averaged normalized right eigenvector ${\bf v}$, with respect to $d^2z\,d\mu_H({\bf v})$, is
\begin{equation}\label{MainJPDinterpolating}
{\cal P}^{(\tau)}_N(z,{\bf v})=\frac{D^{-N/2}}{\pi \Gamma(N)}
\exp{-\frac{\tau u}{2D}\left(2\tau |z|^2-z^2-\overline{z}^2\right)}
\left\{\frac{1-\tau^2}{D}\,\Gamma(N,|z|^2)
\right.
\end{equation}
\[
\left.
+\Gamma(N-1,|z|^2)\frac{\tau(1-u)}{D}
\left[\tau |z|^2\left(1+\frac{1-u}{D}\right)-(z^2+\overline{z}^2)\right]
\right\},
\qquad
u:=|{\bf v}^T{\bf v}|^2,
\]
where $D:=1-\tau^2u$. The incomplete Gamma function is defined by
\begin{equation}\label{gamma_incom}
\Gamma(N,a):=\int_{a}^{\infty}dt t^{N-1}e^{-t}=\Gamma(N)e^{-a}\sum_{n=0}^{N-1}\frac{a^n}{n!}
\end{equation}
The displayed formula is algebraically equivalent to the form derived in Section~\ref{interpol}. It is manifestly regular at $\tau=0$, where it reduces directly to the complex Ginibre intensity, and it extends to $\tau=1$ for $\Im z\ne0$.
\end{theorem}

The proof of Theorem \ref{Theorem1} is given in the Section 3.

\subsubsection{Mean density of complex eigenvalues for the interpolating ensemble}
  The mean eigenvalue density is by definition a marginal of the JPD obtained by performing the ${\bf v}-$integration against the normalized Haar's measure  $d\mu_{H}({\bf v})$.
Note  that the JPD Eq.(\ref{MainJPDinterpolating}) depends on the right eigenvector only via the combination $u=|{\bf v}^T{\bf v}|^2\in [0,1]$.
For such a type of the dependence the ${\bf v}-$integration can be accomplished using the following
\begin{lemma}\label{IntLemma}
 The identity
\begin{equation}\label{sphereintegral}
\int_{\mathbf{S}^c_N}  f(|{\bf v}^T{\bf v}|^2)\,d\mu_H({\bf v})=\frac{N-1}{2}\int_0^1(1-u)^{(N-3)/2}f(u)\,du,
\end{equation}
holds true for any function $f(u), \, u\in \mathbb{R}$ such that the above integrals exist, assuming $N\ge 2$.
\end{lemma}
For a proof see Appendix D. With the help of the  Lemma \ref{IntLemma}  performing integration  over the vector ${\bf v}$ in Eq.(\ref{MainJPDinterpolating})  is straightforward. After further changing the associated integration variable as $u\to \frac{1-w^2}{1-\tau^2w^2}$ one arrives at the following
\begin{corollary}
For matrices from the interpolating Ginibre ensemble Eq.(\ref{def_interpolating})
the mean eigenvalue density $p^{(\tau)}_{N}(x,y)$ in the complex plane $z=x+iy$ is given by
\begin{equation}\label{density_interpolating}
p^{(\tau)}_{N}(x,y)=\frac{1}{\pi \Gamma(N-1)\sqrt{1-\tau^2}}\displaystyle{\int_0^1\frac{dw w^{N-2}}{\sqrt{1-\tau^2w^2}}}\,\,  e^{\tau(1-w^2)\left(\frac{x^2}{1+\tau}-\frac{y^2}{1-\tau}\right)}
\end{equation}
\[
\times \left\{\Gamma(N,x^2+y^2)(1-\tau^2w^2) +\Gamma(N-1,x^2+y^2)\tau w^2 \left[\tau (x^2+y^2)(1+w^2)-2(x^2-y^2)\right]\right\}
 \]
\end{corollary}

\begin{remark}
In the limit $\tau=0$ one immediately recovers the well-known expression for the mean eigenvalue density of the  complex Ginibre ensemble:
 \begin{equation}\label{denGinUE}
p_N^{(GinUE)}(z)=\frac{1}{\pi} e^{-|z|^2}\sum_{n=0}^{N-1}\frac{|z|^{2n}}{n!}.
\end{equation}

The opposite case $\tau\to 1$ requires slightly more care. The limit  can be most conveniently performed after first changing $w\to 1-u(1-\tau), \, u\in[0,1/(1-\tau)]$. Assuming  $y=\Im z\ne 0$ one then can set $\tau \to 1$ straightforwardly and reproduce the mean density of the complex eigenvalues of
real Ginibre matrices (see eq.(7.47) in \cite{ByunFor_book}):
 \begin{equation}\label{denGinOEcomp}
p_N^{(GinOE,c)}(z)=\sqrt{\frac{2}{\pi}} |y|\erfc\left(\sqrt{2}\,|y|\right)\,e^{2y^2}\frac{\Gamma(N-1,|z|^2)}{\Gamma(N-1)},
\end{equation}
where $\erfc{(a)}=\frac{2}{\sqrt{\pi}}\int_a^{\infty}e^{-t^2}dt$. If however one considers the limit $\tau\to 1$ at $y=0$, one observes a divergence.
Such divergence reflects the presence of the $\delta-$function concentrated on the real axis:
  \begin{equation}\label{denGinOEfull}
p_N^{(GinOE)}(z)=p_N^{(GinOE,c)}(z)+\delta(y)\,\,p_N^{(GinOE,r)}(x)
\end{equation}
with the density $p_N^{(GinOE,r)}(x)$ of real eigenvalues known due to Edelman, Kostlan and Shub \cite{EKS},\cite{Ed:97}
 (see eq.(7.35) in \cite{ByunFor_book}):
 \begin{equation}\label{denGinOEreal}
p_N^{(GinOE,r)}(x)=\frac{1}{\sqrt{2\pi}\Gamma(N-1)}\left[\Gamma(N-1,x^2)+|x|^{N-1}e^{-\frac{x^2}{2}}\int_0^{|x|} u^{N-2}e^{-\frac{u^2}{2}}\,du\right].
\end{equation}
This singular part of the density is not accessible from Eq.(\ref{density_interpolating}) by simply letting $\tau\to1$, the reason being that the Gaussian
regulator $\epsilon$ introduced in Eq.(\ref{Gaudelta_vec}) cannot be removed in the sector $|{\bf v}^T{\bf v}|=1$ at $\tau=1$; see the discussion following
Eq.(\ref{quad_form_inverse}). That obstruction is however an artefact of insisting on the {\it complex} unit sphere: a real eigenvalue of a real matrix has a
{\it real} eigenvector, and the singular sector is recovered rigorously, and for every finite $N$, by running the same Kac-Rice argument on the real sphere
instead. This is carried out in Section~\ref{realsector} below, where the exact identity
\[
p_N^{(GinOE,r)}(x)=
\frac{e^{-x^2/2}}{2^{N/2}\,\Gamma\!\left(\frac{N}{2}\right)}\,
\mathbb{E}\left|\det\left(G^{(R)}_{N-1}-x{\bf 1}_{N-1}\right)\right|
\]
is recovered directly from the Kac--Rice representation. This identity goes
back to Edelman, Kostlan and Shub \cite{EKS}; evaluating the
expectation reproduces Eq.~(\ref{denGinOEreal}).

There is also an alternative, purely limiting, way of recovering this singular contribution. Namely, for any $\tau<1$ the mean eigenvalue density  $p^{(\tau)}_{N}(x,y)$  given by  Eq.(\ref{density_interpolating})
is non-singular everywhere in the complex plane, and on approaching $\tau\to 1$ describes gradual formation of the $\delta-$ functional
singularity around the real axis. Therefore by first fixing a small but nonzero value of $1-\tau$ and  integrating that density  over $y=\Im z$ in a narrow interval $|y|<\Delta_y$ whose width satisfies $\sqrt{1-\tau}\ll \Delta_y\ll 1$,   and after that taking the limit $\tau \to 1$ one recovers the singular contribution Eq.(\ref{denGinOEreal}).
In the scaling limit this is carried out explicitly in Proposition \ref{Propscaled} and Eq.(\ref{weaknonRden_equiv_lim}) below, where the coefficient
$1/\sqrt{2\pi}$ of $\delta(y)$ is reproduced; the corresponding computation at finite $N$ is left for a reader as a useful exercise. Note that
this coefficient agrees with the exact finite-$N$ value $p_N^{(GinOE,r)}(0)=1/\sqrt{2\pi}$, valid for every $N\ge 2$, see Corollary \ref{realzero} below.

\end{remark}

\begin{figure}
 \includegraphics[width=80mm]{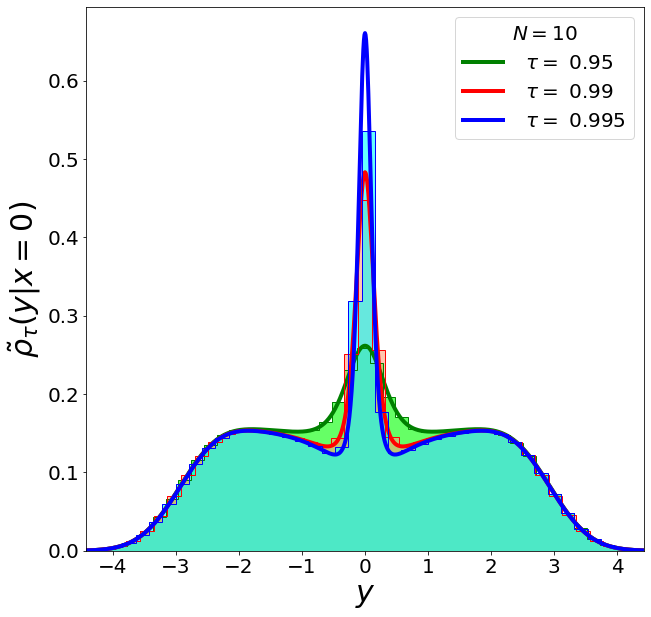}\includegraphics[width=80mm]{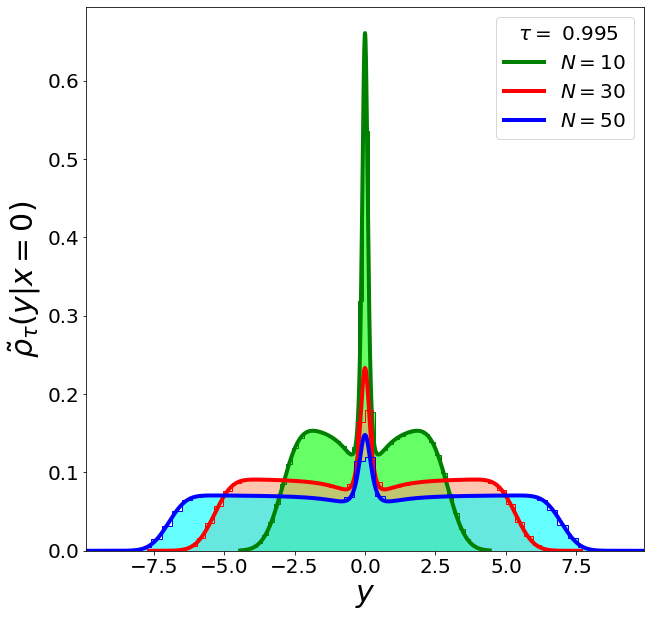}

 \caption{Formation of the peak in the eigenvalue density around the real line on approaching the real Ginibre limit. The function
  $\tilde{\rho}_{\tau}(y|x=0):=p_N^{\tau}(0,y)/\int_{-\infty}^{\infty}p_N^{\tau}(0,y)\,dy$ is plotted for a fixed matrix size $N$
  and changing $\tau$ or at a fixed $\tau$ and changing $N$.}
 \label{peakfig}
\end{figure}

It is instructive to plot the density $p^{(\tau)}_{N}(x,y)$ for $1-\tau\ll 1$ and changing $N$, or alternatively
fixing $N$ and changing $\tau$, see Fig.\ref{peakfig}. These plots suggest the existence in the limit $N\to \infty$ of a new scaling regime of  {\it weak non-reality}, characterized by the parameter
 $\lim_{N\to \infty}(1-\tau)N=\delta^2<\infty$ while keeping the distance $y=\Im z$ to the real axis of the order of unity.
This is proved in the following

\begin{proposition}\label{Propscaled}
%Consider a point with coordinates $x$ on the real axis, satisfying the condition $|x|<\sqrt{N}(1-\epsilon)$ for some fixed $\epsilon>0$.
%In the scaling regime of weak non-reality the density of eigenvalues in the complex plane in the vicinity of such a point is given by
For every fixed $\delta>0$, $|\tilde{x}|<1$ and $y\in \mathbb{R}$
\[
p^{(s)}_{\delta}(\tilde{x},y):=\lim_{N\to \infty} p^{\left(\tau=1-\frac{\delta^2}{N}\right)}_{N}\left(x=\sqrt{N}\tilde{x},y\right)
\]
\begin{equation}\label{weaknonRden}
=\frac{2}{\pi} \int_1^{\infty}dt\,e^{(1-t^2)\left(2y^2+\delta_x^2\right)}\left(\delta_x^2t^2+2y^2\right),  \quad
\delta_x=\delta\sqrt{1-\tilde{x}^2}.
\end{equation}
\end{proposition}
\begin{remark}
For $\delta=0$ the same formula may be taken pointwise as long as $y\ne0$.  At the singular point $\delta=0$, $y=0$ the limiting spectral measure has a component supported on the real axis, so no ordinary two-dimensional density exists there; the correct statement is the distributional limit in Eq.~\eqref{weaknonRden_equiv_lim} below.  Eq.(\ref{weaknonRden}) gives the density of eigenvalues in the bulk near the real line in the scaling limit of 'weak non-reality'. Integrating by parts one can further  represent $p^{(s)}_{\delta}(\tilde{x},y)$ in an alternative form as
\begin{equation}\label{weaknonRden_equiv}
p^{(s)}_{\delta}(\tilde{x},y)=\frac{1}{\pi}\frac{\delta^2_x}{2y^2+\delta^2_x}+\frac{1}{\sqrt{\pi}}
\left[\frac{2y^2}{\sqrt{2y^2+\delta^2_x}}+\frac{\delta^2_x}{2(2y^2+\delta^2_x)^{3/2}}\right]e^{2y^2+\delta^2_x}\erfc\left(\sqrt{2y^2+\delta^2_x}\right).
\end{equation}
\end{remark}

The expression Eq.(\ref{weaknonRden_equiv}) clearly describes in the limit $\delta_x\to 0$ the formation of the Dirac $\delta-$function on the real axis,
namely:
\begin{equation}\label{weaknonRden_equiv_lim}
\lim_{\delta \to 0} p^{(s)}_{\delta}(\tilde{x},y)=\sqrt{\frac{2}{\pi}}\,|y|
e^{2y^2}\erfc\left(\sqrt{2}|y|\right)+\frac{1}{\sqrt{2\pi}}\delta(y),
\end{equation}
where we used the distributional identity
\[
\lim_{\delta_x \to 0}\frac{\delta^2_x}{(2y^2+\delta^2_x)^{3/2}}=\sqrt{2}\delta(y),
\]
whose constant follows from $\int_{-\infty}^{\infty}(u^2+1)^{-3/2}du=2$ after the substitution $y=\delta_xu/\sqrt{2}$. Note that the first, Lorentzian, term of
Eq.(\ref{weaknonRden_equiv}) does {\it not} contribute a second $\delta-$function: it carries the total mass
$\frac{1}{\pi}\int_{-\infty}^{\infty}\frac{\delta_x^2\,dy}{2y^2+\delta^2_x}=\frac{\delta_x}{\sqrt{2}}$, which vanishes as $\delta_x\to0$.
The limiting formula Eq.(\ref{weaknonRden_equiv_lim}) matches exactly the $N\to \infty$ limit of Eq.(\ref{denGinOEfull}), taken at
 scaled $x=\sqrt{N}\tilde{x}$ and fixed $y$.

\begin{remark}
It is natural to expect  the density profile in the  scaling regime described in the Proposition \ref{Propscaled} to be universal in the limit $N\to \infty$
 for a class of "weakly non-real" random matrices of the form $X=X_1+i\frac{\delta}{\sqrt{2N}}X_2$, where the parameter $\delta\in \mathbb{R}$ is fixed and  $X_1$ and $X_2$ are two $N\times N$ random real-valued {\it independent entry} matrices as defined e.g. in the paper \cite{TaoVu2015}.   Moreover, one may further conjecture  that in such a scaling regime not only the mean density, but also the higher spectral correlation functions should be described by expressions nontrivially interpolating between those typical for real eigenvalues of real Ginibre matrices and those typical for the complex eigenvalues in the spectral bulk.  Even in the case of interpolating Ginibre ensemble it is not yet clear if the corresponding eigenvalues will be described statistically by an integrable structure, e.g. of a Pfaffian process. Shedding further light on the spectral correlations in such a regime  seems currently a challenging problem, and will certainly require new insights for further progress.

\end{remark}

\subsubsection{Right eigenvector statistics for the interpolating ensemble}
 In the limiting case of the complex Ginibre ensemble $G_{\tau=0}$ the statistics of right eigenvectors is as trivial as for classical Hermitian GUE ensemble.
Indeed, Eq.(\ref{MainJPDinterpolating})  implies that $D=1$ for  $\tau=0$, and JPD becomes completely independent of ${\bf v}$
 so that the right eigenvectors are simply uniformly distributed over the unit sphere. This fact reflects the unitary-conjugation invariance of the complex Ginibre ensemble, $G_{\tau=0}\mapsto UG_{\tau=0}U^*$ with $U\in U(N)$. Note that in this case for large $N$ we have typically $|{\bf v}^T{\bf v}|^2=O(N^{-1})$.

 For any $0<\tau\le 1$ the unitary invariance is violated and the right eigenvectors for finite $N$ become non-trivially distributed over the complex unit sphere.
Still one may naturally expect that as long as $\tau<1$ is fixed as $N\to \infty$,  for $z$ belonging to the support of the limiting spectral density the quantity $\tilde{u}=N|{\bf v}^T{\bf v}|^2$ will remain asymptotically of the order of unity. To this end we scale $z=\sqrt{N}\tilde{z}$ and define the limiting JPD of $\tilde{u}$ and $\tilde{z}$ via
 \begin{equation}\label{right_eigenvec_stat_lim_def1}
 {\cal P}^{(\tau)}(\tilde{z},\tilde{u})=\lim_{N\to \infty}\int_{\mathbf{S}^{(c)}_N} {\cal P}^{(\tau)}_N(\sqrt{N}\tilde{z},{\bf v})\delta(\tilde{u}-N|{\bf v}^T{\bf v}|^2)d\mu_H({\bf v}).
 \end{equation}
\begin{equation}\label{right_eigenvec_stat_lim_def2}
=\lim_{N\to \infty}\frac{N-1}{2N}\left(1-\frac{\tilde{u}}{N}\right)^{\frac{N-3}{2}}{\cal P}^{(\tau)}_N(\sqrt{N}\tilde{z},{\bf v})\left|_{|{\bf v}^T{\bf v}|^2= \frac{\tilde{u}}{N}}\right.
 \end{equation}
where in the second line we have used the identity Eq.(\ref{sphereintegral}). Exploiting Eq.(\ref{MainJPDinterpolating})
 the limit can be straightforwardly evaluated yielding
 \begin{proposition}
For fixed $0\le\tau<1$ and $|\tilde z|<1$, the limiting joint intensity of $\tilde z$ and $\tilde u=N|{\bf v}^T{\bf v}|^2$ is
\begin{equation}\label{right_eigenvec_stat_lim}
 {\cal P}^{(\tau)}(\tilde{z},\tilde{u})=
 \frac{a_\tau(\tilde{z})}{2\pi}\,
 \exp{-\frac{a_{\tau}(\tilde{z})}{2}\,\tilde u},
 \qquad
 a_{\tau}(\tilde{z})=1-2\tau\Re \left(\tilde{z}^2\right)+\tau^2\left(2|\tilde{z}|^2-1\right).
\end{equation}
Consequently, conditionally on an eigenvalue at the macroscopic point $\tilde z$, the variable $\tilde u$ has density
$\frac12a_\tau(\tilde z)e^{-a_\tau(\tilde z)\tilde u/2}$ on $\mathbb R_+$.  At $\tau=0$ this reduces to the Haar-sphere law $\frac12e^{-\tilde u/2}$.
\end{proposition}

 \begin{remark}
In fact the above expression is also valid for $\tau=1$ as long as $\Im \tilde{z}\ne 0$, in which case $a_{\tau=1}(\tilde{z})=4\left[\Im \tilde{z}\right]^2>0$.
At the same time   $a_{\tau=1}(\tilde{z})\to 0$ when approaching the real axis signalling that the typical values of the variable $\tilde{u}$ become indefinitely large.
 This naturally suggests that for the scaling regime of  {\it weak non-reality} with $\tau=1-\delta^2/N$ close to the real axis, i.e.
$|\Re \tilde{z}|<1,\, y=\Im z=O(1)$ the variable $u=|{\bf v}^T{\bf v}|^2$ is typically of the order of unity rather than $O(1/N)$ as elsewhere in the complex plane.
A straightforward treatment of Eq.(\ref{MainJPDinterpolating}) in such a regime shows that the corresponding right eigenvector distribution is described by the following
\begin{proposition}\label{weak nonreal_eigenvector}
For fixed $\delta>0$, $|\tilde{x}|<1$ and $y\in \mathbb{R}$ the limiting joint intensity of the variable $u=|{\bf v}^T{\bf v}|^2\in [0,1]$ and the eigenvalue
at $z=\sqrt{N}\tilde{x}+iy$ is given as $N\to \infty$ by
\[
{\cal P}^{(s)}_{\delta}(\tilde{x},y;u):=\lim_{N\to \infty} \int_{\mathbf{S}^{(c)}_N} \mathcal{P}^{\left(\tau=1-\frac{\delta^2}{N}\right)}_{N}\left(z=\sqrt{N}\tilde{x}+iy,{\bf v}\right)\delta(u-|{\bf v}^T{\bf v}|^2)d\mu_H({\bf v}).
\]
\begin{equation}\label{weaknonvec}
=\frac{1}{\pi(1-u)^{3/2}}\,e^{-\frac{u}{1-u}\left(2y^2+\delta_x^2\right)}\left(\frac{\delta_x^2}{1-u}+2y^2\right),  \quad
\delta_x=\delta\sqrt{1-\tilde{x}^2}.
\end{equation}
\end{proposition}
For $\delta=0$ and $y\ne0$ the formula has the corresponding pointwise real-Ginibre limit.  At $\delta=0$, $y=0$ it should again be understood only through the singular limiting measure rather than as an ordinary density in $(y,u)$.
It is easy to check that integrating ${\cal P}^{(s)}_{\delta}(\tilde{x},y;u)$ over $u\in[0,1]$ reproduces the eigenvalue density Eq.(\ref{weaknonRden}).
It is also instructive to consider the real Ginibre limit $\delta\to 0$ and define the corresponding probability density $p_y(u)$ of the variable $u=|{\bf v}^T{\bf v}|^2\in[0,1]$ at the point with distance $y\ne 0$ from the real axis as:
\begin{equation}\label{deltazero}
p_y(u)=\lim_{\delta\to 0}\frac{{\cal P}^{(s)}_{\delta}(\tilde{x},y;u)}{p^{(s)}_{\delta}(\tilde{x},y)}=\frac{e^{-\frac{2y^2}{1-u}}}{2A(y)(1-u)^{3/2}}, \quad A(y)=\int_1^{\infty}e^{-2y^2t^2}\,dt.
\end{equation}
We then can see that on approaching the real axis holds  $\lim_{y\to 0}p_y(u)=\delta(u-1)$, indicating that, by orthogonal invariance, the corresponding right eigenvectors become uniformly distributed on the real unit sphere.
\end{remark}

\subsection{ Case 2 - ensemble of additive deformations of complex Ginibre matrices}
Given an arbitrary fixed $N\times N$ matrix $A$ consider matrices $G_{A}$
\begin{equation}\label{def_additive_def}
G_{A}=G^{(C)}+A
\end{equation}
where $G^{(C)}$ is taken from the ensemble of complex Ginibre matrices corresponding to the choice $\tau=0$ in Eq. (\ref{def_interpolating}).
The resulting ensemble for $G_{A}$ is known as the {\it additive deformation} of the complex Ginibre ensemble, and various properties of the corresponding matrices attracted considerable attention over the years, predominantly for $N\to \infty$,  see e.g. \cite{SS_deformed_22,Alt_Krueger_deform1,Alt_Krueger_deform2,LiuZhang2024,LiuZhang2025,Sarapin25} and references therein
(for additive deformations of real asymmetric random matrices see e.g. \cite{Khor1996}).
The deformation breaks the mentioned residual unitary rotational symmetry present in the complex Ginibre matrices. As such it may in general result not only in non-rotationally-invariant
shape of the limiting density of complex eigenvalues, but also generically ensures
a nontrivial dependence of the corresponding JPD on the eigenvectors. Our main result is summarized in the following

   \begin{theorem}\label{Theorem2}
   Let $A_z:=A-z{\bf 1}_N$. The one-point joint intensity of an eigenvalue $z$ and the corresponding phase-averaged normalized right eigenvector ${\bf v}$, with respect to $d^2z\,d\mu_H({\bf v})$, for matrices taken from the additively deformed complex Ginibre ensemble Eq.(\ref{def_additive_def})
  is given by
\begin{equation}\label{MainJPDadditivedef_intro}
{\cal P}^{(A)}_N(z,{\bf v})=\frac{1}{\pi \Gamma(N)}\,e^{-{\bf v}^*A_z^*A_z{\bf v}}\int_0^{\infty}dR\, e^{-R}\,\det{\left[R\,{\bf 1}_N+A_z^*A_z\right]}
\end{equation}
\[
\times \left[R\left({\bf v}^*\frac{1}{R\,{\bf 1}_N+A_z^*A_z}{\bf v}\right)\left({\bf v}^*\frac{1}{R\,{\bf 1}_N+A_zA^*_z}{\bf v}\right)+
\left|\left({\bf v}^*A_z\frac{1}{R\,{\bf 1}_N+A_z^*A_z}{\bf v}\right)\right|^2\right]
\]

\end{theorem}

The proof of Theorem \ref{Theorem2} is given in the Section 4.

\begin{remark}

Assuming the perturbation matrix $A$ to be {\it normal}, that is commuting with its adjoint: $A_z^*A_z=A_zA_z^*$, the expression Eq.(\ref{MainJPDadditivedef_intro}) %Eq.(\ref{mean_dens_deformed_gen})
 takes a somewhat simpler form. Namely, the matrix $A$ can be diagonalized by a unitary rotation: $A=U^{-1}\mbox{diag}(a_1,\ldots,a_N)\,U$.
  To this end, denoting $d_{n}:=z-a_n$ and $\tilde{\bf v}:=U{\bf v}$ and further introducing $g_i=|d_i|^2$ and  $t_i=\overline{\tilde{v}}_i \tilde{v}_i$ and
\begin{equation}
F(g_1,\ldots,g_N)=\int_0^{\infty}e^{-R}\prod_{n=1}^N(R+g_n)\,dR=\sum_{n=0}^N(N-n)!e_n(g_1,\ldots,g_N),
  \end{equation}
  where $e_n(g_1,\ldots,g_N)$ are elementary symmetric polynomials, and
  remembering the normalization $\sum_it_i=1$ one obtains after straightforward algebraic manipulations
 \begin{eqnarray}\label{mainJPD_norm}
 {\cal P}^{(A,norm)}_N(z,{\bf v})=\frac{1}{\pi \Gamma(N)}e^{-\sum_{i=1}^Ng_it_i}\left[\sum_{k=1}^Nt_k\frac{\partial F}{\partial g_k}-\sum_{k<l}^Nt_kt_l|d_k-d_l|^2\frac{\partial^2 F}{\partial g_k\partial g_l}\right]\\
 =-\frac{1}{\pi \Gamma(N)}\left[\sum_{k=1}^N\frac{\partial F}{\partial g_k}\frac{\partial}{\partial g_k}+\sum_{k<l}^N|d_k-d_l|^2\frac{\partial^2 F}{\partial g_k\partial g_l}\frac{\partial^2 }{\partial g_k\partial g_l}\right]
 e^{-\sum_{i=1}^Ng_it_i}
  \end{eqnarray}
Here and below the derivatives $\partial/\partial g_k$ appearing to the right of the coefficients $\partial F/\partial g_k$ and $\partial^2F/\partial g_k\partial g_l$ act only on the exponential factor, those coefficients being held fixed.
The last representation makes the integration over the eigenvector ${\bf v}$ against the Haar's measure $d\mu_H({\bf v})$ especially simple, as
it amounts to the special rank-one case of the Harish--Chandra--Itzykson--Zuber (HCIZ) formula~\cite{HC1956,IZ}, see Eq.(\ref{IZHC_rank1}). In this way one
arrives at the following explicit expression for the mean eigenvalue density for complex Ginibre matrices additively deformed by a normal matrix $A$:
   \begin{equation}\label{mean_dens_deformed_norm}
p_N^{(A,norm)}(z)=-\frac{1}{\pi}\left[\sum_{k=1}^N\frac{\partial F}{\partial g_k}\frac{\partial}{\partial g_k}+\sum_{k<l}^N|d_k-d_l|^2\frac{\partial^2 F}{\partial g_k\partial g_l}\frac{\partial^2 }{\partial g_k\partial g_l}\right]G(g_1,\ldots,g_N),
\end{equation}
where we defined
\begin{equation}
G(g_1,\ldots,g_N)=\sum_{i=1}^N \frac{e^{-g_i}}{\prod_{j\ne i}^N(g_j-g_i)}.
\end{equation}
When some of the $g_i$ coincide, this expression and the formulas containing it are understood by their unique confluent divided-difference continuation; no distinctness assumption is imposed on the final density.
In particular, for the simplest non-trivial case $N=2$ we have 
\[
F(g_1,g_2)=2+g_1+g_2+g_1g_2, \quad G(g_1,g_2)=\frac{e^{-g_2}-e^{-g_1}}{g_1-g_2}
\]
so that the formula Eq.(\ref{mean_dens_deformed_norm}) takes the form
 \begin{equation}\label{mean_dens_deformed_norm N=2}
p_2^{(A,norm)}(z)=\frac{1}{\pi}\left\{\frac{g_1e^{-g_2}-g_2e^{-g_1}}{g_1-g_2}-|d_1-d_2|^2\left[\frac{e^{-g_1}+e^{-g_2}}{(g_1-g_2)^2}+
2\frac{e^{-g_1}-e^{-g_2}}{(g_1-g_2)^3}\right]\right\}. \end{equation}\\[1ex]

Hikami and Pnini \cite{HP1996}, using a completely different method, derived for the normal additive perturbation case an alternative form of the mean eigenvalue density.  After converting their unit-radius normalization to the variance-one Ginibre convention used here, the formula reads
\begin{equation}\label{HikamiPnini}
p_N^{(HP)}(z)=\frac{1}{\pi}\frac{\partial^2}{\partial z\partial \overline{z}}\int_0^1\frac{dp}{p^2}\oint_{\gamma}
\frac{dt}{2\pi i}\frac{e^{-pt}}{t}\prod_{n=1}^N\left(1-\frac{pt}{t-g_n}\right),
\end{equation}
where $\gamma$ encircles $0,g_1,\ldots,g_N$.  The formula is understood with a lower cutoff $p\ge\eta>0$, the derivatives being applied before $\eta\downarrow0$.  The divergent terms are independent of $z$ and therefore disappear; Appendix~E gives an explicitly finite subtraction.

On the open set where the $g_n$ are pairwise distinct the contour integral can be evaluated by residues.  Appendix~E proves that the Hikami--Pnini formula is exactly equivalent, for every finite $N$, to the Kac--Rice expression Eq.~\eqref{mean_dens_deformed_norm}; it also explains the confluent continuation across the collision curves $g_i=g_j$.  The explicit $N=2$ expression Eq.~\eqref{mean_dens_deformed_norm N=2} is simply the first non-trivial special case of that general identity.

\end{remark}

Coming back to the general case of non-normal matrices, we integrate the exact JPD Eq.~(\ref{MainJPDadditivedef_intro}) over the complex sphere.  Here one needs the weighted, rather than the bare, fourth moment on the sphere.  It is convenient to keep the same Fourier--Bromwich representation which already appeared above.  With $H=A_z^*A_z$, $Q_s=(is{\bf 1}_N+H)^{-1}$ and arbitrary matrices $B,C$, one has
\[
\int_{\mathbf S_N^c}e^{-{\bf v}^*H{\bf v}}({\bf v}^*B{\bf v})({\bf v}^*C{\bf v})\,d\mu_H({\bf v})
=\Gamma(N)\int_{\mathbb R-i0}e^{is}\frac{\operatorname{Tr}(Q_sB)\operatorname{Tr}(Q_sC)+\operatorname{Tr}(Q_sBQ_sC)}{\det(is{\bf 1}_N+H)}\,\frac{ds}{2\pi}.
\]
Indeed, this follows by inserting the Fourier representation of $\delta({\bf v}^*{\bf v}-1)$ and performing the ordinary complex Gaussian fourth-moment contraction.  The first term is the disconnected contraction and the second one is the connected contraction; the latter is essential for a general non-normal perturbation.  The familiar unweighted sphere identity is recovered by setting $H=0$.  To write the result compactly set
\[
H:=A_z^*A_z,\qquad \widetilde H:=A_zA_z^*,\qquad Q_s:=(is{\bf 1}_N+H)^{-1},
\]
\[
B_R:=(R{\bf 1}_N+H)^{-1},\qquad C_R:=(R{\bf 1}_N+\widetilde H)^{-1},
\]
and, for any two $N\times N$ matrices $X,Y$,
\begin{equation}\label{Xi_def}
\Xi_s(X,Y):=\operatorname{Tr}(Q_sX)\operatorname{Tr}(Q_sY)
+\operatorname{Tr}(Q_sXQ_sY).
\end{equation}
Then one obtains the following exact finite-$N$ marginal.
\begin{corollary}\label{cor_dens_deformed}
The mean density of complex eigenvalues for the deformed Ginibre ensemble is
\begin{equation}\label{mean_dens_deformed_gen}
\begin{split}
p_N^{(A)}(z)=\frac1\pi\int_0^{\infty}e^{-R}\,dR
\int_{\mathbb R-i0} e^{is}
\frac{\det(R{\bf 1}_N+H)}{\det(is{\bf 1}_N+H)}
\Big[&R\,\Xi_s(B_R,C_R)\\
&+\Xi_s(A_zB_R,B_RA_z^*)\Big] \,\frac{ds}{2\pi}.
\end{split}
\end{equation}
\end{corollary}
As a check, for $A=0$ Eq.~\eqref{mean_dens_deformed_gen} reduces to
\[
p_N^{(0)}(z)=\frac{\Gamma(N,|z|^2)}{\pi\Gamma(N)}
=\frac{e^{-|z|^2}}{\pi}\sum_{k=0}^{N-1}\frac{|z|^{2k}}{k!},
\]
the standard finite-$N$ complex Ginibre density.

\begin{remark}

The expression Eq.(\ref{mean_dens_deformed_gen}) is ideally suited for extracting the limiting eigenvalue density as $N\to \infty$ by the saddle-point method: under appropriate conditions on existence of the limiting normalized eigenvalue counting measure $\nu_z$ for the matrices $A_zA_z^*$ the saddle-point in variables $R$ and $s$ is at $R=is=R_0>0$, with $R_0$ solving the equation
$1=\int\frac{d\nu_z(\lambda)}{(\lambda+R_0)}$. The corresponding computation is standard and yields the expression coinciding with one obtained by two different methods, first in
\cite{Zhong_Ping21} and then more recently in \cite{Sarapin25}, see Theorem~1.1 and Corollary~1.2 in the latter paper, for a class of matrices $G$ with i.i.d. complex entries, under appropriate conditions (see also a detailed study of possible edge singularities of the limiting density in \cite{Alt_Krueger_deform2}). For convenience of the reader we present below the corresponding density statement for the deformed Ginibre case.

\begin{theorem*} ({\bf Sarapin}, \cite{Sarapin25})\label{SarapinTh}
Define ${\cal Y}_0(z)=A_zA_z^*$ and $\tilde{\cal Y}_0(z)=A^*_zA_z$. Assume that the normalized counting measure of ${\cal Y}_0(z)$ converges weakly as $N\to \infty$ (in the almost sure sense) to some deterministic measure $\nu_z$ for almost all $z\in \mathbb{C}$. Denote $\sigma_0=\{z|0\in \mbox{supp}\nu_z\}$.
Then, under the additional technical assumptions of \cite{Sarapin25}, one can prove that for each $R>0$ and $z\in \mathbb{C}$ there exist continuous functions defined by the following limits.  We stress the normalization because it is essential here: $\operatorname{tr}_N:=N^{-1}\operatorname{Tr}$.
\begin{equation}
T_1(z,R)=\lim_{N\to \infty}\operatorname{tr}_N A_z\left({\cal Y}_0(z)+R\right)^{-2}, \quad T_2(z,R)=\lim_{N\to \infty}\operatorname{tr}_N \left({\cal Y}_0(z)+R\right)^{-1}
\left(\tilde{{\cal Y}}_0(z)+R\right)^{-1}.
\end{equation}
The normalized counting measure of complex eigenvalues of the deformed complex Ginibre Ensemble converges weakly in almost sure sense to the deterministic measure $\mu$
supported on the domain $D=\sigma_0\bigcup\{z\in \mathbb{C}\backslash\sigma_0: \int \frac{d\nu_z(\lambda)}{\lambda}\ge 1\}$ with the density
\begin{equation}
\rho_{\mu}(z)=\frac{1}{\pi}\left(R_0\cdot T_2(z,R_0)+\frac{|T_1(z,R_0)|^2}{\int\frac{d\nu_z(\lambda)}{(\lambda+R_0)^2}} \right)
\end{equation}
where $R_0>0$ satisfies the equation  $\int\frac{d\nu_z(\lambda)}{(\lambda+R_0)}=1$.
\end{theorem*}

\end{remark}

\subsubsection{Non-Hermitian Rosenzweig-Porter model}
In this section we present results for a special case of additive diagonal perturbation of interest in physics of disordered systems. This is the so-called non-Hermitian
Rosenzweig-Porter (NHRP) model introduced and discussed at physical level of rigour by De Tomasi and Khaymovich in \cite{TomasiKhaimovich_NHRP22}.
We warn the reader that, in contrast with the rest of the paper, the objects appearing in this subsection are normalized to unit total mass rather than to
the total mass $N$ of an intensity; the normalizations are stated explicitly in Eq.(\ref{NHRP_finite_normalized_densities}) below. This case is covered by the following

\begin{theorem}\label{NHRP}
Consider
\[
G_{NHRP}=\frac{1}{\sqrt N}G^{(C)}+A,
\qquad A=\operatorname{diag}(a_1,\ldots,a_N),
\]
where $G^{(C)}$ is standard complex Ginibre and the $a_i$ are i.i.d. centered isotropic complex Gaussians in the convention
\[
\mathbb P(a_i\in d^2a)=\frac{1}{2\pi\sigma^2}e^{-|a|^2/(2\sigma^2)}d^2a,
\]
so that each real component has variance $\sigma^2$. Keep $\sigma^2>0$ fixed as $N\to\infty$. Let $z_a^{(N)}$ denote the eigenvalues and let ${\bf v}_a$ be phase-averaged unit right eigenvectors. Define the normalized mean eigenvalue density and the normalized marked density of $p=N|({\bf v}_a)_1|^2$ by
\begin{equation}\label{NHRP_finite_normalized_densities}
\rho_N^{NHRP}(w):=\frac1N\,\mathbb E\sum_{a=1}^N\delta^{(2)}(w-z_a^{(N)}),
\qquad
\widehat{\mathcal P}_N^{NHRP}(w,p):=\frac1N\,\mathbb E\sum_{a=1}^N
\delta^{(2)}(w-z_a^{(N)})\delta\!\left(p-N|({\bf v}_a)_1|^2\right).
\end{equation}
Thus $\int_{\mathbb C}\rho_N^{NHRP}(w)d^2w=1$ and
$\int_0^\infty\widehat{\mathcal P}_N^{NHRP}(w,p)dp=\rho_N^{NHRP}(w)$. Whenever $\rho_N^{NHRP}(w)>0$, the conditional density at the spectral point $w$ is understood in the density-ratio, or Palm, sense,
\[
\pi_N(p\mid w):=\frac{\widehat{\mathcal P}_N^{NHRP}(w,p)}{\rho_N^{NHRP}(w)}.
\]

For fixed $w\in\mathbb C$, let $R_*=R_*(w,\sigma)>0$ be the unique solution of
\begin{equation}\label{NHRosenzweigPorter_R}
1=\frac{1}{2\sigma^2} e^{-\frac{|w|^2}{2\sigma^2}} \int_0^{\infty}\frac{1}{1+t}I_0\left(\frac{|w|}{\sigma}\sqrt{\frac{tR_*}{\sigma^2}}\right)e^{-\frac{R_*t}{2\sigma^2}}\,dt,
\end{equation}
and for $\nu\ge0$ put
\begin{equation}\label{NHRosenzweigPorter_T}
T_{\nu}(w,R_*)=\int_0^{\infty}\frac{t^{\nu/2}}{(1+t)^2}I_{\nu}\left(\frac{|w|}{\sigma}\sqrt{\frac{tR_*}{\sigma^2}}\right)e^{-\frac{R_*t}{2\sigma^2}}\,dt.
\end{equation}
Then, as $N\to\infty$ at fixed $(w,\sigma)$,
\begin{equation}\label{NHRosenzweigPorter2}
\rho_N^{NHRP}(w)\longrightarrow
\rho^{NHRP}_{\infty}(w)=\frac{T_0(w,R_*)}{2\pi\sigma^2}e^{-\frac{|w|^2}{2\sigma^2}}
\left(1+\left(\frac{T_1(w,R_*)}{T_0(w,R_*)}\right)^2\right).
\end{equation}
Moreover $\pi_N(\,\cdot\mid w)$ converges weakly, as a probability measure on $\mathbb R_+$, to the measure with density
\begin{equation}\label{NHRosenzweigPorter1}
\pi_{\infty}(p\mid w)=-\frac{d}{dp}\Phi_w(p),
\qquad
\Phi_w(p):=\frac{1}{1+2p\sigma^2}\,e^{-p\left(R_*+\frac{|w|^2}{1+2p\sigma^2}\right)},
\qquad p\ge0.
\end{equation}
For every fixed $q>-1$ the corresponding conditional moments converge as well, and
\begin{equation}\label{NHRosenzweigPorter1mom}
\lim_{N\to\infty}\int_0^\infty p^q\pi_N(p\mid w)\,dp
=\mathbb E_{\infty}[p^q\mid w]
=\Gamma(q+1)\frac{e^{-\frac{|w|^2}{2\sigma^2}}}{2\sigma^2R_*^{q-1}}
\int_0^{\infty}\frac{1}{(1+t)^q}I_0\left(\frac{|w|}{\sigma}\sqrt{\frac{tR_*}{\sigma^2}}\right)e^{-\frac{R_*t}{2\sigma^2}}\,dt.
\end{equation}
\end{theorem}

   Proof of this Theorem is given in the Section \ref{NHRPproof}.

\begin{remark}
The above Theorem complements and essentially extends findings by De Tomasi and Khaymovich presented in \cite{TomasiKhaimovich_NHRP22}.
 Those authors parametrized the variance of diagonal entries of the perturbation matrix as $\sigma^2=\frac{1}{\lambda}N^{\gamma-1}, \, \lambda>0, \gamma>0$
 and provided theoretical and numerical arguments that all eigenvectors are localized for $\gamma>1$ and delocalized/extended for $\gamma<1$.
 The case of $N-$independent $\sigma^2<\infty$ in their terminology is exactly the critical case $\gamma=1$ where the methods used in
 \cite{TomasiKhaimovich_NHRP22} do not seem to work. In this sense our study of the problem presented above is exactly complementary and provides the missing information
 about this critical behaviour of the model. In fact the overall picture for large $N\gg 1$ turns out to be even more nuanced, see the analysis below the Remark \ref{logNHRZ}.
\end{remark}

\vspace{0.2cm}

   Now we proceed with analyzing the solutions $R_*$ to Eq.(\ref{NHRosenzweigPorter_R}) as the function of parameters $(w,\sigma^2)$ in the two limiting cases: the unperturbed Ginibre limit $\sigma^2\to 0$  and the opposite case of strong diagonal perturbation $\sigma^2 \gg 1$.

 In the first case one should exploit the large-argument asymptotic for the Bessel function: $I_{\nu}(x)\approx \frac{e^{x}}{\sqrt{2\pi x}}$ and finally use that $\forall a\in \mathbb{R}$ holds $\lim_{\sigma \to 0}\frac{1}{\sigma\sqrt{2\pi}}e^{-a^2/2\sigma^2}=\delta(a)$. This procedure reduces Eq.(\ref{NHRosenzweigPorter_R}) to  $1=(R_*+|w|^2)^{-1}$, which has the positive solution  $R_*=1-|w|^2$ as long as $|w|<1$; equivalently, and more directly, $F(R)=\left\langle(R+|a-w|^2)^{-1}\right\rangle\to(R+|w|^2)^{-1}$ as $\sigma\to0$. For these values of $|w|$  the mean density then follows from Eq.(\ref{fin_denA}): in the same limit $G_*\to(R_*+|w|^2)^{-2}=1$ and $\left\langle d(R_*+|d|^2)^{-2}\right\rangle\to w$, so that
\[
\rho^{NHRP}_{\infty}(w)=\frac{G_*}{\pi}\left[R_*+\left|\frac{1}{G_*}\left\langle\frac{d}{(R_*+|d|^2)^2}\right\rangle\right|^2\right]
\longrightarrow\frac{1}{\pi}\left[\left(1-|w|^2\right)+|w|^2\right]=\frac{1}{\pi}.
\]
Note that the circular law arises here from the two terms of Eq.(\ref{fin_denA}) combining, and not from either of them separately. We thus reproduced the standard result for the density of unperturbed Ginibre matrix. Not surprisingly,  the
 Eq. (\ref{NHRosenzweigPorter1}) in this case implies the limiting probability density for the variable $p=N|v_1|^2$  given by $\mathcal{P}^{NHRP}(p;w)=e^{-p}$. Such density with the moments $\mathbb{E}(p^q)=\Gamma(q+1)$ is characteristic of the right eigenvector  ${\bf v}$ uniformly distributed over the complex unit sphere.

\begin{figure}
 \includegraphics[width=40mm]{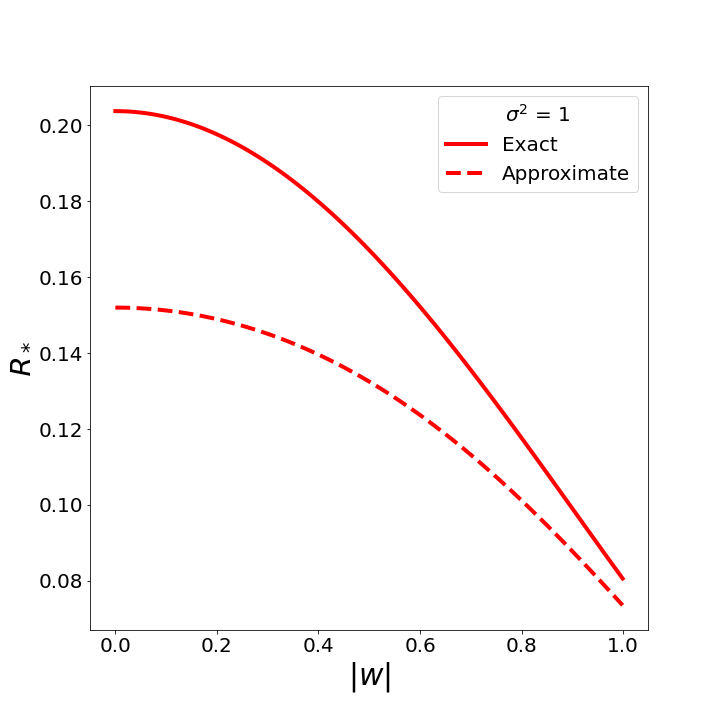}\includegraphics[width=40mm]{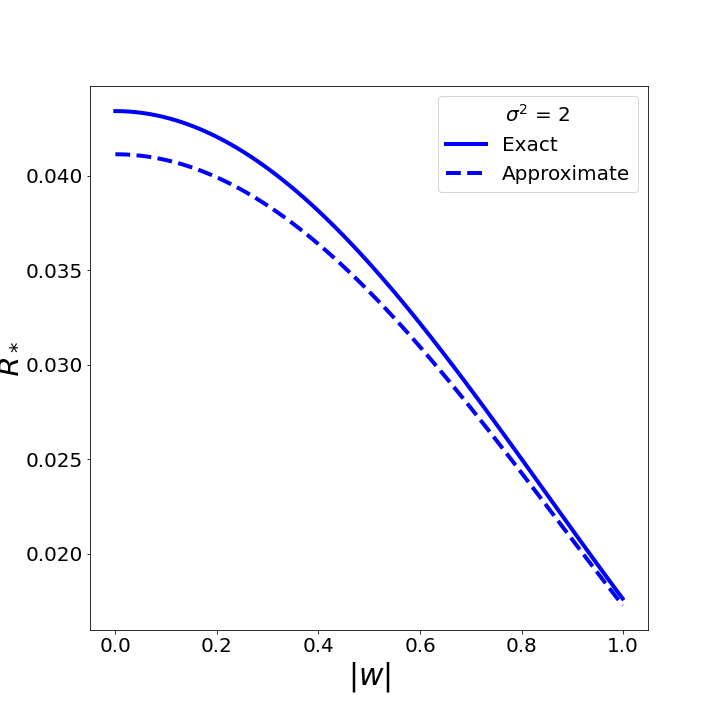}\includegraphics[width=40mm]{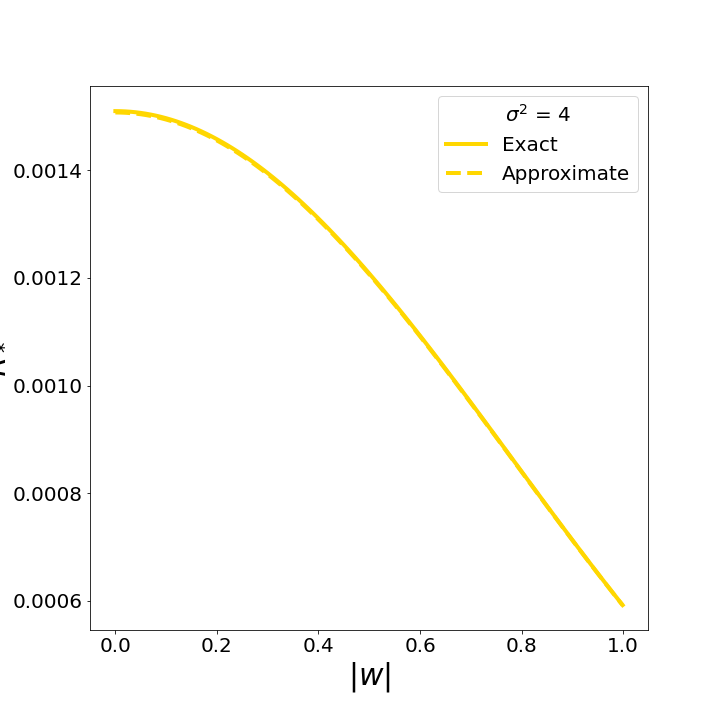}
 \includegraphics[width=40mm]{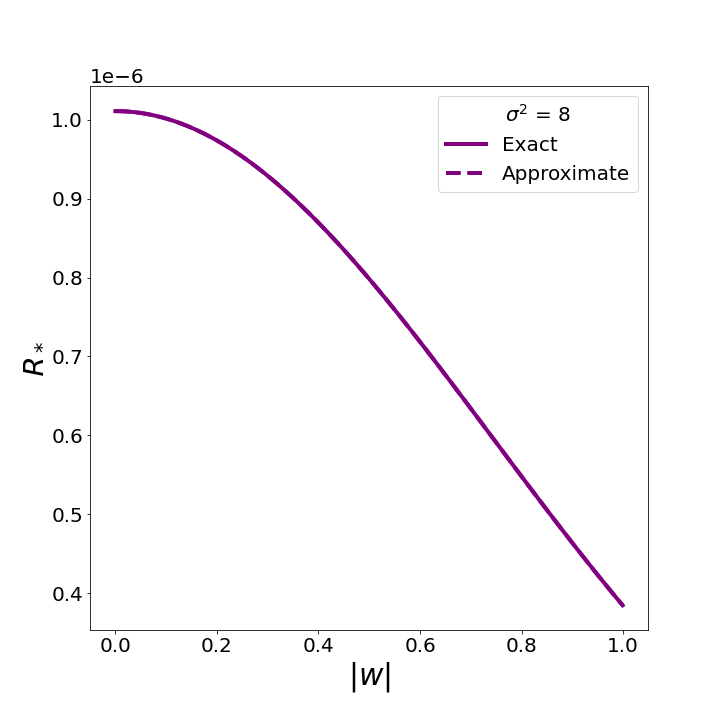}

 \caption{ Comparison between the asymptotic approximation Eq.(\ref{NHRosenzweigPorter_R_asy}) and the exact numerical solution of the equation (\ref{NHRosenzweigPorter_R})
  for different values of parameter $\sigma^2$.}

  \label{asy_vs_exact}
\end{figure}

We next take a second limit, after the fixed-$\sigma$ limit $N\to\infty$ of Theorem~\ref{NHRP}: namely $\sigma\to\infty$, with $|w|/\sigma$ kept bounded. Thus the notation $\sigma\gg1$ below never means an $N$-dependent variance unless this is stated explicitly.
It is expedient to introduce the parameter $a=\frac{R_*}{2\sigma^2}$, anticipating $a\to 0$ as $\sigma^2\to \infty$.
We also find it to be convenient to shift variables $t=\frac{x}{a}-1$ in Eq.(\ref{NHRosenzweigPorter_R}),  rewriting the latter as
\begin{equation}\label{NHRosenzweigPorter_R1}
 1=\frac{1}{2\sigma^2} e^{-\frac{|w|^2}{2\sigma^2}+a}\phi_w(a), \quad \phi_w(a):=\int_a^{\infty}I_0\left(\frac{|w|}{\sigma}\sqrt{2(x-a)}\right)\,e^{-x}\,\frac{dx}{x}
  \end{equation}
Noticing that $\phi_w(a)$ is logarithmically diverging as $a\to 0$, we
  subtract from it  $\phi_0(a)=\int_a^{\infty}e^{-x}\frac{dx}{x}:=E_1(a)$  whose asymptotic behaviour at $a\to 0$ is given by $E_1(a)=-\ln{a}-\gamma+O(a)$, where $\gamma\approx 0.577$ is the Euler-Mascheroni constant. We then see that the following limit exists:
 \begin{equation}\label{zerolimit}
 \lim_{a\to 0}\left[\phi_w(a)-\phi_0(a)\right]=\int_0^{\infty}\left[I_0\left(\frac{|w|}{\sigma}\sqrt{2x}\right)-1\right]e^{-x}\frac{dx}{x}
 %=2\int_{0}^{\infty}\left[I_0\left(\frac{|w|}{\sigma}t\right)-1\right]e^{-\frac{t^2}{2}}\frac{dt}{t}
 \end{equation}
 and after a few manipulations with the above integral one finds that
  \begin{equation}\label{zerolimit1}
 \lim_{a\to 0}\left[\phi_w(a)-\phi_0(a)\right]=\Delta_w, \quad \Delta_w:=\int_0^{\frac{|w|^2}{2\sigma^2}}\frac{e^{t}-1}{t}\,dt.
  \end{equation}
Such a relation implies the leading asymptotic behaviour for $\phi_w(a)$ as $a\to 0$.  In the regime considered here the remainder below is uniform when $b=|w|^2/(2\sigma^2)$ stays in a fixed compact set:
 \begin{equation}\label{zerolimit_phi}
 \phi_w(a)=-\gamma+\Delta_w-\ln{a}+O\!\left(a\log\frac1a\right)=\ln{\left(\frac{e^{-\gamma+\Delta_w}}{a}\right)}+O\!\left(a\log\frac1a\right).
  \end{equation}
 Substituting the above to the equation
Eq.(\ref{NHRosenzweigPorter_R1}) and keeping both the leading and subleading terms as $a\to 0$ brings the latter equation to the form
   \begin{equation}\label{NHRosenzweigPorter_R_equ_asy}
 1=\frac{1}{2\sigma^2} e^{-\frac{|w|^2}{2\sigma^2}} \ln{\left(\frac{2\sigma^2e^{-\gamma+\Delta_w}}{R_*}\right)}+O\!\left(a\log\frac1a\right),
  \end{equation}
which is solved by
    \begin{equation}\label{NHRosenzweigPorter_R_asy}
 R_* = 2\sigma^2e^{-(\gamma-\Delta_w)} \exp{-2\sigma^2e^{\frac{|w|^2}{2\sigma^2}}\left(1+O\!\left(a\log\frac1a\right)\right)}.
  \end{equation}
Here exponentiating the remainder is still harmless.  In the present second limit $a=R_*/(2\sigma^2)$ is exponentially small, and therefore $\sigma^2a\log(1/a)\to0$; the correction generated in the exponent is $o(1)$ and produces only a relative factor $1+o(1)$.
  In the figure Fig. \ref{asy_vs_exact} we compare the asymptotic formula Eq.(\ref{NHRosenzweigPorter_R_asy}) with a direct numerical solution of the equation (\ref{NHRosenzweigPorter_R})
  for different values of $\sigma^2$. It appears the asymptotic provides a very precise approximation already for $\sigma^2=4$.

Put $b=|w|^2/(2\sigma^2)$ and recall $a=R_*/(2\sigma^2)$.  The first corrections are non-analytic in $a$.  After changing variables $x=at$ in Eq.~\eqref{NHRosenzweigPorter_T}, one finds
\begin{equation}\label{NHRosenzweigPorter_T_asy}
T_0(w,R_*)=1+(b-1)a\log\frac1a+O(a),
\qquad
T_1(w,R_*)=\sqrt{ab}\log\frac1a+O(\sqrt a).
\end{equation}
For $T_0$ this follows conveniently by integrating by parts using $a(a+x)^{-2}=\frac{d}{dx}[x/(a+x)]$; for $T_1$ the logarithm comes from $I_1(2\sqrt{bx})\sim\sqrt{bx}$ as $x\downarrow0$.  The leading density is therefore unchanged, but the error term is
\begin{equation}\label{NHRosenzweigPorter2_denasy}
\rho^{NHRP}(w)=\frac{1}{2\pi\sigma^2}e^{-\frac{|w|^2}{2\sigma^2}}
\left(1+O\!\left(a\log^2\frac1a\right)\right).
\end{equation}
 corresponding to the eigenvalue density of the pure diagonal perturbation without the Ginibre background.

 As to the eigenvector component, we find it instructive to look at the limiting conditional moments $\mathbb{E}_{\infty}[p^q\mid w]$ given in
 Eq.(\ref{NHRosenzweigPorter1mom})  yielding  in the limit $a\to 0$ the leading-order behaviour
  \begin{equation} \label{eigenvec_mom_approx}
  \mathbb{E}_{\infty}[p^q\mid w]\approx \frac{\Gamma(q+1)}{q-1}\frac{1}{2\sigma^2e^{\frac{|w|^2}{2\sigma^2}}R_*^{q-1}}, \quad q>1.
  \end{equation}
   Upon substituting here the relation Eq. (\ref{NHRosenzweigPorter_R_asy}) this can be further rewritten explicitly as
 \begin{equation} \label{eigenvec_mom_asy}
 \mathbb{E}_{\infty}[p^q\mid w]\approx \frac{q\Gamma(q-1)}{2\sigma^2e^{\frac{|w|^2}{2\sigma^2}}}
 \left(\frac{\exp{2\sigma^2e^{\frac{|w|^2}{2\sigma^2}}}}{2\sigma^2e^{-(\gamma-\Delta_w)}}\right)^{q-1}, \quad q>1, \quad \sigma^2\gg 1.
 \end{equation}
  Thus, as long as $\sigma^2$ is arbitrarily large but fixed while $N\to\infty$, the Palm law of a fixed component stays on the scale $p=N|v_1|^2=O(1)$, or $|v_1|^2=O(1/N)$.  In the usual physical language this is the extended/ergodic side of the problem.  Strictly speaking, the one-component statement by itself is not a separate concentration theorem for a whole-vector inverse participation ratio.
  At the same time the moments show a huge exponential enhancement as the perturbation strength $\sigma^2$ grows.  Physically this is a clear precursor of localization, even though the fixed-$\sigma$ theorem still keeps every marked component on the $N^{-1}$ scale.

 \begin{remark}\label{logNHRZ}
 Although Theorem \ref{NHRP} only covers the case when the limit $N\to \infty$ is taken at a fixed value of the variance $\sigma^2<\infty$, it is natural to expect the
  analysis still largely applies if  $\sigma^2$ is growing with $N$ slowly enough. We thus
    conjecture that the formulas for the regime $\sigma^2\gg 1$, Eq. (\ref{NHRosenzweigPorter2_denasy}) and Eq.(\ref{eigenvec_mom_asy}),  retain their validity for
     the logarithmic growth of the variance: $\sigma^2=\frac{\mu}{2}\log{N}$. Under such a conjecture one finds the rich picture of a localization-type transition
    in the eigenvector statistics as the parameter $\mu$ changes.  Such a picture essentially extends, completes and quantifies the localization scenario
    proposed by  De Tomasi and Khaymovich in  \cite{TomasiKhaimovich_NHRP22}.

     In the logarithmic regime it is convenient to rescale the eigenvalue positions as to keep $\tilde{w}=w/\sqrt{\log{N}}$ finite. The limiting mean eigenvalue density in the $\tilde w$-plane will then be simply given by $\rho_{NHRP}(\tilde w)=\frac{1}{\pi\mu}e^{-\frac{|\tilde{w}|^2}{\mu}}$.
 The parameter $R_*$ of the theory in such a regime becomes $N-$dependent and up to a multiplicative factor of order unity given by (cf. Eq.(\ref{NHRosenzweigPorter_R_asy}))
\begin{equation}\label{R_log}
R_* \asymp \frac{\log{N}}{N^{\mu_{eff}}}, \quad \mu_{eff}=\mu e^{\frac{|\tilde{w}|^2}{\mu}}.
\end{equation}
Closer inspection shows that consistency of the procedure requires this parameter to satisfy $R_*\gg 1/N$, which implies the consistency condition $\mu_{eff}\leq1$; at $\mu_{eff}=1$ one still has $R_*\asymp(\log N)/N\gg N^{-1}$.
 In order to investigate the behaviour of the moments of the eigenvector components in this case it is convenient
  to evaluate them directly by averaging Eq.(\ref{MainJPDadditivedef_proof3_moments}) over the diagonal entries of the perturbation.
   One finds by inspection that as long as $\mu_{eff}\leq1$, after appropriate rescaling and a formal saddle-point computation the leading contribution to moments comes from the last line in Eq.(\ref{MainJPDadditivedef_proof3_moments}) and coincides with results for the moments found by us earlier, with appropriate replacement of $R_*$
    with its asymptotic Eq.(\ref{R_log}). This yields the leading-order moments asymptotic which is fully consistent with Eq.(\ref{eigenvec_mom_asy}):
   \begin{equation} \label{eigenvec_mom_asy_log}
 \mathbb{E}_{\rm ann}[p^q\mid w]\asymp \frac{1}{(\mu\log{N})^{q}}\, N^{\mu_{eff} (q-1)}, \quad q>1, \quad \mu_{eff}=\mu e^{\frac{|\tilde{w}|^2}{\mu}}.
 \end{equation}
  The quantity controlled by this computation is the \emph{annealed conditional} inverse participation ratio
\[
\overline I_q(w):=\mathbb E\!\left[\sum_{i=1}^N|v_i|^{2q}\,\middle|\,w\right].
\]
By permutation symmetry of the diagonal disorder,
$\overline I_q(w)=\mathbb E[p^q\mid w]/N^{q-1}$, so the formal asymptotic above predicts
\[
\frac{\log \overline I_q(w)}{\log N}\longrightarrow(\mu_{eff}-1)(q-1),\qquad q>1.
\]
For $\mu_{eff}<1$ this annealed exponent is consistent with a nonergodic extended picture with effective dimension $D=1-\mu_{eff}$; $\mu_{eff}=1$ is the predicted crossover boundary. This is not, by itself, a quenched statement about a typical eigenvector: rare samples can affect annealed moments, and proving concentration of $I_q$ would require additional work. Likewise, no statement on local eigenvalue statistics follows from the present one-point calculation. It is natural to ask whether the predicted nonergodic extended regime has a non-Hermitian analogue of the universal local statistics familiar from the Hermitian Rosenzweig--Porter model; this remains open here.

We stress that this prediction does not contradict the obstruction to a fractal {\it phase} established for a generic complex diagonal potential in
\cite{TomasiKhaimovich_NHRP22}. That obstruction concerns the power-law window $\sigma^2\propto N^{\gamma-1}$ with $\gamma\ne1$, whereas the nonergodic
extended behaviour predicted here occupies only a logarithmic window sitting exactly at the critical point $\gamma=1$, which is precisely the case their
methods do not reach.

Under the logarithmic-scaling conjecture, for any fixed $\mu\in(0,1)$ the condition $\mu_{eff}<1$ corresponds to
$|\tilde w|<\sqrt{\mu\log(1/\mu)}$. The normalized Gaussian eigenvalue density then predicts an annealed fraction $1-\mu$ of eigenstates in this region. Outside it localization is expected, and for variances growing faster than $(1/2)\log N$ the prediction is that all eigenstates become localized.
\end{remark}

\subsubsection{Rank-one additive perturbation of complex Ginibre matrices}
In this section we discuss results for another special case of a  non-trivial additive deformation of complex Ginibre matrices allowing more explicit treatment for all $N$
and eventually in the limit $N\to \infty$.  This is a {\it rank-one non-normal} perturbation parametrized by a single complex eigenvalue $a\in \mathbb{C}$
and two complex vectors: the left ${\bf l}$ and the right ${\bf r}$ so that
\begin{equation}\label{rank_one_gen}
 A=a\,{\bf r}\otimes {\bf l}^*, \quad \mbox{with}\quad a\in \mathbb{C}\ \quad \mbox{and s.t.}\,\, {\bf l}^*{\bf r}=1.
 \end{equation}
For such deformations of Ginibre matrices one can show after straightforward but lengthy manipulations (see Appendix F for more detail)  the validity of the following
\begin{proposition}\label{cor2.10}
Let $N\ge3$ and let $A$ be given as in Eq.(\ref{rank_one_gen}). The corresponding one-point joint intensity (JPD, in the terminology used throughout the paper) then takes the form
\begin{equation}\label{rank_one_JPD}
{\cal P}^{(r1)}_N(z,{\bf v})=\frac{1}{\pi\Gamma(N)} e^{-|a|^2\,{\bf r}^*{\bf r}|({\bf v}^*{\bf l})|^2+a\overline{z}({\bf l}^*{\bf v})({\bf v}^*{\bf r})+
\overline{a}z({\bf r}^*{\bf v})({\bf v}^*{\bf l})}\left\{\Gamma(N,|z|^2)+\right.
\end{equation}
\[
\left.+\Gamma(N-1,|z|^2)\left[|a|^2\left({\bf r}^*{\bf r}-|{\bf r}^*{\bf v}|^2\right)\left({\bf l}^*{\bf l}-|{\bf l}^*{\bf v}|^2\right)
-a\overline{z}\left(1-({\bf l}^*{\bf v})({\bf v}^*{\bf r})\right)-\overline{a}z\left(1-({\bf r}^*{\bf v})({\bf v}^*{\bf l})\right)\right]\right.
\]
\[
\left.-|a|^2|z|^2\Gamma(N-2,|z|^2)\left[\left({\bf r}^*{\bf r}-|{\bf r}^*{\bf v}|^2\right)\left({\bf l}^*{\bf l}-|{\bf l}^*{\bf v}|^2\right)
-|1-({\bf r}^*{\bf v})({\bf v}^*{\bf l})|^2\right]\right\}
\]
\end{proposition}
\begin{remark}
The restriction $N\ge3$ is only a matter of writing Eq.~\eqref{rank_one_JPD} without meaningless products such as $0\cdot\Gamma(0,0)$.  For $N=2$ the Gram determinant multiplying the last line of Eq.~\eqref{rank_one_JPD} vanishes identically, so that this whole line must first be cancelled and the remaining two terms give the regular answer.  For $N=1$ the model is scalar and Theorem~\ref{Theorem2} gives simply $\pi^{-1}e^{-|z-a|^2}$.
\end{remark}
\begin{remark}
The above JPD depends on the vector ${\bf v}$ only via the two complex combinations $q_1={\bf r}^*{\bf v}$ and $q_2={\bf l}^*{\bf v}$.  When $N\ge3$ and ${\bf l}$ and ${\bf r}$ are linearly independent, so that $\det\mathcal B>0$, one can pass to the JPD of $z,q_1,q_2$ using the identity:
\begin{equation}\label{rank_one_nononrm_JPD q1q2}
P_N(q_1,q_2):=\int_{|{\bf v}|=1} \delta^{(2)}\left(q_1-{\bf r}^*{\bf v}\right)\delta^{(2)}\left(q_2-{\bf l}^*{\bf v}\right)d\mu_H({\bf v})=
\frac{(N-1)(N-2)}{\pi^2} \frac{(1-B)^{N-3}}{\det{\mathcal{B}}}{\bf 1}_{B\in [0,1]},
\end{equation}
where we denoted
\begin{equation}\label{def_B}
B=(\overline{q_1},\overline{q_2})\mathcal{B}^{-1}\left(\begin{array}{c}q_1\\ q_2\end{array}\right), \quad \mathcal{B}=\left(\begin{array}{cc}{\bf r}^*{\bf r} & 1\\ 1 &
{\bf l}^*{\bf l}\end{array}\right)
\end{equation}
and used the condition $ {\bf l}^*{\bf r}=1$.  The factor $\pi^{-2}$ is fixed by normalization with respect to $d^2q_1d^2q_2$.  For $N=2$ the push-forward is instead supported on the boundary ellipsoid $B=1$ and is not a four-dimensional Lebesgue density. One can further write down an explicit formula for the ensuing mean eigenvalue density, but it is not very transparent.
We will not pursue this topic further here, but rather concentrate on a somewhat simpler normal rank-one deformation case.
\end{remark}

\begin{corollary}
For $N\ge2$, in the particular case of normal rank-one deformation ${\bf r}={\bf l}, \, \, {\bf r}^*{\bf r}=1$ the JPD further simplifies to
\begin{equation}\label{rank_one_norm JPD}
{\cal P}^{(r1,norm)}_N(z,{\bf v}) =\frac{1}{\pi\Gamma(N)} e^{-|({\bf v}^*{\bf r})|^2\left(|a|^2-a\overline{z}-
\overline{a}z\right)}\left\{\Gamma(N,|z|^2)+\right.
\end{equation}
\[
\left.+\Gamma(N-1,|z|^2)\left(1-|{\bf r}^*{\bf v}|^2\right)\left[|a|^2\left(1-|({\bf r}^*{\bf v})|^2\right)
-a\overline{z}-\overline{a} z\right]\right\}
\]
\end{corollary}

\vspace{1cm}

As the JPD Eq.(\ref{rank_one_norm JPD}) depends on the  vector ${\bf v}$ only via the variable $q=|({\bf r}^*{\bf v})|^2$ it is natural to pass to the JPD of $z$ and $q$ by
\begin{equation}\label{rank_one_norm JPD_q}
{\cal P}^{(r1,norm)}_N(z,q)=\int_{|{\bf v}|=1} \delta\left(q-|({\bf r}^*{\bf v})|^2\right){\cal P}^{(r1,norm)}_N(z,{\bf v})\,d\mu_H({\bf v})
  \end{equation}
For $N\ge2$ the above can be computed using the following relation, valid as long as ${\bf r}^*{\bf r}=1$:
\begin{equation}\label{identity_rankonenormal}
P_N(q)=\int_{|{\bf v}|=1} \delta\left(q-|({\bf r}^*{\bf v})|^2\right)d\mu_H({\bf v})=(N-1)(1-q)^{N-2}\, {\bf 1}_{q\in [0,1]}.
\end{equation}
Finally, the associated mean eigenvalue density is found as the marginal:
\begin{equation}\label{mean den rank_one_norm}
p^{(r1,norm)}_N(z)=\int_0^1{\cal P}^{(r1,norm)}_N(z,q)\,dq=\frac{(N-1)}{\pi\Gamma(N)} \left\{\Gamma(N,|z|^2)I^{(0)}(N,\varkappa) \right.
 \end{equation}
 \[
 +\left.\Gamma(N-1,|z|^2)\left[|a|^2I^{(2)}(N,\varkappa)+(\varkappa-|a|^2)I^{(1)}(N,\varkappa)\right]\right\}
 \]
where we defined
\begin{equation}\label{delta}
\varkappa:=|a|^2-(a\overline{z}+\overline{a}z), \quad I^{(k)}(N,\varkappa):=\int_0^1\,e^{-q\varkappa}(1-q)^{N+k-2}\,dq
\end{equation}

\begin{remark}

The paper \cite{GrelaGuhr16} provided some computation of the mean eigenvalue density for a general non-normal rank-one deformation of complex Ginibre matrices in a some
 complicated form, which looks more amenable to numerical rather than analytical consideration. The formula for rank-one normal perturbation equivalent to  Eq.(\ref{mean den rank_one_norm}) does not seem to appear there explicitly. Some formulae for general correlation functions for finite-rank deformations of complex Ginibre
matrices have been recently given in  \cite{LiuZhang2024,LiuZhang2025}.

\end{remark}

%\begin{remark}
The formula Eq.(\ref{mean den rank_one_norm}) can be used to study in great detail how an outlier is formed in the large-$N$ limit with the growth of
the perturbation strength. In such a limit one needs to rescale $z=\sqrt{N} w, \quad a=\sqrt{N} \alpha$ and keep the parameters $w,\alpha$ fixed as $N\to \infty$.

\begin{proposition}\label{asydenprop}
Inside the unit disk the limiting mean eigenvalue density is uniform for every fixed $\alpha$:
\begin{equation}\label{inside}
\lim_{N\to\infty}p_N^{(r1,\,norm)}(\sqrt N\,w)=\frac1\pi,
\qquad |w|<1.
\end{equation}
In the exterior $|w|>1$ there are two distinct large-deviation branches, separated by
\[
\Delta:=|\alpha|^2-(\alpha\bar w+\bar\alpha w)=-1.
\]
If $\Delta<-1$, then
\begin{equation}\label{asydenout}
p_N^{(r1,\,norm)}(\sqrt N\,w)\sim \frac1\pi e^{-N{\cal F}(w)}
\left(\frac{|\Delta|}{|w|^2}-
\frac{|\alpha|^2(1-|\Delta|^{-1})}{(|w|^2-1)|w|^2}\right),
\end{equation}
where
\[
{\cal F}(w)=|w-\alpha|^2+\log\left|1-\frac{|w-\alpha|^2}{|w|^2}\right|.
\]
If $\Delta>-1$, then
\begin{equation}\label{asydenout_gin_branch}
p_N^{(r1,\,norm)}(\sqrt N\,w)
\sim
\frac{e^{-N I_{\rm Gin}(|w|^2)}}{\pi\sqrt{2\pi N}}
\frac{|w-\alpha|^2}{|w|^2(|w|^2-1)(1+\Delta)},
\end{equation}
with $I_{\rm Gin}(r):=r-1-\log r$.  The transition surface $\Delta=-1$ is a separate crossover regime and is not covered by these two outer asymptotics.
\end{proposition}
Proof of these statements can be found in Appendix G.

\begin{remark}
 On the outlier-controlled branch $\Delta<-1$, the rate ${\cal F}(w)$ has the unique minimum at $w_{*}=\alpha$, and ${\cal F}(w_{*})=0$.
 The condition $|w|>1$ implies the minimum is only possible for $|\alpha|>1$, which in turn implies $\Delta=-|\alpha|^2<-1$ as required.
  We conclude that the density is exponentially small except for a  vicinity of the point $w=\alpha$ such that $|w-\alpha|=O(N^{-1/2})$.
 In such a vicinity the Eq.(\ref{asydenout})
  can be further  simplified yielding in the leading order  the Gaussian profile:
   \begin{equation}\label{asydenout_Gau}
  p^{(r1,\, norm)}_N(\sqrt{N}w)\sim \frac{1}{\pi \sigma}e^{-\frac{N}{\sigma}|w-\alpha|^2}, \,\,\sigma=\frac{|\alpha|^2}{|\alpha|^2-1}>0.
 \end{equation}
 Such a density describes the formation of an outlier at the point $w=\alpha$ for $|\alpha|>1$, as was first predicted by Tao, see \cite{Tao_outl}
 and since then attracted a lot of interest, see e.g. \cite{BR2016,Bordenave-Captaine2016,O'RourkeRenfew14,LiuZhang2024,LiuZhang2025}.
     Such a phenomenon is a certain non-Hermitian analogue of the famous BBP transition \cite{BBP} which, in particular, found numerous applications in studying the "noise plus signal" phenomenon in pattern recognition, with signal modelled by a rank-one perturbation $a{\bf r}\otimes {\bf r}^*$. Recently similar questions have been addressed in the non-selfadjoint setting
     \cite{CCF2021,BCL25}. In such a context it is important also to characterize the overlap
 between the signal vector ${\bf r}$ and the eigenvector corresponding to the outlier, which in our context is exactly the variable $q=|({\bf r}^*{\bf v})|^2$.
 Using the same approach as above it is straightforward to compute from Eq.(\ref{rank_one_norm JPD_q}) the corresponding asymptotic of associated JPD at $z=a$, which turns out
 to have the Large Deviation form as well:
 \begin{equation}\label{rank_one_norm JPD_q_LD}
{\cal P}^{(r1, norm)}_N(z=\sqrt{N}\alpha,q)\sim \sqrt{\frac{N}{2\pi}} \frac{\left(\frac{q}{1-q}\right)^2}{\pi (|\alpha|^2-1)}\exp{-N\mathcal{G}(q)},
\quad {\cal G}(q)=|\alpha|^2(1-q)-\ln\left[|\alpha|^2(1-q)\right]-1.
  \end{equation}

It is easy to check that the large-deviation rate function $\mathcal{G}(q)$ has a unique minimizer at $q_*=1-|\alpha|^{-2}$ and $\mathcal{G}(q_*)=0$. The value $q_*$ provides
the typical value for the overlap $q=|({\bf r}^*{\bf v})|^2$ for the outlier. The overlap vanishes at the threshold $|\alpha|=1$ when the outlier
merges to the unit disk and the associated eigenvector stops to bear any useful information on the perturbation.

\end{remark}

\begin{remark}
As was noticed by Tao \cite{Tao_outl} a special (random) choice of vectors ${\bf r}$ and ${\bf l}$ in the rank-one {\it additive} deformations Eq.(\ref{rank_one_gen}) of complex Ginibre matrices provides in the limit $N\to \infty$ an interesting connection to the subject of zeroes of random Gaussian Analytic Functions. This observation motivated a line of research in \cite{ForIps19}, though from  a somewhat different perspective of {\it multiplicative} subunitary rank-one deformations of CUE matrices.  Most recently this subject was further developed in \cite{FKP24}.

\end{remark}

\subsection{Discussion of the method, further developments and open problems.}

In the present paper we aim to develop an approach inspired by Kac--Rice counting formulas which provides
an access to the joint statistics of an eigenvalue $z$ and the associated normalized right eigenvector ${\bf v}$ for general non-Hermitian random matrices.
By a combination of the Grassmann integral representation for the involved determinants and Fourier representation for the regularized $\delta$-functions
the method was demonstrated to be quite powerful and produced results which seem less immediately accessible via alternative techniques,
such as the incomplete Schur decomposition or the Girko Hermitization trick.
Here it is however necessary to mention that so far all examples of successful application of the method developed in the present paper remained restricted to matrices
with Gaussian-distributed entries. The important problem of finding a way of extending the applicability of the method to matrices with entries distributed
according to more general laws poses currently a considerable technical challenge and remains outstanding.

The approach looks nevertheless promising for addressing
more general classes of non-Hermitian matrices with Gaussian-distributed entries, in particular those with banded structure. Such studies might help to
understand better an intriguing interplay between Anderson localization phenomena and non-Hermiticity, which attracted much interest in theoretical physics community
in the recent years, see \cite{HuangShklov2020a,HuangShklov2020b,KawabataRyu21,Ghosh_etal_2023} but remains poorly understood mathematically beyond the particular case of Hatano-Nelson model \cite{HN1996,HN1998}, see \cite{GoldKhor98,GoldKhor00} for the mathematical treatment of the latter model. We hope to be able to explore some of these directions in our future works, with the first results  concerning eigenvectors for a pair of coupled complex Ginibre blocks relegated to a separate publication
with Margherita Disertori \cite{ergod26}.

At the same time the knowledge of statistics of the right eigenvectors in general is not enough to shed light on such important characteristics
as the matrix of overlaps ${\cal O}_{ab}$ between right-and left eigenvectors featuring in Chalker-Mehlig correlators discussed in the beginning of the present paper. There is however one special case when the knowledge of right eigenvectors suffices: the case of {\it complex symmetric} matrices, as in that case the left eigenvectors are simply related to transposed right eigenvectors. In a separate joint paper with Gernot Akemann and Dmitry Savin \cite{Akemann4} we considered that case for Gaussian matrices with i.i.d. entries and derive the probability density of the corresponding diagonal  overlaps ${\cal O}_{aa}$, thus generalizing the results of  works by
 Bourgade and Dubach \cite{BourgadeDubach}  and Fyodorov \cite{Fyo2018} to the complex symmetric case. That computation further demonstrates the power and utility of the approach
developed in the present paper, the more remarkably that the complex symmetric matrices are notoriously difficult case to study by standard tools,
and even the mean eigenvalue density for such an ensemble has been not known beyond $N=2$ case\footnote{By the time of writing the present version, September 2026, a remarkable progress has been reported in \cite{XiaoChenLiuRyu2026}.}, despite considerable interest in matrices of such symmetry motivated by
applications in physics of quantum chaotic dissipative systems, see e.g. \cite{Pandey_etal_2019,Hamazaki_etal_2020}.

\subsubsection{The joint empirical measure of left and right eigenvectors}\label{Towards}
The same transversality idea gives a rigorous extension to a left--right marked eigenpair.  For every simple eigenvalue $z_a$ choose a unit right eigenvector ${\bf v}_a$ and the unique left eigenvector ${\bf u}_a$ satisfying
\[
X{\bf v}_a=z_a{\bf v}_a,\qquad
X^*{\bf u}_a=\bar z_a{\bf u}_a,\qquad
{\bf u}_a^*{\bf v}_a=1.
\]
Under a phase change ${\bf v}_a\mapsto e^{i\phi}{\bf v}_a$, the corresponding normalized left eigenvector changes as ${\bf u}_a\mapsto e^{i\phi}{\bf u}_a$.  We therefore define the phase-averaged left--right marked empirical measure.
In the customary physics notation, regarding $\Pi_X^{LR}$ as a
distributional density with respect to
$d^2z\,d\mu_H({\bf v})\,d^{2N}{\bf u}$, we write
\begin{equation}\label{empir_right_left_eigenv}
\Pi_X^{LR}(z,{\bf v},{\bf u})
=
\sum_{a=1}^N\delta^{(2)}(z-z_a)
\int_0^{2\pi}
\delta_H\!\left({\bf v}-e^{i\phi}{\bf v}_a\right)
\delta^{(2N)}\!\left({\bf u}-e^{i\phi}{\bf u}_a\right)
\,\frac{d\phi}{2\pi}.
\end{equation}

For ${\bf v}\in\mathbf S_N^c$, let $\delta^{(2N-2)}_{{\bf v}^\perp}$ denote the Euclidean Dirac distribution on the complex space ${\bf v}^\perp$, as defined in Appendix~C.

\begin{theorem}[Left--right transverse Kac--Rice identity]\label{thm_left_right}
Let $X$ have simple spectrum and put $p_X(z)=\det(X-z{\bf 1}_N)$.  Then $\Pi_X^{LR}$ has, with respect to $d^2z\,d\mu_N({\bf v})\,d^{2N}{\bf u}$, the distributional density
\begin{equation}\label{left_right_KacRice}
\frac1\pi |p_X'(z)|^4
\delta^{(2N)}\!\left((X-z{\bf 1}_N){\bf v}\right)
\delta^{(2N-2)}_{{\bf v}^\perp}\!\left((X-z{\bf 1}_N)^*{\bf u}\right)
\delta^{(2)}({\bf u}^*{\bf v}-1).
\end{equation}
The identity is understood against test functions in
$C_c^\infty(\mathbb C\times\mathbf S_N^c\times\mathbb C^N)$.
\end{theorem}

\begin{proof}
The identity is understood as an iterated distribution, the ${\bf u}$-integration being performed first.  For fixed $(z,{\bf v})\in\mathbb C\times\mathbf S_N^c$ the last two Dirac factors impose exactly $2N$ real constraints on ${\bf u}$, namely the vanishing of the map
\[
{\bf u}\longmapsto \left(V_{\bf v}^*(X-z{\bf 1}_N)^*{\bf u},\ \overline{{\bf u}^*{\bf v}-1}\right)\in\mathbb C^{N-1}\times\mathbb C ,
\]
which is complex linear in ${\bf u}$ after the indicated conjugation of the second component.  Its real Jacobian is therefore
$J(z,{\bf v})=\left|\det\left(V_{\bf v}^*(X-z{\bf 1}_N)^*\,\big|\,{\bf v}^*\right)\right|^2$, where the determinant is that of the $N\times N$ matrix whose first $N-1$ rows are those of $V_{\bf v}^*(X-z{\bf 1}_N)^*$ and whose last row is ${\bf v}^*$.  A different choice of the isometry $V_{\bf v}$ changes this determinant by a unimodular factor only, so $J$ is well defined, and it is a continuous function of $(z,{\bf v})$ on $\mathbb C\times\mathbf S_N^c$.  Consequently $|p_X'(z)|^4/J(z,{\bf v})$ is continuous there and may legitimately be multiplied against the singular measure of Theorem~\ref{MainMetaTheorem}.

At an eigenpair, $(z,{\bf v})=(z_a,e^{i\phi}{\bf v}_a)$, Appendix~C shows that $J(z_a,e^{i\phi}{\bf v}_a)=|p_X'(z_a)|^2$; the ${\bf u}$-integration of the last two Dirac factors multiplied by $|p_X'(z_a)|^2$ therefore produces exactly the Dirac mass at the uniquely normalized left eigenvector ${\bf u}_a$, see Eq.~\eqref{left_transverse_delta}.  The remaining factor $\frac1\pi|p_X'(z)|^2\delta^{(2N)}((X-z{\bf 1}_N){\bf v})$ is precisely the phase-averaged right-eigenvector measure of Theorem~\ref{MainMetaTheorem}, and the phases of ${\bf v}_a$ and ${\bf u}_a$ are locked together by the normalization ${\bf u}^*{\bf v}=1$, as noted before Eq.~\eqref{empir_right_left_eigenv}.
\end{proof}

Integrating Eq.~\eqref{left_right_KacRice} first over ${\bf u}$ gives Theorem~\ref{MainMetaTheorem}, and integrating further over ${\bf v}$ gives Eq.~\eqref{Kac-Rice_charpol}.  Thus the marginal consistency is automatic and does not require applying an invertible-matrix Schur complement on the singular support.  Whether this left--right representation can be turned into an efficient computational tool for non-orthogonality observables is left for future work.

Appendix~H gives a second derivation based on the source-regularized
equations, which provide a natural gauge-fixing regularization of the
left--right eigenvector constraints.
\[
(X-z{\bf 1}_N){\bf v}=\sigma {\bf h}_1,\qquad
(X-z{\bf 1}_N)^*{\bf u}=\sigma {\bf h}_2,\qquad
{\bf u}^*{\bf v}=1.
\]
For $\sigma>0$ the source terms break the complex scaling degeneracy and replace every gauge orbit by two isolated solutions.  Their finite-$\sigma$ Kac--Rice Jacobian can be evaluated exactly, and after the compensating factor $1/2$ and phase averaging the regularized measure converges to Eq.~\eqref{empir_right_left_eigenv}.  This provides a direct mathematical realization of the gauge-breaking picture while leaving the transverse formula~\eqref{left_right_KacRice} as the rigorous distributional identity at $\sigma=0$.

%%%%%%%%%%%%%%%%%%%%
%%%%%%%%%%%%%%%%%%%%%

\section{Proofs for the ensemble interpolating between complex and real Ginibre}\label{interpol}

In this section we prove Theorem~\ref{Theorem1}, which gives the one-point joint intensity ${\cal P}_N(z,{\bf v})$ of a complex eigenvalue $z$ and its phase-averaged normalized right eigenvector ${\bf v}$ for the interpolating Ginibre ensemble defined in Eq.(\ref{def_interpolating}). The starting point is to substitute the integral representations Eq. (\ref{Gaudelta_vec}) and Eq.(\ref{Berezin_int}) into the identity
Eq.(\ref{main_practical})  resulting in
\begin{equation}\label{JPDinter1}
{\cal P}_N(z,{\bf v})=\frac{1}{\pi}\lim_{|w-z|\to 0}\frac{\partial^2}{\partial w\partial \overline{w}}\lim_{\epsilon\to 0}
\int_{\mathbb{C}^N}\frac{d{\bf k}d{\bf k}^*}{(2\pi)^{2N}}e^{-\epsilon\frac{{\bf k}^*{\bf k}}{2}-\frac{i}{2}\left[z{\bf k}^*{\bf v}+\overline{z}
{\bf v}^*{\bf k}\right]}
\end{equation}
\[
\times \int \exp{-\frac{i}{2}\left[w\left(\mathbf{\Psi}_{+}^T\mathbf{\Phi}_{-}\right)
  +\overline{w} \left(\mathbf{\Psi}_{-}^T\mathbf{\Phi}_{+}\right)\right]}\,{\cal D}(\mathbf{\Psi},\mathbf{\Phi}) \,\left\langle e^{\frac{i}{2}\mbox{\small Tr }\left(XB_{-+}+X^*B_{+-}\right)}\right\rangle_{X},
\]
where we introduced two $N\times N$ matrices $B_{-+},B_{+-}$ via
\begin{equation}\label{Bmat}
B_{-+}={\bf v}\otimes {\bf k}^*-\mathbf{\Phi}_{-}\otimes \mathbf{\Psi}^T_{+}, \quad B_{+-}={\bf k}\otimes {\bf v}^*-\mathbf{\Phi}_{+}\otimes \mathbf{\Psi}^T_{-}.
\end{equation}
Now use the elementary Gaussian identity
\begin{equation}\label{RGin_ave}
\left\langle e^{\mbox{Tr}(GC)}\right\rangle_G
=\exp{\frac{1}{2}\mbox{Tr}(CC^T)},
\end{equation}
for a real Ginibre matrix $G$.  In the present application the entries of $C$ belong to the finite Grassmann algebra generated by the variables introduced above.  Equation~\eqref{RGin_ave} remains valid coefficientwise: expand the exponential in the Grassmann generators, apply the ordinary Gaussian moment formula to each coefficient, and resum the resulting finite Grassmann polynomial.  Equivalently, for ordinary matrices $B_1,B_2$ one has
\[
\left\langle e^{\frac{i}{2}\operatorname{Tr}(GB_1+G^TB_2)}\right\rangle_G
=\exp{-\frac{1}{8}\operatorname{Tr}\left(B_1^TB_1+B_2^TB_2+2B_1B_2\right)},
\]
and the same identity holds coefficientwise for the Grassmann-valued matrices used below.  Recalling the definition Eq.~(\ref{def_interpolating}) of the ensemble of matrices $X$, one obtains
\[
\left\langle e^{\frac{i}{2}\mbox{\small Tr }\left(XB_{-+}+X^*B_{+-}\right)}\right\rangle_{X}=
\left\langle e^{\frac{i}{2}\sqrt{\frac{1+\tau}{2}}\mbox{\small Tr }\left(G_1B_{-+}+G_1^TB_{+-}\right)}\right\rangle_{G_1}
\left\langle e^{-\frac{1}{2}\sqrt{\frac{1-\tau}{2}}\mbox{\small Tr }\left(G_2B_{-+}-G_2^TB_{+-}\right)}\right\rangle_{G_2}
\]
\begin{equation}
=e^{-\frac{1}{8}\tau\mbox{\small Tr }\left(B^T_{-+}B_{-+}+B^T_{+-}B_{+-}\right)-\frac{1}{4}\mbox{\small Tr }\left(B_{-+}B_{+-}\right)}
\end{equation}
with the parameter $\tau\in[0,1]$ controlling the interpolation. The definitions Eq.(\ref{Bmat}) imply that
\[
\mbox{\small Tr }\left(B^T_{-+}B_{-+}\right)=2\left({\bf k}^*\mathbf{\Psi}_{+}\right)\left(\mathbf{\Phi}^T_{-}{\bf v}\right)+\overline{\left({\bf k}^T{\bf k}\right)}\left({\bf v}^T{\bf v}\right), \quad \mbox{\small Tr }\left(B^T_{+-}B_{+-}\right)=2\left({\bf v}^*\mathbf{\Psi}_{-}\right) \left(\mathbf{\Phi}^T_{+}{\bf k}\right)+
\left({\bf k}^T{\bf k}\right)\overline{\left({\bf v}^T{\bf v}\right)}
\]
and after recalling the normalization ${\bf v}^*{\bf v}=1$ also
\[
\mbox{\small Tr }\left(B_{-+}B_{+-}\right)=\left({\bf k}^*{\bf k}\right)-
\left({\bf v}^*\mathbf{\Phi}_{-}\right)\left(\mathbf{\Psi}^T_{+}{\bf k}\right)-\left({\bf k}^*\mathbf{\Phi}_{+}\right)\left(\mathbf{\Psi}_{-}^T{\bf v}\right)
-\left(\mathbf{\Psi}_{+}^T\mathbf{\Phi}_{+}\right)\left(\mathbf{\Psi}_{-}^T\mathbf{\Phi}_{-}\right).
\]
Collecting all the factors one arrives at
\begin{equation}\label{JPDinter2}
{\cal P}_N(z,{\bf v})=\frac{1}{\pi}\lim_{|w-z|\to 0}\frac{\partial^2}{\partial w\partial \overline{w}}\lim_{\epsilon\to 0}
 \int \, {\cal D}(\mathbf{\Psi},\mathbf{\Phi})e^{-\frac{i}{2}\left[w\left(\mathbf{\Psi}_{+}^T\mathbf{\Phi}_{-}\right)
  +\overline{w} \left(\mathbf{\Psi}_{-}^T\mathbf{\Phi}_{+}\right)\right]+\frac{1}{4}\left(\mathbf{\Psi}_{+}^T\mathbf{\Phi}_{+}\right)\left(\mathbf{\Psi}_{-}^T\mathbf{\Phi}_{-}\right)}
  \end{equation}
\begin{equation}\label{JPDinter2a}
\times \int_{\mathbb{C}^N}\frac{d{\bf k}d{\bf k}^*}{(2\pi)^{2N}}e^{-\left(\frac{1}{2}+\epsilon\right)\frac{{\bf k}^*{\bf k}}{2}-\frac{\tau}{8}\left[\left({\bf k}^T{\bf k}\right)\overline{\left({\bf v}^T{\bf v}\right)}+\overline{\left({\bf k}^T{\bf k}\right)}\left({\bf v}^T{\bf v}\right)\right]+\frac{1}{4}\left[{\bf k}^*{\bf b}_1+{\bf b}_2^T{\bf k}\right]}\,,
\end{equation}
where we defined two vectors ${\bf b}_{1,2}$ by
\begin{equation}\label{def_b}
{\bf b}_1:=-2iz{\bf v}+\mathbf{\Phi}_{+}\left(\mathbf{\Psi}^T_{-}{\bf v}\right)-\tau \mathbf{\Psi}_{+}\left(\mathbf{\Phi}^T_{-}{\bf v}\right), \quad
{\bf b}^T_2:=-2i\overline{z}{\bf v}^*+\left({\bf v}^*\mathbf{\Phi}_{-}\right)\mathbf{\Psi}_{+}^T-\tau \left({\bf v}^*\mathbf{\Psi}_{-}\right)\mathbf{\Phi}_{+}^T.
\end{equation}
The integral over ${\bf k}, {\bf k}^*$ in Eq.(\ref{JPDinter2a}) is obviously Gaussian, and can be most conveniently evaluated after presenting the integration going
over two real vectors ${\bf k}_{1,2}$ defined by ${\bf k}={\bf k}_1+i{\bf k}_2$, so that
\[
{\bf k}^*{\bf k}={\bf k}_1^T{\bf k}_1+{\bf k}_2^T{\bf k}_2 \,\, \mbox{and} \,\, {\bf k}^T{\bf k}={\bf k}_1^T{\bf k}_1-{\bf k}_2^T{\bf k}_2+2i{\bf k}_1^T{\bf k}_2.
\]
Further introducing $2N-$component real vector $\tilde{\bf k}=({\bf k}_1,{\bf k}_2)^T$ one finds that
\begin{equation}\label{quad_form1}
-\left(\frac{1}{2}+\epsilon\right)\frac{{\bf k}^*{\bf k}}{2}-\frac{\tau}{8}\left[\left({\bf k}^T{\bf k}\right)\overline{\left({\bf v}^T{\bf v}\right)}+\overline{\left({\bf k}^T{\bf k}\right)}\left({\bf v}^T{\bf v}\right)\right]
+\frac{1}{4}\left[{\bf k}^*{\bf b}_1+{\bf b}_2^T{\bf k}\right]=-\frac{1}{2}\tilde{\bf k}^TC_{\epsilon}\tilde{\bf k}
+\frac{1}{4}\tilde{\bf k}^T{\bf b},
\end{equation}
where, writing $s={\bf v}^T{\bf v}$, we denoted
\begin{equation}\label{quad_form2}
C_{\epsilon}=\frac{1}{2}\left(\begin{array}{cc} 1+2\epsilon+\tau \Re s & \tau \Im s\\
\tau \Im s & 1+2\epsilon-\tau \Re s\end{array}\right)\otimes {\bf 1}_N, \quad
 {\bf b}=\left(\begin{array}{c}{\bf b}_1+{\bf b}_2\\ i\left({\bf b}_2-{\bf b}_1\right)\end{array}\right).
\end{equation}
The two eigenvalues of the $2\times2$ factor in $C_\epsilon$, each repeated $N$ times, are
\[
\lambda_\pm^{(\epsilon)}=\frac12\left(1+2\epsilon\pm\tau|s|\right).
\]
Thus $C_\epsilon$ is positive definite for every $\epsilon>0$.  If $0\leq\tau<1$, then $\lambda_-^{(0)}\geq(1-\tau)/2>0$ uniformly on the whole complex sphere, so one may safely set $\epsilon=0$.  At $\tau=1$ and $\Im z\ne0$, a right eigenvector of the real Ginibre matrix cannot be phase-equivalent to a real vector, since that would force its eigenvalue to be real; hence $|s|<1$ on the support of the complex-eigenvalue contribution and again $C_0$ is positive definite.  In contrast, for the real-eigenvalue sector at $\tau=1$ one may have $|s|=1$, in which case $\lambda_-^{(\epsilon)}=\epsilon$ and the regulator must be retained.  We do not remove the regulator in that singular sector here.  As explained in Section~\ref{realsector} below, this degeneration is not a defect of the method but a signal that the real sector must be treated on the real unit sphere rather than the complex one; that section derives the corresponding density exactly, at finite $N$ and without any regularization.  An independent recovery of the same singular contribution by the limiting argument described above Eq.~\eqref{denGinOEreal} and in Proposition~\ref{Propscaled} is also available.

The Gaussian integration over $\tilde{\bf k}$ can now be performed directly.  Since
\[
\det C_0=\left(\frac{D}{4}\right)^N,\qquad D=1-\tau^2|s|^2,
\]
one obtains, up to the overall normalization restored below,
\[
(\det C_0)^{-1/2}\exp{\frac{1}{32}{\bf b}^TC^{-1}_{0}{\bf b}}
=  {\left(\frac{4}{D}\right)^{N/2}}
 \exp{-\frac{\tau}{8D}\left[{\bf b}_1^T{\bf b}_1\overline{s}+{\bf b}_2^T{\bf b}_2s\right]
 +\frac{1}{4D}{\bf b}_1^T{\bf b}_2}.
\]
Here we used the inverse of $C_0=\lim_{\epsilon\to0}C_\epsilon$, namely
  \begin{equation}\label{quad_form_inverse}
  C^{-1}_{0}=\frac{2}{D}\left(\begin{array}{cc} 1-\tau \Re({\bf v}^T{\bf v}) & -\tau \Im({\bf v}^T{\bf v})\\
-\tau \Im({\bf v}^T{\bf v}) & 1+\tau \Re({\bf v}^T{\bf v})\end{array}\right)\otimes {\bf 1}_N, \quad D=1-\tau^2|{\bf v}^T{\bf v}|^2.
\end{equation}
Substituting here the expressions for ${\bf b}_{1,2}$ from Eq.(\ref{def_b}) and introducing the following 3 matrices of size $N\times N$:
\begin{equation}\label{def_M}
{\cal M}=\tau^2 \left({\bf v}^T{\bf v}\right)\,\overline{{\bf v}\otimes {\bf v}^{T}}+\tau^2 \overline{ \left({\bf v}^T{\bf v}\right)}\,
{\bf v}\otimes {\bf v}^{T}-\tau^2\overline{{\bf v}\otimes {\bf v}^*}-{\bf v}\otimes {\bf v}^*
\end{equation}
\begin{equation}\label{def_K12}
{\cal K}_{12}=w{\bf 1}_N+\frac{\overline{z}\tau}{D} \left({\bf v}^T{\bf v}\right)\,\overline{{\bf v}\otimes {\bf v}^{T}}+
\frac{z\tau^2}{D} \overline{ \left({\bf v}^T{\bf v}\right)}\,{\bf v}\otimes {\bf v}^{T}-
\frac{1}{D}\left(\overline{z}\tau\overline{{\bf v}\otimes {\bf v}^*}+z{\bf v}\otimes {\bf v}^*\right)
\end{equation}
and
\begin{equation}\label{def_K21}
{\cal K}_{21}=\overline{w}{\bf 1}_N+\frac{z\tau}{D}\overline{ \left({\bf v}^T{\bf v}\right)}\,{\bf v}\otimes {\bf v}^{T}+
\frac{\overline{z}\tau^2}{D}  \left({\bf v}^T{\bf v}\right)\,\overline{{\bf v}\otimes {\bf v}^{T}}-
\frac{1}{D}\left(z\tau\overline{{\bf v}\otimes {\bf v}^*}+\overline{z}{\bf v}\otimes {\bf v}^*\right)
\end{equation}
allows us to represent our main object of interest as
\[
{\cal P}_N(z,{\bf v})\propto  {\left(\frac{4}{D}\right)^{N/2}}e^{-\frac{1}{D}\left[|z|^2-\frac{1}{2}\tau(z^2+\overline{z}^2)|{\bf v}^T{\bf v}|^2\right]}
\]
\begin{equation}\label{JPDinter3}
\times \lim_{|w-z|\to 0}\frac{\partial^2}{\partial w\partial \overline{w}}
 \int \, {\cal D}(\mathbf{\Psi},\mathbf{\Phi})e^{-\frac{i}{2}\left[\left(\mathbf{\Psi}_{+}^T{\cal K}_{12}\mathbf{\Phi}_{-}\right)
  + \left(\mathbf{\Psi}_{-}^T{\cal K}_{21}\mathbf{\Phi}_{+}\right)\right]+\frac{1}{4}\left(\mathbf{\Psi}_{+}^T\mathbf{\Phi}_{+}\right)\left(\mathbf{\Psi}_{-}^T\left\{{\bf 1}_N+\frac{\cal M}{D}\right\}\mathbf{\Phi}_{-}\right)},
  \end{equation}
  where here and below we do not write explicitly constant multiplicative factors which will be restored only in the final expressions.
The last term in the exponent is quartic in anticommuting variables, so it is expedient to employ a variant of the Hubbard-Stratonovich transformation:
\begin{equation}\label{HS_trafo1}
e^{\frac{1}{4}\left(\mathbf{\Psi}_{+}^T\mathbf{\Phi}_{+}\right)\left(\mathbf{\Psi}_{-}^T\left\{{\bf 1}_N+\frac{\cal M}{D}\right\}\mathbf{\Phi}_{-}\right)}
=\int e^{-q\overline{q}-\frac{q}{2} \left(\mathbf{\Psi}_{+}^T\mathbf{\Phi}_{+}\right)-\frac{\overline{q}}{2} \left(\mathbf{\Psi}_{-}^T\left\{{\bf 1}_N+\frac{\cal M}{D}\right\}\mathbf{\Phi}_{-}\right)}\,\frac{dqd\overline{q}}{\pi}.
\end{equation}
Substituting this relation back to Eq.(\ref{JPDinter3}) allows to perform the (by now, Gaussian) integral over anticommuting variables, eventually resulting in
\begin{equation}\label{JPDinter4}
{\cal P}_N(z,{\bf v})\propto  {\left(\frac{4}{D}\right)^{N/2}}e^{-\frac{1}{D}\left[|z|^2-\frac{1}{2}\tau(z^2+\overline{z}^2)|{\bf v}^T{\bf v}|^2\right]}
 \lim_{|w-z|\to 0}\frac{\partial^2}{\partial w\partial \overline{w}}
 \int e^{-q\overline{q}}\det{{\cal T}_N(|q|^2)},
  \end{equation}
  where we defined
  \begin{equation}\label{Tmat}
   {\cal T}_N(|q|^2)=|q|^2\left[{\bf 1}_N+\frac{\cal M}{D}\right]+{\cal K}_{21}{\cal K}_{12}\,.
   \end{equation}
The definitions of ${\cal M},{\cal K}_{12},{\cal K}_{21}$ show that
\[
{\cal T}_N(|q|^2)-\left(|q|^2+|w|^2\right){\bf 1}_N
\]
has range contained in $\operatorname{span}\{{\bf v},\overline{\bf v}\}$ and therefore has rank at most two.  Hence $|q|^2+|w|^2$ is an eigenvalue of multiplicity at least $N-2$.  On the open dense set $|{\bf v}^T{\bf v}|<1$, the vectors ${\bf v}$ and $\overline{\bf v}$ are linearly independent, and the remaining two eigenvalues $\lambda_{1,2}$ are obtained from the restriction to this two-dimensional space.  Thus
\begin{equation}\label{det_T_interm}
\det{{\cal T}_N(|q|^2)}=\left(|q|^2+|w|^2\right)^{N-2}\lambda_1\lambda_2.
\end{equation}
If $|{\bf v}^T{\bf v}|=1$, the span is one-dimensional and the complementary
eigenspace has dimension $N-1$. For $\tau<1$ the same determinant identity
then follows by continuity, since $D\geq1-\tau^2>0$.

To determine $\lambda_1\lambda_2$ on the two-dimensional set we compute the action of ${\cal T}_N(|q|^2)$ in the basis $\{{\bf v},\overline{\bf v}\}$. Introducing the notations
\begin{equation}\label{notations}
\delta w=w-z, \quad \zeta=z-\tau \overline{z}
\end{equation}
we first find that
\begin{equation}\label{actionK12}
{\cal K}_{12}{\bf v}=\delta w{\bf v}, \quad {\cal K}_{12}\overline{\bf v}=-\left(1-\tau^2\right)\frac{z}{D}\overline{\left({\bf v}^T{\bf v}\right)}\,\,{\bf v}
+\left(\delta w+\frac{\zeta+\tau|{\bf v}^T{\bf v}|^2 \overline{\zeta}}{D}\right)\overline{\bf v}
\end{equation}
and
\begin{equation}\label{actionK21}
{\cal K}_{21}{\bf v}=\left(\overline{\delta w}+\frac{\tau|{\bf v}^T{\bf v}|^2 \zeta}{D}\right){\bf v}-\frac{\tau\left({\bf v}^T{\bf v}\right) \zeta}{D} \overline{\bf v},
\quad
{\cal K}_{21}\overline{\bf v}=-\frac{\overline{\zeta}}{D}\overline{\left({\bf v}^T{\bf v}\right)}\,\,{\bf v}+\left(\overline{\delta w}+\frac{\overline{\zeta}}{D}\right)\,\overline{\bf v},
\end{equation}
in turn implying
\begin{equation}\label{actionK21K12v}
{\cal K}_{21}{\cal K}_{12}{\bf v}=\delta w\left[\left(\overline{\delta w}+\frac{\tau|{\bf v}^T{\bf v}|^2 \zeta}{D}\right){\bf v}-\frac{\tau\left({\bf v}^T{\bf v}\right) \zeta}{D} \overline{\bf v}\right]
\end{equation}
and
\begin{equation}\label{actionK21K12v*}
{\cal K}_{21}{\cal K}_{12}\overline{\bf v}=-\frac{\overline{\left({\bf v}^T{\bf v}\right)}}{D}\left[\overline{\zeta}\left(\delta w+\frac{\zeta+\tau|{\bf v}^T{\bf v}|^2 \overline{\zeta}}{D}\right)+\left(1-\tau^2\right)z\left(\overline{\delta w}+\frac{\tau|{\bf v}^T{\bf v}|^2 \zeta}{D}\right)\right]{\bf v}
\end{equation}
\[
+\left[\left(\delta w+\frac{\zeta+\tau|{\bf v}^T{\bf v}|^2 \overline{\zeta}}{D}\right)\left(\overline{\delta w}+\frac{\overline{\zeta}}{D}\right)+\frac{\tau(1-\tau^2)|{\bf v}^T{\bf v}|^2 z \zeta}{D^2}
\right]\overline{\bf v}.
\]
We also have that
\begin{equation}\label{actionM}
\left({\bf 1}_N+\frac{\cal M}{D}\right){\bf v}=0, \quad \left({\bf 1}_N+\frac{\cal M}{D}\right)\overline{\bf v}=-\frac{1-\tau^2}{D}
\left[\overline{\left({\bf v}^T{\bf v}\right)}{\bf v}-\overline{\bf v}\right].
\end{equation}
Combining all these expressions one may find the characteristic equation $\det{\left({\cal F}-\lambda{\bf 1}_2\right)}=0$, where the matrix ${\cal F}$ is specified below:
\begin{equation}\label{calF}
\left(\begin{array}{cc}\left(\overline{\delta w}+\frac{\tau|{\bf v}^T{\bf v}|^2 \zeta}{D}\right)\delta w & -\frac{\overline{\left({\bf v}^T{\bf v}\right)}}{D}
\left[|q|^2(1-\tau^2)+\overline{\zeta}\left(\delta w+\frac{\zeta+\tau|{\bf v}^T{\bf v}|^2 \overline{\zeta}}{D}\right)+(1-\tau^2)z
\left(\overline{\delta w}+\frac{\tau|{\bf v}^T{\bf v}|^2 \zeta}{D}\right)\right]\\-\frac{\tau \zeta}{D}\left({\bf v}^T{\bf v}\right)\delta w  &
\frac{1-\tau^2}{D}\left(|q|^2+\frac{\tau\zeta z}{D}|{\bf v}^T{\bf v}|^2\right)+\left(\delta w+\frac{\zeta+\tau|{\bf v}^T{\bf v}|^2 \overline{\zeta}}{D}\right)
\left(\overline{\delta w}+\frac{\overline{\zeta}}{D}\right) \end{array}\right)
\end{equation}
The two roots of the characteristic equations are precisely the two eigenvalues $\lambda_1$ and $\lambda_2$, hence $\lambda_1\lambda_2=\det{\cal F}(\delta w, \overline{\delta w})$.
At this point it is useful to look back at Eq.(\ref{JPDinter4}) and Eq.(\ref{det_T_interm}) and recall that for our goals we need to evaluate in fact the following limit:
\begin{equation}\label{limit}
\lim_{|\delta w|\to 0}\frac{\partial^2}{\partial (\delta w)\,\partial \overline{(\delta w)}}\,\left\{
\left[|q|^2+(z+\delta w)(\overline{z}+\overline{\delta w})\right]^{N-2}\det{\cal F}(\delta w, \overline{\delta w})\right\}.
\end{equation}
A direct $2\times2$ determinant calculation gives the exact identity
\begin{equation}\label{detcalF}
\det{\cal F}(\delta w, \overline{\delta w})
=|\delta w|^2\left[
F+\delta w\,\frac{\overline{\zeta}+\tau |s|^2\zeta}{D}
+\overline{\delta w}\,\frac{\zeta+\tau |s|^2\overline{\zeta}}{D}
+|\delta w|^2\right],
\end{equation}
where $s={\bf v}^T{\bf v}$ and
\begin{equation}\label{detcalF_coeff}
F=\frac{1-\tau^2}{D}|q|^2+\frac{\left|\zeta+\tau |s|^2\overline{\zeta}\right|^2}{D^2}.
\end{equation}
This gives the complete expansion to the stated order.  Since the prefactor $\left[|q|^2+(z+\delta w)(\overline z+\overline{\delta w})\right]^{N-2}$ is smooth at $\delta w=0$, Eq.~\eqref{limit} is immediately equal to $\left(|q|^2+|z|^2\right)^{N-2}F$. Combining
all factors and definitions together, restoring the correct normalization and using $R=|q|^2$ as the integration variable one finally arrives at an explicit expression for the one-point joint intensity of an eigenvalue $z$ and its phase-averaged normalized right eigenvector ${\bf v}$
in the form:
\[
{\cal P}_N(z,{\bf v})= \frac{1}{\pi \Gamma(N)} \left(\frac{1}{D}\right)^{N/2}\,e^{-\frac{1}{D}\left(|z|^2-\frac{1}{2}\tau(z^2+\overline{z}^2)|{\bf v}^T{\bf v}|^2\right)}
\]
\begin{equation}\label{JPDfin1}
\times  \int_0^{\infty} dR e^{-R}\left[R+|z|^2\right]^{N-2}\left(\frac{1-\tau^2}{D}R+\frac{\left|z-\tau \overline{z}+\tau (\overline{z}-\tau z)|{\bf v}^T{\bf v}|^2\right|^2}{D^2}\right),  \quad D=1-\tau^2|{\bf v}^T{\bf v}|^2.
  \end{equation}
The overall multiplicative constant, which was not tracked through the intermediate steps above, is fixed once and for all by the case $\tau=0$, in which the omitted factors cannot depend on $\tau$, $z$ or ${\bf v}$. Indeed, at $\tau=0$ one has $D=1$, the round bracket in
Eq.(\ref{JPDfin1}) reduces to $R+|z|^2$, and
\[
\int_0^{\infty}dR\,e^{-R}\left(R+|z|^2\right)^{N-1}=e^{|z|^2}\int_{|z|^2}^{\infty}dv\,e^{-v}v^{N-1}=e^{|z|^2}\Gamma\left(N,|z|^2\right),
\]
so that the right-hand side of Eq.(\ref{JPDfin1}) becomes $\Gamma(N,|z|^2)/\left(\pi\Gamma(N)\right)$. By the unitary invariance of the complex Ginibre
ensemble the joint intensity at $\tau=0$ must be independent of ${\bf v}$ and equal to the mean eigenvalue density Eq.(\ref{denGinUE}), which is exactly this
expression. Hence the prefactor displayed in Eq.(\ref{JPDfin1}) is the correct one.

For $\tau>0$, shifting the integration variable $R+|z|^2\to v\in [|z|^2, \infty)$ and collecting also the factor $e^{|z|^2}$ generated by $e^{-R}$ gives
\[
{\cal P}_N(z,{\bf v})= \frac{1}{\pi \Gamma(N)} \left(\frac{1}{D}\right)^{N/2}\, e^{\left(1-\frac{1}{D}\right)\frac{2\tau |z|^2-(z^2+\overline{z}^2)}{2\tau}}\int_{|z|^2}^{\infty} dv\, e^{-v}v^{N-2}
\]
\begin{equation}\label{JPDfin2}
\times  \left\{\frac{1-\tau^2}{D}v+|z|^2\left(1- \frac{1-\tau^2}{D}\right)-\tau(z^2+\overline{z}^2)\frac{1-|{\bf v}^T{\bf v}|^2}{D}+\tau^2|z|^2\left(\frac{1-|{\bf v}^T{\bf v}|^2}{D}\right)^2\right\}.
  \end{equation}
The case $\tau=0$ is read directly from Eq.~\eqref{JPDfin1} and agrees with the continuous $\tau\downarrow0$ limit of the final formula.

Using the definition Eq.(\ref{gamma_incom}) and the identity
\[\frac{1-|{\bf v}^T{\bf v}|^2}{D}=\frac{1}{\tau^2}\left(1-\frac{1-\tau^2}{D}\right)\]
the form of Eq.(\ref{JPDfin2}) is finally reduced to one
presented in the Theorem (\ref{Theorem1}), see Eq.(\ref{MainJPDinterpolating}).

\subsection{The real-eigenvalue sector at $\tau=1$}\label{realsector}

The computation above was carried out on the complex unit sphere and, as explained after Eq.(\ref{quad_form_inverse}), it degenerates at $\tau=1$ precisely on the
locus $|{\bf v}^T{\bf v}|=1$. That locus is exactly where the real eigenvalues live: if $X$ is real and $X{\bf v}=z{\bf v}$ with $z\in\mathbb{R}$, then ${\bf v}$
may be chosen real, so that $|{\bf v}^T{\bf v}|=1$; conversely, if $|{\bf v}^T{\bf v}|=1$ for a unit vector ${\bf v}$, then by the equality case in the
Cauchy-Schwarz inequality ${\bf v}$ is a real vector times a phase, and $X{\bf v}=z{\bf v}$ with $X$ real forces $z\in\mathbb{R}$. The degeneration therefore indicates that the real sector is naturally
formulated on the real sphere. We now treat this sector directly, by an exact
finite-$N$ computation rather than by continuation from $\tau<1$.

\subsubsection{The real Kac-Rice identity}

Let ${\bf S}^{\mathbb R}_N=\{{\bf v}\in\mathbb{R}^N:{\bf v}^T{\bf v}=1\}$ and write
\[
d\sigma_N({\bf v})=\delta\left({\bf v}^T{\bf v}-1\right)d^N{\bf v}=\frac{1}{2}dS({\bf v}),\qquad
\int_{{\bf S}^{\mathbb R}_N}d\sigma_N=\frac{\pi^{N/2}}{\Gamma\left(\frac{N}{2}\right)}.
\]
The relevant scalar identity is the real analogue of Eq.(\ref{main_integral_identity}) recorded in the Remark at the end of Appendix \ref{main_integral_identity_induction}:
for $B\in\mathcal{M}_N(\mathbb{R})$ of corank at most one,
\begin{equation}\label{real_sphere_delta}
\int_{\mathbb{R}^N}\delta^{(N)}(B{\bf v})\,\delta\left({\bf v}^T{\bf v}-1\right)d^N{\bf v}=\delta\left(\det B\right).
\end{equation}
Note the absence of a factor $\pi$, in contrast with Eq.(\ref{main_integral_identity}).

\begin{theorem}[Sign-averaged real Kac--Rice formula]\label{RealMetaTheorem}
Let $X\in\mathcal{M}_N(\mathbb{R})$ have simple eigenvalues, let $x_1,\ldots,x_m$ be those of them which are real, and let ${\bf v}_a\in{\bf S}^{\mathbb R}_N$ be a
corresponding real unit eigenvector. Set $p_X(x)=\det\left(X-x{\bf 1}_N\right)$. Then for every $F\in C_c^{\infty}(\mathbb{R}\times{\bf S}^{\mathbb R}_N)$,
\begin{equation}\label{real_KacRice_test}
\int_{\mathbb{R}}\int_{{\bf S}^{\mathbb R}_N}F(x,{\bf v})\,\delta^{(N)}\!\left(\left(X-x{\bf 1}_N\right){\bf v}\right)\left|p'_X(x)\right|\,dx\,d\sigma_N({\bf v})
=\sum_{a=1}^{m}\frac{F(x_a,{\bf v}_a)+F(x_a,-{\bf v}_a)}{2}.
\end{equation}
Equivalently, the sign-averaged marked empirical measure of the real eigenvalues and their real unit eigenvectors has the distributional density
$\left|p'_X(x)\right|\delta^{(N)}\left(\left(X-x{\bf 1}_N\right){\bf v}\right)$ with respect to $dx\,d\sigma_N({\bf v})$.
\end{theorem}

\begin{proof}
Consider $\mathcal{F}_X:\mathbb{R}\times{\bf S}^{\mathbb R}_N\to\mathbb{R}^N$, $\mathcal{F}_X(x,{\bf v})=(X-x{\bf 1}_N){\bf v}$. Both the domain and the target
have dimension $N$, so in contrast with the complex case of Theorem \ref{MainMetaTheorem} the zero set is discrete: it consists exactly of the $2m$ points
$(x_a,\pm{\bf v}_a)$, there being no continuous gauge orbit but only the residual sign. Writing $B_a=X-x_a{\bf 1}_N$, denoting by $s_1,\ldots,s_{N-1}$ its
nonzero singular values and by ${\boldsymbol\ell}_a$ a real unit left null vector, the differential
\[
D\mathcal{F}_X(\dot x,\dot{\bf v})=B_a\dot{\bf v}-\dot x\,{\bf v}_a,\qquad \dot{\bf v}\perp{\bf v}_a,
\]
is a real linear isomorphism $\,{\bf v}_a^{\perp}\oplus\mathbb{R}\to\mathbb{R}^N$ whose determinant is, exactly as in Eq.(\ref{normal_jacobian_eigenpair}) but
without squaring, the product of the volume $\prod_{j=1}^{N-1}s_j$ of the image of ${\bf v}_a^{\perp}$ under $B_a$ with the component
$|{\boldsymbol\ell}_a^T{\bf v}_a|$ of ${\bf v}_a$ orthogonal to $\mbox{Ran}\,B_a={\boldsymbol\ell}_a^{\perp}$. By Eq.(\ref{pprime_singular}), in its real form,
\begin{equation}\label{real_normal_jacobian}
J_{\mathcal{F}_X}(x_a,\pm{\bf v}_a)=\left(\prod_{j=1}^{N-1}s_j\right)\left|{\boldsymbol\ell}_a^T{\bf v}_a\right|=\left|p'_X(x_a)\right|>0,
\end{equation}
so that $0$ is a regular value of $\mathcal{F}_X$ along its zero set. Lemma \ref{lemma_coarea_delta} with $k=N$, $\dim M=N$ and
$\mathcal{H}^0$ the counting measure now gives, using $d\sigma_N=\frac12 dS$,
\[
\int F\,\delta^{(N)}\left(\mathcal{F}_X\right)\left|p'_X\right|dx\,d\sigma_N
=\frac12\sum_{a=1}^m\sum_{\varsigma=\pm1}F(x_a,\varsigma{\bf v}_a),
\]
which is Eq.(\ref{real_KacRice_test}).
\end{proof}

Setting $F\equiv1$ in Eq.(\ref{real_KacRice_test}) and performing the ${\bf v}-$integration first by means of Eq.(\ref{real_sphere_delta}) returns
$\int_{\mathbb{R}}\left|p'_X(x)\right|\delta\left(p_X(x)\right)dx$, the one-dimensional Kac-Rice count of the real zeros of $p_X$, as it must. In particular the
${\bf v}-$marginal of Eq.(\ref{real_KacRice_test}) is the empirical density of the real eigenvalues,
\begin{equation}\label{real_marginal}
\rho^{\mathbb R}_X(x)=\left|p'_X(x)\right|\delta\left(p_X(x)\right)=\left|p'_X(x)\right|\int_{{\bf S}^{\mathbb R}_N}\delta^{(N)}\!\left(\left(X-x{\bf 1}_N\right){\bf v}\right)d\sigma_N({\bf v}).
\end{equation}

\subsubsection{Averaging over the real Ginibre ensemble}

\begin{proposition}\label{GinOErealprop}
For $N\ge2$ the mean density of real eigenvalues of the real Ginibre ensemble is
\begin{equation}\label{GinOEreal_master}
p_N^{(GinOE,r)}(x)=\frac{e^{-x^2/2}}{2^{N/2}\,\Gamma\!\left(\frac{N}{2}\right)}\;
\mathbb{E}\left|\det\left(G^{(R)}_{N-1}-x{\bf 1}_{N-1}\right)\right|,
\end{equation}
where $G^{(R)}_{N-1}$ is a real Ginibre matrix of size $N-1$.
\end{proposition}

\begin{proof}
Averaging Eq.(\ref{real_marginal}) over $X=G^{(R)}$ we have
\[
p_N^{(GinOE,r)}(x)=\int_{{\bf S}^{\mathbb R}_N}\mathbb{E}\left[\left|p'_X(x)\right|\delta^{(N)}\!\left(\left(X-x{\bf 1}_N\right){\bf v}\right)\right]d\sigma_N({\bf v}).
\]
The law of $G^{(R)}$ is invariant under $X\mapsto OXO^T$ for $O$ real orthogonal, while $\left(X-x\right){\bf v}\mapsto O\left(X-x\right)O^T{\bf v}$ and
$p_X$ is unchanged; since $\delta^{(N)}$ is invariant under the orthogonal $O$, the expectation in the integrand does not depend on ${\bf v}$ and may be evaluated
at ${\bf v}={\bf e}_1$. Hence
\begin{equation}\label{GinOEreal_e1}
p_N^{(GinOE,r)}(x)=\frac{\pi^{N/2}}{\Gamma\left(\frac{N}{2}\right)}\,
\mathbb{E}\left[\left|p'_X(x)\right|\,\delta\left(X_{11}-x\right)\delta^{(N-1)}\left({\bf c}\right)\right],
\end{equation}
where we wrote $X{\bf e}_1=\left(X_{11},{\bf c}\right)$ with ${\bf c}\in\mathbb{R}^{N-1}$ the remaining part of the first column.

The Dirac factors do the rest of the work. On their support one has ${\bf c}=0$, so that $X$ is block triangular,
\[
X=\left(\begin{array}{cc}x & {\bf b}^T\\ 0 & X_{N-1}\end{array}\right),\qquad
p_X(t)=\left(x-t\right)\det\left(X_{N-1}-t{\bf 1}_{N-1}\right),
\]
whence, differentiating and setting $t=x$,
\begin{equation}\label{pprime_triangular}
p'_X(x)=-\det\left(X_{N-1}-x{\bf 1}_{N-1}\right).
\end{equation}
Equation~(\ref{pprime_triangular}) is exact on the support of the Dirac
factors. Since $X_{11}\sim\mathcal{N}(0,1)$ and
${\bf c}\sim\mathcal{N}(0,{\bf 1}_{N-1})$ are independent of $X_{N-1}$,
which is itself a real Ginibre matrix of size $N-1$, the expectation in
Eq.~(\ref{GinOEreal_e1}) factorizes into the Gaussian densities of $X_{11}$
and ${\bf c}$, evaluated respectively at $X_{11}=x$ and ${\bf c}=0$, and
the remaining characteristic-polynomial average:
\[
p_N^{(GinOE,r)}(x)
=
\frac{\pi^{N/2}}{\Gamma\left(\frac{N}{2}\right)}
\cdot\frac{e^{-x^2/2}}{\sqrt{2\pi}}
\cdot\frac{1}{(2\pi)^{(N-1)/2}}
\cdot
\mathbb{E}\left|
\det\left(G^{(R)}_{N-1}-x{\bf 1}_{N-1}\right)
\right|.
\]
Using $\pi^{N/2}(2\pi)^{-N/2}=2^{-N/2}$ gives
Eq.~(\ref{GinOEreal_master}).
\end{proof}

\begin{remark}
It is worth stressing what has happened. The Dirac factor $\delta^{(N)}\left((X-x{\bf 1}_N){\bf v}\right)$ of the Kac--Rice representation imposes exactly the
statement that ${\bf v}$ is an eigenvector, and in the basis adapted to ${\bf v}$ this is the statement that $X$ is block triangular. The Kac--Rice route therefore
reproduces the {\it partial real Schur decomposition} of Edelman, Kostlan and Shub \cite{EKS,Ed:97} --- the second of the two methods listed in the Introduction ---
with no Jacobian computation at all, the Jacobian being supplied automatically by Eq.(\ref{real_normal_jacobian}). This is the real counterpart of the observation,
made after Eq.(\ref{main_practical}), that in the complex case the same factor produces $|p'_X(z)|^2$.
\end{remark}

Formula (\ref{GinOEreal_master}) reduces the singular sector to a single classical Gaussian expectation, namely the expected modulus of the characteristic
polynomial of a real Ginibre matrix at a real point. That expectation is exactly the quantity evaluated by Edelman, Kostlan and Shub, and reads
\begin{equation}\label{Edet_classical}
\mathbb{E}\left|\det\left(G^{(R)}_{M}-x{\bf 1}_{M}\right)\right|
=\frac{2^{M/2}\Gamma\!\left(\frac{M+1}{2}\right)}{\sqrt{\pi}\,\Gamma(M)}
\left[e^{x^2/2}\Gamma\left(M,x^2\right)+|x|^{M}\int_0^{|x|}u^{M-1}e^{-u^2/2}\,du\right],\quad M\ge1 .
\end{equation}
Substituting Eq.(\ref{Edet_classical}) with $M=N-1$ into Eq.(\ref{GinOEreal_master}), the two Gamma functions $\Gamma\left(\frac{M+1}{2}\right)=\Gamma\left(\frac{N}{2}\right)$
cancel and the powers of two combine into $2^{(N-1)/2}/2^{N/2}=2^{-1/2}$, so that the prefactor becomes $1/(\sqrt{2\pi}\,\Gamma(N-1))$ while the factor $e^{-x^2/2}$
cancels the $e^{x^2/2}$ of the first term inside the square bracket. One recovers in this way precisely Eq.(\ref{denGinOEreal}):
\[
p_N^{(GinOE,r)}(x)=\frac{1}{\sqrt{2\pi}\,\Gamma(N-1)}\left[\Gamma\left(N-1,x^2\right)+|x|^{N-1}e^{-x^2/2}\int_0^{|x|}u^{N-2}e^{-u^2/2}\,du\right].
\]
Thus the singular contribution to Eq.(\ref{denGinOEfull}) is indeed obtained by the methods of the present paper, at finite $N$ and without any regularization,
the only external input being the classical Gaussian expectation Eq.(\ref{Edet_classical}).

\begin{remark}
The absolute characteristic-polynomial average in
Eq.~(\ref{Edet_classical}) can also be evaluated by supersymmetry methods
using half-integer powers of characteristic polynomials arising in the
real Ginibre problem, see \cite{Fyo2018}. Here we use the direct finite-$N$
reduction above, which is particularly transparent for the present purpose.
\end{remark}

\begin{corollary}\label{realzero}
For every $N\ge2$ one has $p_N^{(GinOE,r)}(0)=\frac{1}{\sqrt{2\pi}}$, independently of $N$.
\end{corollary}
\begin{proof}
At $x=0$ the matrix $G^{(R)}_{N-1}$ has independent standard Gaussian rows, so by the Gram--Schmidt (chi) decomposition of the volume
$\left|\det G^{(R)}_{M}\right|=\prod_{k=1}^{M}\chi_k$ with independent $\chi_k$ having $k$ degrees of freedom, whence
\[
\mathbb{E}\left|\det G^{(R)}_{M}\right|=\prod_{k=1}^{M}\frac{\sqrt2\,\Gamma\left(\frac{k+1}{2}\right)}{\Gamma\left(\frac{k}{2}\right)}
=\frac{2^{M/2}\Gamma\left(\frac{M+1}{2}\right)}{\Gamma\left(\frac12\right)}
=\frac{2^{M/2}\Gamma\left(\frac{M+1}{2}\right)}{\sqrt{\pi}},
\]
the product telescoping. Inserting this with $M=N-1$ into Eq.(\ref{GinOEreal_master}) gives
$p_N^{(GinOE,r)}(0)=2^{-N/2}\Gamma\left(\frac{N}{2}\right)^{-1}\cdot2^{(N-1)/2}\Gamma\left(\frac{N}{2}\right)/\sqrt{\pi}=1/\sqrt{2\pi}$.
\end{proof}
\begin{remark}
The same value follows immediately by setting $x=0$ in the classical
finite-$N$ density formula of Edelman, Kostlan and Shub \cite{EKS}.
\end{remark}
This last identity provides a sharp consistency check between the finite-$N$ singular sector and the scaling regime of weak non-reality: the constant
$1/\sqrt{2\pi}$ of Corollary \ref{realzero} is exactly the coefficient of $\delta(y)$ in Eq.(\ref{weaknonRden_equiv_lim}), which was obtained there by a
completely independent route, from the $\delta\to0$ limit of Proposition \ref{Propscaled}.

\section{ Proofs for additively deformed complex Ginibre matrices}

In this section we prove the Theorem \ref{Theorem2} by applying our method to the ensemble of an additive deformation of the complex Ginibre ensemble
\begin{equation}\label{additive_def}
G_{A}=G^{(C)}+A
\end{equation}
where $G^{(C)}$ is taken from the ensemble of complex $N\times N$ Ginibre matrices and $A$ is an arbitrary fixed matrix of size $N$.

We start with proving

  \begin{proposition}\label{proposition_deformation}
    The one-point joint intensity of an eigenvalue $z$ and the corresponding phase-averaged normalized right eigenvector ${\bf v}$, with respect to $d^2z\,d\mu_H({\bf v})$, for matrices taken from the additively deformed complex Ginibre ensemble Eq.(\ref{additive_def}) is given by
\begin{equation}\label{MainJPDadditivedef}
{\cal P}^{(A)}_N(z,{\bf v})=\frac{1}{\pi \Gamma(N)}\,e^{-{\bf v}^*A_z^*A_z{\bf v}}\int_0^{\infty}\frac{dR}{R}e^{-R}\,\det{\left[R\,{\bf 1}_N+T_A({\bf v})\right]},
\end{equation}
where we denoted $A_z:=A-z{\bf 1}_N$ and
\begin{equation}\label{TAV}
T_A({\bf v})=A_z^*\left({\bf 1}_N-{\bf v}\otimes{\bf v}^*\right)\,A_z\left({\bf 1}_N-{\bf v}\otimes{\bf v}^*\right).
\end{equation}
\end{proposition}

\begin{proof}
Following the same initial steps as in the previous section we have the following representation for the JPD of eigenvalue $z$ and
normalized right eigenvector ${\bf v}$ with respect to the measure $d^2z\,d\mu_N({\bf v})$ (cf. Eq.(\ref{JPDinter1}):
\begin{equation}\label{JPDadditive1}
{\cal P}^{(A)}_N(z,{\bf v})=\frac{1}{\pi}\lim_{|w-z|\to 0}\frac{\partial^2}{\partial w\partial \overline{w}}\lim_{\epsilon\to 0}
\int_{\mathbb{C}^N}\frac{d{\bf k}d{\bf k}^*}{(2\pi)^{2N}}e^{-\epsilon\frac{{\bf k}^*{\bf k}}{2}+\frac{i}{2}\left[{\bf k}^*(A-z{\bf 1}){\bf v}+
{\bf v}^*(A^*-\overline{z}{\bf 1}){\bf k}\right]}
\end{equation}
\[
\times \int \exp{-\frac{i}{2}\left[\left(\mathbf{\Psi}_{+}^*(A-w{\bf 1})\mathbf{\Phi}_{-}\right)
  + \left(\mathbf{\Psi}_{-}^*(A^*-\overline{w}{\bf 1})\mathbf{\Phi}_{+}\right)\right]}\,{\cal D}(\mathbf{\Psi},\mathbf{\Phi}) \,\left\langle e^{\frac{i}{2}\mbox{\small Tr }\left(XB_{-+}+X^*B_{+-}\right)}\right\rangle_{X}
\]
where we renamed $\mathbf{\Psi}^T_{+,-}\to \mathbf{\Psi}^*_{+,-}$ so that the matrices $B_{+-}$ and $B_{-+}$ are now defined as
\begin{equation}\label{Bmat_additive}
B_{-+}={\bf v}\otimes {\bf k}^*-\mathbf{\Phi}_{-}\otimes \mathbf{\Psi}^*_{+}, \quad B_{+-}={\bf k}\otimes {\bf v}^*-\mathbf{\Phi}_{+}\otimes \mathbf{\Psi}^*_{-}
\end{equation}
The ensemble average is now performed with the help of Eq.(\ref{RGin_ave}) where we put $\tau=0$ yielding
\begin{equation}\label{ans_ave}
 \left\langle e^{\frac{i}{2}\mbox{\small Tr }\left(XB_{-+}+X^*B_{+-}\right)}\right\rangle_{X}=e^{-\frac{1}{4}\left[{\bf k}^*{\bf k}-
 \left({\bf k}^*\mathbf{\Phi}_{+}\right)\left(\mathbf{\Psi}_{-}^*{\bf v}\right)- \left({\bf v}^*\mathbf{\Phi}_{-}\right)\left(\mathbf{\Psi}_{+}^*{\bf k}\right)
 -\left(\mathbf{\Psi}_{+}^*\mathbf{\Phi}_{+}\right)\left(\mathbf{\Psi}_{-}^*\mathbf{\Phi}_{-}\right)\right]}.
\end{equation}
Substituting the above back to Eq.(\ref{JPDadditive1}) we see that the exponential becomes
 a Hermitian quadratic form $\frac{1}{2}\left(\epsilon+\frac{1}{2}\right){\bf k}^*{\bf k}$ which is positive definite even for $\epsilon=0$. Hence the $\epsilon-$regularization can be safely  omitted, with the corresponding Gaussian integral readily evaluated, yielding the factor proportional to:
 \begin{equation}\label{factor}
 e^{-\mathbf{v}^*\left(A^*-\overline{z}{\bf 1}\right)\left(A-z{\bf 1}\right){\bf v}-\frac{i}{2}\mathbf{\Psi}_{+}^*\left[(A-z{\bf 1}\right]{\bf v}\otimes {\bf v}^*\mathbf{\Phi}_{-} -\frac{i}{2}\mathbf{\Psi}_{-}^*\left[{\bf v}\otimes {\bf v}^*(A^*-\overline{z}{\bf 1}\right]\mathbf{\Phi}_{+}-\frac{1}{4}\left(\mathbf{\Psi}_{+}^*\mathbf{\Phi}_{+}\right)\left(\mathbf{\Psi}_{-}^*\left({\bf v}\otimes {\bf v}^*\right)\mathbf{\Phi}_{-}\right)}
  \end{equation}
Substituting this back to Eq.(\ref{JPDadditive1}) and combining with the last exponential factor in Eq.(\ref{ans_ave}) we arrive at
\[
{\cal P}^{(A)}_N(z,{\bf v})=\frac{1}{\pi^{N+1}}e^{-\mathbf{v}^*\left(A^*-\overline{z}{\bf 1}\right)\left(A-z{\bf 1}\right){\bf v}}
\]
\begin{equation}\label{JPDadditive2}
\times \lim_{|w-z|\to 0}\frac{\partial^2}{\partial w\partial \overline{w}}  \int e^{\frac{i}{2}\left(\mathbf{\Psi}_{+}^*,\mathbf{\Psi}_{-}^*\right)\mathcal H(w,\overline{w})\left(\begin{array}{c}\mathbf{\Phi}_{+}\\ \mathbf{\Phi}_{-}\end{array}\right)+
\frac{1}{4}\left(\mathbf{\Psi}_{+}^*\mathbf{\Phi}_{+}\right)\left(\mathbf{\Psi}_{-}^*\left({\bf 1}_N-{\bf v}\otimes {\bf v}^*\right)\mathbf{\Phi}_{-}\right)}\,{\cal D}(\mathbf{\Psi},\mathbf{\Phi}),
 \end{equation}
where we defined the following matrix
\begin{equation}
{\cal H}(w,\overline{w})=\left(\begin{array}{cc} 0 & \left(A-w{\bf 1}\right)-\left(A-z{\bf 1}\right){\bf v}\otimes {\bf v}^*
 \\ \left(A^*-\overline{w}{\bf 1}\right)-{\bf v}\otimes {\bf v}^*\left(A^*-\overline{z}{\bf 1}\right) &0 \end{array}\right).
\end{equation}
To facilitate the further evaluation, we again employ a variant
 of the Hubbard-Stratonovich transformation, cf. (\ref{HS_trafo1}) :
\begin{equation}\label{HS_trafo2}
e^{\frac{1}{4}\left(\mathbf{\Psi}_{+}^*\mathbf{\Phi}_{+}\right)\left(\mathbf{\Psi}_{-}^*\left({\bf 1}_N-{\bf v}\otimes {\bf v}^*\right)\mathbf{\Phi}_{-}\right)}
=\int e^{-q\overline{q}+\frac{q}{2} \left(\mathbf{\Psi}_{+}^*\mathbf{\Phi}_{+}\right)+\frac{\overline{q}}{2} \left(\mathbf{\Psi}_{-}^*\left({\bf 1}_N-{\bf v}\otimes {\bf v}^*\right)\mathbf{\Phi}_{-}\right)}\,\frac{dqd\overline{q}}{\pi}.
\end{equation}
Substituting the above back to Eq.(\ref{JPDadditive2}) and performing the Gaussian integral over anticommuting variables yields
\[
{\cal P}^{(A)}_N(z,{\bf v})=\frac{1}{\pi^{N+1}}e^{-\mathbf{v}^*\left(A^*-\overline{z}{\bf 1}\right)\left(A-z{\bf 1}\right){\bf v}}
\]
\begin{equation}\label{JPDadditive2a}
\times \lim_{|w-z|\to 0}\frac{\partial^2}{\partial w\partial \overline{w}}  \int  e^{-q\overline{q}}\, \det{\left[\left(\begin{array}{cc}q{\bf 1}_N & 0\\ 0& \overline{q}\left({\bf 1}_N-{\bf v}\otimes {\bf v}^*\right)\end{array}\right)+i{\cal H}(w,\overline{w})\right]}\,\frac{dqd\overline{q}}{\pi}.
 \end{equation}
To evaluate the remaining limit we set the notations
\begin{equation}\label{def Proj}
 w:=z+\delta w,\quad  \overline{w}:=\overline{z}+\overline{\delta w}, \quad P({\bf v}):={\bf 1}_N-{\bf v}\otimes {\bf v}^*=P^*({\bf v}),\,\, A_z:=A-z{\bf 1}_N
\end{equation}
and after applying the Schur's formula (cf. Eq.(\ref{Schur})) for the determinant of block matrices aim at computing
\[
\lim_{|\delta w|\to 0}\frac{\partial^2}{\partial (\delta w)\partial (\overline{\delta w})}\det{\left\{|q|^2 P({\bf v})+\left(A_z P({\bf v})-\delta w{\bf 1}_N\right)^*\left(A_z P({\bf v})-\delta w{\bf 1}_N\right) \right\}}.
\]
 Further introduce a unitary matrix $U_{\bf v}=({\bf v}|{\bf v}_1|\ldots|{\bf v}_{N-1})$, where the columns  of the matrix starting from the second are vectors ${\bf v}_{i},\,i=1,\ldots,N-1$ forming an orthonormal basis spanning the subspace in $\mathbb{C}^N$  orthogonal to the vector ${\bf v}$. It is then obvious that $P({\bf v})U_{\bf v}=(0|{\bf v}_1|\ldots|{\bf v}_{N-1})$.
 As $\det{\left\{\ldots\right\}}=\det{U^*_{\bf v}\left\{\ldots \right\}U_{\bf v}}$ we arrive at the task of evaluating
 \begin{equation}
\det{\left[U^*_{\bf v}\left\{|q|^2 P({\bf v})+P({\bf v})A^*_zA_z P({\bf v})-\delta w P({\bf v})A^*_z-\overline{\delta w} A_zP({\bf v})+|\delta w|^2{\bf 1}_N\right\}U_{\bf v}\right]}.
\end{equation}
Explicit computation shows that the matrix under the determinant in the above expression has the following block structure:
\begin{equation}\label{block}
\left(\begin{array}{cc}|\delta w|^2 & -\overline{\delta w}\, {\bf t}^*\\  -\delta w\, {\bf t} & {\cal R}(\delta w, \overline{\delta w})\end{array}\right),
\end{equation}
where we introduced the vector ${\bf t}\in \mathbb{C}^{(N-1)}$ with components ${\bf t}_i={\bf v}^*_iA_z^*{\bf v}$ and the $(N-1)\times (N-1)$ matrix ${\cal R}(\delta w, \overline{\delta w})$ with entries
\begin{equation}\label{Rmat}
\left[{\cal R}(\delta w, \overline{\delta w})\right]_{ij}=(|q|^2+|\delta w|^2)\delta_{ij}+{\bf v}_i^*A_z^*A_z{\bf v}_j-\delta w {\bf v}^*_iA_z^*{\bf v}_j-
\overline{\delta w}{\bf v}^*_iA_z{\bf v}_j.
\end{equation}
The structure of the matrix in Eq.(\ref{block}) implies that its determinant can be represented by the Schur determinant formula as:
\begin{equation}\label{blockSchur}
\det{\left[{\cal R}(\delta w, \overline{\delta w})\right]}\left(1-{\bf t}^*{\cal R}^{-1}(\delta w, \overline{\delta w}){\bf t}\right)|\delta w|^2.
\end{equation}
Now it is obvious that
\begin{equation}\label{lim}
\lim_{|\delta w|\to 0}\frac{\partial^2}{\partial (\delta w)\partial (\overline{\delta w})}
\left\{\det{\left[{\cal R}(\delta w, \overline{\delta w})\right]}\left(1-{\bf t}^*{\cal R}^{-1}(\delta w, \overline{\delta w}){\bf t}\right)|\delta w|^2\right\}
\end{equation}
\begin{equation}
=\det{\left[{\cal R}(0,0)\right]}\left(1-{\bf t}^*{\cal R}^{-1}(0,0){\bf t}\right)=\det{\left({\cal R}(0,0)-{\bf t}\otimes {\bf t}^*\right)}.
\end{equation}
Combining the expressions
\[
\left[{\cal R}(0,0)\right]_{ij}=|q|^2\delta_{ij}+{\bf v}^*_iA^*_zA_z{\bf v}_j \quad \mbox{and} \quad \left[{\bf t}\otimes {\bf t}^*\right]_{ij}=\left({\bf v}^*_iA^*_z{\bf v}\right)
\left({\bf v}^*A_z{\bf v}_j\right)
\]
we have
\begin{equation}\label{def T}
\left[{\cal R}(0,0)-{\bf t}\otimes {\bf t}^*\right]_{ij}=|q|^2\delta_{ij}+T_{ij}, \quad T_{ij}={\bf v}^*_iA^*_zP({\bf v})A_z {\bf v}_j,
\end{equation}
resulting in the explicit formula for the JPD of an eigenvalue $z$ and the eigenvector ${\bf v}$ (with respect to the measure $d^2zd\mu_N({\bf v})$) for the additively perturbed complex Ginibre ensemble
 \begin{equation}\label{JPDadditive_fin}
{\cal P}^{(A)}_N(z,{\bf v})=\frac{1}{\pi^{N+1}}e^{-\mathbf{v}^*A^*_zA_z{\bf v}} \int  e^{-|q|^2}\, \det{\left[|q|^2
{\bf 1}_{N-1}+T({\bf v})\right]}\,\frac{dqd\overline{q}}{\pi},
\end{equation}
where we recall that
\begin{equation}\label{def_not_proj}
  \left[T({\bf v})\right]_{ij}={\bf v}^*_iA^*_zP({\bf v})A_z {\bf v}_j \quad  \mbox{where} \quad P({\bf v}):={\bf 1}_N-{\bf v}\otimes {\bf v}^*=P^*({\bf v}), \quad A_z:=A-z{\bf 1}_N.
\end{equation}
Note that $T({\bf v})$ is of size $(N-1)\times(N-1)$.  Let $V:\mathbb C^{N-1}\to\mathbb C^N$ be the isometry whose columns are ${\bf v}_1,\ldots,{\bf v}_{N-1}$, so that $VV^*=P({\bf v})$ and $V^*V={\bf1}_{N-1}$.  Then
\[
T({\bf v})=V^*A_z^*P({\bf v})A_zV,
\]
whereas
\begin{equation}\label{TAVproof}
T_A({\bf v})=A_z^*P({\bf v})A_zP({\bf v})
=\big(A_z^*P({\bf v})A_zV\big)V^*.
\end{equation}
Thus $T({\bf v})=BA$ and $T_A({\bf v})=AB$ with $A=A_z^*P({\bf v})A_zV$ and $B=V^*$.  Sylvester's determinant identity gives, for every $R>0$,
\begin{equation}\label{TAVient}
\det\!\left[R{\bf1}_N+T_A({\bf v})\right]
=R\,\det\!\left[R{\bf1}_{N-1}+T({\bf v})\right].
\end{equation}
Substituting Eq.~\eqref{TAVient} into Eq.~\eqref{JPDadditive_fin} and using radial integration $R=|q|^2$ gives Eq.~\eqref{MainJPDadditivedef}.  This argument does not require $T_A({\bf v})$ to be self-adjoint and avoids any expansion of $\log\det$.
\end{proof}

\begin{remark}
It is worth recording the consistency check provided by the unperturbed case $A=0$, which in particular confirms the $R^{-1}$ in the measure of
Eq.(\ref{MainJPDadditivedef}). There $A_z=-z{\bf 1}_N$, so $T_A({\bf v})=|z|^2P({\bf v})P({\bf v})=|z|^2P({\bf v})$ because $P({\bf v})$ is a projector, and
since $P({\bf v})$ has the eigenvalue $0$ once and the eigenvalue $1$ with multiplicity $N-1$,
\[
\det\left[R\,{\bf 1}_N+T_A({\bf v})\right]=R\left(R+|z|^2\right)^{N-1}.
\]
Eq.(\ref{MainJPDadditivedef}) then gives ${\cal P}^{(0)}_N(z,{\bf v})=\frac{e^{-|z|^2}}{\pi\Gamma(N)}\int_0^{\infty}dR\,e^{-R}\left(R+|z|^2\right)^{N-1}
=\frac{\Gamma(N,|z|^2)}{\pi\Gamma(N)}$, which is ${\bf v}$-independent, as it must be by unitary invariance, and coincides with the finite-$N$ complex Ginibre
density Eq.(\ref{denGinUE}).
\end{remark}

Now we are in position to prove Theorem~\ref{Theorem2}.

\begin{proof}
Introducing the following notations:
\begin{equation}\label{2vectors}
{\bf a}:=A_z^*{\bf v}, \qquad
{\bf b}:=A_z^*A_z{\bf v}-\left({\bf v}^*A_z{\bf v}\right)A_z^*{\bf v},
\qquad Y^{(A)}_0(R)=R\,{\bf 1}_N+A_z^*A_z
\end{equation}
and exploiting the identity $\det{({\bf 1}-AB)}=\det{({\bf 1}-BA)}$ for any pair of matrices $A,B$ one may notice that
\begin{equation}\label{det_rewrite1}
\det{\left[R\,{\bf 1}_N+T_A({\bf v})\right]}=\det{\left[Y^{(A)}_0(R)-{\bf a}\otimes \bf{a}^*-{\bf b}\otimes{\bf v}^*\right]}
\end{equation}
\[
=\det{\left[Y^{(A)}_0(R)-{\bf a}\otimes \bf{a}^*\right]}\det{\left({\bf 1}_N-\left[Y^{(A)}_0(R)-{\bf a}\otimes \bf{a}^*\right]^{-1}{\bf b}\otimes{\bf v}^*\right)}
\]
\begin{equation}\label{det_rewrite2}
=\det{Y^{(A)}_0(R)}\,\left(1-{\bf a}^*[Y^{(A)}_0(R)]^{-1}{\bf a}\right)\left(1-{\bf v}^*[Y^{(A)}_0(R)-{\bf a}\otimes \bf{a}^*]^{-1} \bf{b}\right),
\end{equation}
For $R>0$, $Y^{(A)}_0(R)$ is positive definite.  Moreover,
\[
1-{\bf a}^*[Y^{(A)}_0(R)]^{-1}{\bf a}
=R{\bf v}^*(R{\bf1}_N+A_zA_z^*)^{-1}{\bf v}>0,
\]
so $Y^{(A)}_0(R)-{\bf a}\otimes{\bf a}^*$ is invertible as well.  We may therefore apply the Sherman--Morrison formula:
\begin{equation}\label{ShermanMorrison}
[Y^{(A)}_0(R)-{\bf a}\otimes {\bf a}^*]^{-1}=[Y^{(A)}_0(R)]^{-1}+\frac{1}{1-{\bf a}^*[Y^{(A)}_0(R)]^{-1}{\bf a}}[Y^{(A)}_0(R)]^{-1}{\bf a}\otimes {\bf a}^*[Y^{(A)}_0(R)]^{-1}
\end{equation}
to bring Eq.(\ref{det_rewrite2}) to the following form:
\begin{equation}\label{det_rewrite3}
=\det{Y^{(A)}_0(R)}\,\left\{\left(1-{\bf a}^*[Y^{(A)}_0(R)]^{-1}{\bf a}\right)\left(1-{\bf v}^*[Y^{(A)}_0(R)]^{-1}{\bf b}\right)-\left({\bf v}^*[Y^{(A)}_0(R)]^{-1}{\bf a}\right)\,\left(
 {\bf a}^*[Y^{(A)}_0(R)]^{-1}{\bf b}\right)\right\}.
\end{equation}
We then have
\[
\left(1-{\bf v}^*[Y^{(A)}_0(R)]^{-1}{\bf b}\right)=1-{\bf v}^*\left[R\,{\bf 1}_N+A_z^*A_z\right]^{-1}\left(A_z^*A_z-\left({\bf v}^*A_z{\bf v}\right)A^*_z\right){\bf v}
\]
\[
=1-{\bf v}^*\left[R\,{\bf 1}_N+A_z^*A_z\right]^{-1}A_z^*A_z{\bf v}+\left({\bf v}^*A_z{\bf v}\right){\bf v}^*\left[R\,{\bf 1}_N+A_z^*A_z\right]^{-1}A^*_z{\bf v}
\]
\[
=R{\bf v}^*\left[R\,{\bf 1}_N+A_z^*A_z\right]^{-1}{\bf v}+\left({\bf v}^*A_z{\bf v}\right){\bf v}^*\left[R\,{\bf 1}_N+A_z^*A_z\right]^{-1}A^*_z{\bf v}
\]
where in the last line we used ${\bf v}^*{\bf v}=1$. This allows us to write
\begin{equation}\label{first}
\left(1-{\bf a}^*[Y^{(A)}_0(R)]^{-1}{\bf a}\right)\left(1-{\bf v}^*[Y^{(A)}_0(R)]^{-1}{\bf b}\right)
\end{equation}
\[
=R{\bf v}^*\left[R\,{\bf 1}_N+A_z^*A_z\right]^{-1}{\bf v}+\left({\bf v}^*A_z{\bf v}\right)\left({\bf v}^*\left[R\,{\bf 1}_N+A_z^*A_z\right]^{-1}A^*_z{\bf v}\right)
\]
\[
-R\left({\bf v}^*A_z\left[R\,{\bf 1}_N+A_z^*A_z\right]^{-1}A^*_z{\bf v}\right)\left({\bf v}^*\left[R\,{\bf 1}_N+A_z^*A_z\right]^{-1}{\bf v}\right)
\]
\[
-\left({\bf v}^*A_z{\bf v}\right)\left({\bf v}^*\left[R\,{\bf 1}_N+A_z^*A_z\right]^{-1}A^*_z{\bf v}\right)\left( {\bf v}^*A_z\left[R\,{\bf 1}_N+A_z^*A_z\right]^{-1}A^*_z{\bf v}\right).
\]
Similarly
\begin{equation}\label{second}
-\left({\bf v}^*[Y^{(A)}_0(R)]^{-1}{\bf a}\right)\,\left(
 {\bf a}^*[Y^{(A)}_0(R)]^{-1}{\bf b}\right)
\end{equation}
\[
=-\left({\bf v}^*\left[R\,{\bf 1}_N+A_z^*A_z\right]^{-1}A^*_z{\bf v}\right)  \left({\bf v}^*A_z\left[{\bf 1}_N-\frac{R}{R\,{\bf 1}_N+A_z^*A_z}\right]{\bf v}\right)
\]
\[
+\left({\bf v}^*A_z{\bf v}\right)\left({\bf v}^*\left[R\,{\bf 1}_N+A_z^*A_z\right]^{-1}A^*_z{\bf v}\right)\left( {\bf v}^*A_z\left[R\,{\bf 1}_N+A_z^*A_z\right]^{-1}A^*_z{\bf v}\right).
\]
Adding Eq.(\ref{first}) and Eq.(\ref{second}) all terms proportional to $\left({\bf v}^*A_z{\bf v}\right)$ cancel identically, and every surviving term carries a common factor $R$. We find in this way that the expression in curly brackets in Eq.(\ref{det_rewrite3}) takes the form
\begin{equation}\label{third}
R\left({\bf v}^*\left[R\,{\bf 1}_N+A_z^*A_z\right]^{-1}{\bf v}\right)\left(1-\left({\bf v}^*A_z\left[R\,{\bf 1}_N+A_z^*A_z\right]^{-1}A^*_z{\bf v}\right)\right)
\end{equation}
\[
+R\left({\bf v}^*\left[R\,{\bf 1}_N+A_z^*A_z\right]^{-1}A^*_z{\bf v}\right)\left({\bf v}^*A_z\left[R\,{\bf 1}_N+A_z^*A_z\right]^{-1}{\bf v}\right).
\]
Finally we use the identity
\[
A_z\left[R\,{\bf 1}_N+A_z^*A_z\right]^{-1}A^*_z=\left[R\,{\bf 1}_N+A_zA_z^*\right]^{-1}A_zA_z^*={\bf 1}_N-R\left[R\,{\bf 1}_N+A_zA_z^*\right]^{-1}
\]
to see that Eq.(\ref{third}) is equal to
\begin{equation}\label{forth}
R\left\{R\left({\bf v}^*\left[R\,{\bf 1}_N+A_z^*A_z\right]^{-1}{\bf v}\right)\left({\bf v}^*\left[R\,{\bf 1}_N+A_zA_z^*\right]^{-1}{\bf v}\right)\right.
\end{equation}
\[
\left.+\left|\left({\bf v}^*A_z\left[R\,{\bf 1}_N+A_z^*A_z\right]^{-1}{\bf v}\right)\right|^2\right\},
\]
where we used that $\left({\bf v}^*\left[R\,{\bf 1}_N+A_z^*A_z\right]^{-1}A^*_z{\bf v}\right)$ is the complex conjugate of
$\left({\bf v}^*A_z\left[R\,{\bf 1}_N+A_z^*A_z\right]^{-1}{\bf v}\right)$, the matrix $\left[R\,{\bf 1}_N+A_z^*A_z\right]^{-1}$ being Hermitian.
The expression in the curly brackets of Eq.(\ref{forth}) is precisely the second line of Eq.(\ref{MainJPDadditivedef_intro}). The
prefactor $R$ in front of it cancels the factor $R^{-1}$ present in the measure of Eq.(\ref{MainJPDadditivedef}), so that
substituting Eq.(\ref{det_rewrite1})--(\ref{forth}) into Eq.(\ref{MainJPDadditivedef}) yields
\[
{\cal P}^{(A)}_N(z,{\bf v})=\frac{1}{\pi \Gamma(N)}\,e^{-{\bf v}^*A_z^*A_z{\bf v}}\int_0^{\infty}\frac{dR}{R}e^{-R}\,
R\,\det{Y^{(A)}_0(R)}\left\{\ldots\right\}=\mbox{r.h.s. of Eq.(\ref{MainJPDadditivedef_intro})},
\]
thereby completing the proof.
\end{proof}

\subsection{Statistics of the first component of the eigenvector in the case of a diagonal perturbation }\label{firstcomp}
 As the diagonal perturbation is a special case of normal perturbation, we start with presenting Eq.(\ref{mainJPD_norm}) in an equivalent form:
\begin{equation}\label{MainJPDadditivedef_proof1}
{\cal P}^{(A,diag)}_N(z,{\bf v})=\frac{1}{\pi \Gamma(N)}\,e^{-\sum_{n=1}^N|v_n|^2|d_n|^2}\int_0^{\infty}dR\, e^{-R}\,\prod_{n=1}^N\left[R+|d_n|^2\right]
\end{equation}
\[
\times \sum_{k,l=1}^N |v_l|^2|v_k|^2 \frac{R+d_l\overline{d_k}}{(R+|d_l|^2)(R+|d_k|^2)}, \quad  d_n=z-a_n.
\]

 It is also not difficult to integrate in Eq.(\ref{MainJPDadditivedef_proof1}) over all components of the right eigenvector ${\bf v}$ but one, thereby arriving at the JPD of a single eigenvector component, say $v_1$. It is convenient
to start with separating in Eq.(\ref{MainJPDadditivedef_proof1}) explicitly the first eigenvector component,  rewriting it as
\begin{equation}\label{MainJPDadditivedef_proof2}
{\cal P}^{(A)}_N(z,{\bf v})=\frac{1}{\pi \Gamma(N)}\,e^{-|v_1|^2|d_1|^2}e^{-\sum_{n=2}^N|v_n|^2|d_n|^2}\int_0^{\infty}dR\, e^{-R}\,\prod_{n=1}^N\left[R+|d_n|^2\right]
\end{equation}
\[
\times \left[\frac{|v_1|^4}{R+|d_1|^2}+\frac{|v_1|^2}{R+|d_1|^2}\sum_{k=2}^N|v_k|^2\frac{2R+d_1\overline{d_k}+\overline{d_1}d_k}{R+|d_k|^2}\right.
\]
\[
\left.+\sum_{k,l=2}^N |v_l|^2|v_k|^2 \frac{R+d_l\overline{d_k}}{(R+|d_l|^2)(R+|d_k|^2)}\right].
\]

Our next step is to integrate in Eq.~(\ref{MainJPDadditivedef_proof2}) over the complex variables $v_2,\ldots,v_N$ against the Lebesgue measure premultiplied by the Haar factor $\frac{\Gamma(N)}{\pi^N}\delta(|v_1|^2+\cdots+|v_N|^2-1)$.  The following Bromwich representation fixes the normalization needed below.
\begin{lemma}\label{Lemma4.2}
Given $v_1\in\mathbb C$ and non-negative parameters $\boldsymbol\alpha=(\alpha_2,\ldots,\alpha_N)$, define for $q\ge0$
\begin{equation}\label{defIntA}
I_{\boldsymbol\alpha,|v_1|^2}(q)=\frac{1}{\pi^N}
\int_{\mathbb C^{N-1}}e^{-\sum_{n=2}^N\alpha_n|v_n|^2}
\delta\!\left(|v_1|^2+\cdots+|v_N|^2-q\right)
\prod_{n=2}^N d^2v_n.
\end{equation}
Then
\begin{equation}\label{defIntB}
I_{\boldsymbol\alpha,|v_1|^2}(q)=\frac1\pi
\int_{\mathbb R-i0}e^{i(q-|v_1|^2)s}
\prod_{n=2}^N\frac{1}{\alpha_n+is}\,\frac{ds}{2\pi}.
\end{equation}
When one or more $\alpha_n$ vanish, the contour integral is understood as the boundary value obtained by first taking $\alpha_n>0$ and then sending the relevant parameters to zero.  In particular the right-hand side vanishes when $q<|v_1|^2$.
\end{lemma}
\begin{proof}
Assume first that all $\alpha_n>0$. For $\Re\lambda>0$,
\[
\int_0^\infty I_{\boldsymbol\alpha,|v_1|^2}(q)e^{-\lambda q}\,dq
=\frac1\pi e^{-\lambda|v_1|^2}\prod_{n=2}^N\frac1{\alpha_n+\lambda},
\]
because each complex Gaussian integral contributes $\pi/(\alpha_n+\lambda)$ while the definition contains $\pi^{-N}$.  Bromwich inversion gives Eq.~\eqref{defIntB}.  The extension to non-negative $\alpha_n$ follows by continuity from the defining compact-sphere integral (equivalently, by taking the corresponding boundary value of the Bromwich representation).  For example, for $N=2$,
\[
I_{\alpha,|v_1|^2}(q)=\frac1\pi e^{-\alpha(q-|v_1|^2)}{\bf 1}_{q>|v_1|^2}.
\]
\end{proof}

Since the full JPD depends on $v_1$ only through $x:=|v_1|^2$, it is convenient from now on to integrate out its phase and use a density with respect to $d^2z\,dx$.  Thus
\[
{\cal P}^{(A)}_N(z,x):=\int_0^{2\pi}{\cal P}^{(A)}_N(z,\sqrt{x}\,e^{i\theta})\,\frac{d\theta}{2},
\qquad 0\le x\le1,
\]
where $d^2v_1=\frac12\,dx\,d\theta$.  This phase integration supplies the factor $\pi$ which compensates the $1/\pi$ in Lemma~\ref{Lemma4.2}.  Applying the lemma to Eq.~\eqref{MainJPDadditivedef_proof2}, with $\alpha_i=|d_i|^2$ and $q=1$, gives
\begin{equation}\label{MainJPDadditivedef_proof3}
{\cal P}^{(A)}_N(z,x)=\frac{{\bf 1}_{0\le x\le1}}{\pi}
\int_0^{\infty}dR\, e^{-R}\int_{\mathbb R-i0} e^{is}e^{-x(is+|d_1|^2)}
\prod_{n=2}^N\frac{R+|d_n|^2}{is+|d_n|^2}
\end{equation}
\[
\times\left[x^2+x\sum_{k=2}^N\frac{2R+d_1\overline{d_k}+\overline{d_1}d_k}{(R+|d_k|^2)(is+|d_k|^2)}\right.
\]
\[
\left.+(R+|d_1|^2)\sum_{l\ge k\ge2}^N
\frac{2R+d_l\overline{d_k}+\overline{d_l}d_k}
{(R+|d_l|^2)(is+|d_l|^2)(R+|d_k|^2)(is+|d_k|^2)}\right]\frac{ds}{2\pi}.
\]
For later use define the unnormalised marked moment
\[
M^{(A)}_{q,N}(z):=\int_0^1x^q{\cal P}^{(A)}_N(z,x)\,dx,\qquad q>-1.
\]
By the last statement of Lemma \ref{Lemma4.2} the $s$-integrand in Eq.(\ref{MainJPDadditivedef_proof3}) vanishes identically for $x>1$: for $x>q=1$ the factor
$e^{i(q-x)s}$ permits closing the contour in the lower half-plane, where $\prod_{n\ge2}(|d_n|^2+is)^{-1}$ is analytic. The $x$-integration may therefore be
extended from $[0,1]$ to $[0,\infty)$, and performing it by means of $\int_0^{\infty}x^{q+j}e^{-x(is+|d_1|^2)}dx=\Gamma(q+j+1)(is+|d_1|^2)^{-(q+j+1)}$ gives
\begin{equation}\label{MainJPDadditivedef_proof3_moments}
M^{(A)}_{q,N}(z)=\Gamma(q+1)
\int_0^{\infty}dR\,e^{-R}\int_{\mathbb R-i0}e^{is}
\prod_{n=2}^N\frac{R+|d_n|^2}{is+|d_n|^2}
\end{equation}
\[
\times\left[\frac{(q+1)(q+2)}{(is+|d_1|^2)^{q+3}}
+\frac{q+1}{(is+|d_1|^2)^{q+2}}
\sum_{k=2}^N\frac{2R+d_1\overline{d_k}+\overline{d_1}d_k}{(R+|d_k|^2)(is+|d_k|^2)}\right.
\]
\[
\left.+\frac{R+|d_1|^2}{(is+|d_1|^2)^{q+1}}
\sum_{l\ge k\ge2}^N
\frac{2R+d_l\overline{d_k}+\overline{d_l}d_k}
{(R+|d_l|^2)(is+|d_l|^2)(R+|d_k|^2)(is+|d_k|^2)}\right]\frac{ds}{2\pi}.
\]
The conditional moment given an eigenvalue at $z$ is $M^{(A)}_{q,N}(z)/p_N^{(A)}(z)$.

\subsection{Proof of the Theorem \ref{NHRP}}\label{NHRPproof}

We now specialize the diagonal perturbation to the NHRP ensemble.  The entries $a_1,\ldots,a_N$ are i.i.d.; write $d=z-a$, consistently with Section \ref{firstcomp}, and denote expectation over one diagonal entry by $\langle\cdot\rangle$.  Re-averaging the exact component formula Eq.~\eqref{MainJPDadditivedef_proof3} requires separating the diagonal terms $k=l$ from the off-diagonal terms $k<l$.  Put
\[
D:=is+|d|^2,\qquad S:=R+|d|^2,
\]
\[
A:=\left\langle\frac{S}{D}\right\rangle,\quad E_0:=\langle e^{-xD}\rangle,\quad E_S:=\langle S e^{-xD}\rangle,
\]
\[
B_2:=\langle D^{-2}\rangle,\quad B_3:=\langle D^{-3}\rangle,
\quad C_2:=\langle dD^{-2}\rangle,\quad \overline C_2:=\langle\bar dD^{-2}\rangle.
\]

Then the exact finite-$N$ iid average is
\begin{eqnarray}\label{MainJPDadditivedef_proof4}
\left\langle{\cal P}^{(A)}_N(z,x)\right\rangle
&=&\frac{{\bf1}_{0\le x\le1}}{\pi}
\int_0^\infty e^{-R}dR\int_{\mathbb R-i0}e^{is}A^N
\Bigg\{\frac{x^2E_0}{A}
+\frac{(N-1)x}{A^2}{\cal B}_1 \nonumber\\
&&+\frac{2(N-1)E_SB_3}{A^2}
+\frac{(N-1)(N-2)E_S}{A^3}
\big(RB_2^2+C_2\overline C_2\big)\Bigg\}\frac{ds}{2\pi}.
\end{eqnarray}
where
\[
{\cal B}_1=2RB_2E_0+\langle\bar dD^{-2}\rangle\langle d e^{-xD}\rangle
+\langle dD^{-2}\rangle\langle\bar d e^{-xD}\rangle.
\]
The coefficient $(N-1)(N-2)$ counts the off-diagonal index pairs after combining the two conjugate terms; the separate $2(N-1)$ contribution comes from $k=l$.

 To match Theorem~\ref{NHRP}, first apply the exact formula to the
variance-one Ginibre matrix before the overall $N^{-1/2}$ scaling: write
$a=\sqrt N\,\widetilde a$, $z=\sqrt N\,w$,
$(R,s)=N(\widetilde R,\widetilde s)$ and $p=Nx$, and then omit tildes.
The factor $N^{-1}$ converting the one-point intensity into the normalized
marked-eigenpair law is cancelled by $d^2z=N\,d^2w$, while
$dx=dp/N$ supplies the remaining factor $N^{-1}$. Thus, in terms of the
iid-averaged intensity appearing in Eq.~\eqref{MainJPDadditivedef_proof4},
\begin{equation}\label{NHRP_scaled_marked_density}
\widehat{\mathcal P}_N^{NHRP}(w,p)
=
\frac1N
\left\langle
\mathcal P_N^{(A)}(\sqrt N\,w,p/N)
\right\rangle.
\end{equation}

Shift the $s$ contour through $s=-iR$ and write $s=-iR+t$. With
\[
Y:=R+it+|d|^2,\qquad
A_R(t):=\left\langle\frac{R+|d|^2}{Y}\right\rangle,
\]
one obtains
\begin{equation}\label{MainJPDadditivedef_proof5}
\widehat{\mathcal P}_N^{NHRP}(w,p)
=\frac{N}{\pi}\int_0^\infty dR\int_{\mathbb R}
 e^{iNt}A_R(t)^N\,{\cal F}_N(R,t;p)\,\frac{dt}{2\pi}.
\end{equation}
This form avoids introducing a logarithm and hence no branch choice is needed.  The pre-exponential factor is
\begin{equation}\label{MainJPDadditivedef_preexp}
\begin{split}
{\cal F}_N(R,t;p)=&\frac{p^2}{N^2}\frac{\langle e^{-pY}\rangle}{A_R(t)}
+\frac{(N-1)p}{N^2A_R(t)^2}{\cal C}_1\\
&+\frac{2(N-1)}{N^2}\frac{\langle(R+|d|^2)e^{-pY}\rangle}{A_R(t)^2}\langle Y^{-3}\rangle\\
&+\frac{(N-1)(N-2)}{N^2}\frac{\langle(R+|d|^2)e^{-pY}\rangle}{A_R(t)^3}
\left[R\langle Y^{-2}\rangle^2+\langle dY^{-2}\rangle\langle\bar dY^{-2}\rangle\right],
\end{split}
\end{equation}
where
\[
{\cal C}_1=2R\langle Y^{-2}\rangle\langle e^{-pY}\rangle
+\langle\bar dY^{-2}\rangle\langle d e^{-pY}\rangle
+\langle dY^{-2}\rangle\langle\bar d e^{-pY}\rangle.
\]
At fixed $p$, the first three lines of Eq.~\eqref{MainJPDadditivedef_preexp} are $O(N^{-1})$ or smaller, whereas the last line has a nonzero limit.

We next justify the concentration in $(R,t)$ through a one-dimensional local central-limit analysis. For fixed $R>0$ define
\begin{equation}\label{gR_density}
g_R(y):=\left\langle(R+|d|^2)e^{-(R+|d|^2)y}\right\rangle{\bf1}_{y>0}.
\end{equation}
Then $g_R$ is a probability density and
\begin{equation}\label{AR_fourier}
A_R(t)=\int_0^\infty e^{-ity}g_R(y)\,dy.
\end{equation}
Consequently Fourier inversion gives
\begin{equation}\label{convolution_identity}
\frac1{2\pi}\int_{\mathbb R}e^{iNt}A_R(t)^N\,dt=g_R^{*N}(N).
\end{equation}
Its mean and variance are
\begin{equation}\label{gR_moments}
m(R)=F(R):=\left\langle\frac1{R+|d|^2}\right\rangle,
\qquad V(R)=2G(R)-F(R)^2,
\end{equation}
where $G(R):=\langle(R+|d|^2)^{-2}\rangle$.  For the isotropic Gaussian law of Theorem~\ref{NHRP}, $F$ is strictly decreasing, $F(R)\to0$ as $R\to\infty$, and $F(R)\to\infty$ as $R\downarrow0$, hence there is a unique $R_*>0$ satisfying
\begin{equation}\label{saddlepoint}
F(R_*)=1.
\end{equation}
Moreover
\begin{equation}\label{defG*}
F'(R_*)=-G_*,\qquad G_*:=G(R_*)>0.
\end{equation}
Note also that by the Cauchy--Schwarz inequality $G_*=\langle S^{-2}\rangle\ge\langle S^{-1}\rangle^2=F(R_*)^2=1$, where $S=R_*+|d|^2$, so that the variance
$V_*:=V(R_*)=2G_*-1\ge1$ is bounded away from zero; this is used in Eq.(\ref{local_clt_nhrp}) below.
Note first that $g_R$ is \emph{not} smooth at the origin: it is smooth on $(0,\infty)$ but jumps from $0$ to $g_R(0+)=\langle R+|d|^2\rangle$ there. What the Fourier-inversion form of the local central limit theorem requires is not smoothness but boundedness and integrability, and both hold uniformly. Indeed, for $R$ in a compact interval $I$ containing $R_*$ in its interior and contained in $(0,\infty)$, one has $\|g_R\|_{\infty}\le\langle R+|d|^2\rangle$ uniformly on $I$, and the moments $\int_0^{\infty}y^kg_R(y)\,dy=k!\,\langle (R+|d|^2)^{-k}\rangle$ are uniformly bounded for every fixed $k$ because $R+|d|^2\ge R>0$ there. The laws $g_R$ are absolutely continuous, hence non-lattice, and the corresponding Cram\'er condition is uniform on $I$: for every fixed $\delta>0$ one has $\sup_{|t|\geq\delta}|A_R(t)|<1$ uniformly on $I$, while $A_R(t)=O(|t|^{-1})$ uniformly as $|t|\to\infty$, the latter because $\langle R+|d|^2\rangle<\infty$. In particular $g_R\in L^1\cap L^{\infty}$ gives $A_R\in L^2(\mathbb R)$, and off any neighbourhood of the origin $|A_R(t)|^N\le(1-c)^{N-2}|A_R(t)|^2$, so that the inversion integral in Eq.~\eqref{convolution_identity} is absolutely convergent for every $N\ge2$. These bounds give the standard uniform local central limit theorem by Fourier inversion; see, for example, Petrov~\cite[Chap.~VII]{PetrovSums}. Uniformly for $R-R_*=O(N^{-1/2})$,
\begin{equation}\label{local_clt_nhrp}
g_R^{*N}(N)=\frac{1+o(1)}{\sqrt{2\pi NV_*}}
\exp{-\frac{NG_*^2(R-R_*)^2}{2V_*}},
\qquad V_*:=2G_*-1.
\end{equation}
To control the $R$-integration we need an exponential bound not on a tail probability but on the density $g_R^{*N}(N)$ itself; this is supplied by an exponential tilt. For $\theta>-R$ set
\[
M_R(\theta):=\int_0^{\infty}e^{-\theta y}g_R(y)\,dy=\left\langle\frac{R+|d|^2}{R+|d|^2+\theta}\right\rangle,\qquad
h_{R,\theta}:=\frac{e^{-\theta y}g_R(y)}{M_R(\theta)},
\]
so that $h_{R,\theta}$ is again a probability density and $g_R^{*N}(N)=e^{\theta N}M_R(\theta)^N\,h_{R,\theta}^{*N}(N)$. By the same Fourier bounds as above,
$\sup_yh_{R,\theta}^{*N}(y)\le\frac{1}{2\pi}\int_{\mathbb R}|\widehat{h_{R,\theta}}(t)|^N\,dt$ is finite and bounded uniformly in the relevant range of $(R,\theta)$, whence
\[
g_R^{*N}(N)\le C\,e^{N\left(\theta+\log M_R(\theta)\right)}.
\]
Since $\frac{d}{d\theta}\left(\theta+\log M_R(\theta)\right)\big|_{\theta=0}=1-F(R)$, one may choose a small $\theta>0$ on $[0,R_-]$, where $F(R)>1$, and a small $\theta<0$ on $[R_+,\infty)$, where $F(R)<1$; in the latter case the constraint $\theta>-R$, which is what makes $M_R(\theta)$ finite because $\mbox{ess\,inf}(R+|d|^2)=R$, is respected by taking $|\theta|<R/2$, and for large $R$ this yields an even stronger estimate. In both cases $\theta+\log M_R(\theta)<0$ uniformly. Thus the contribution from $|R-R_*|>\delta$ is exponentially small for every fixed $\delta>0$. Consequently
\begin{equation}\label{R_integral_localclt}
\int_0^\infty g_R^{*N}(N)\,dR=\frac{1+o(1)}{NG_*}.
\end{equation}
The same Fourier proof also allows a smooth multiplier. On $|R-R_*|=O(N^{-1/2})$ split the $t$-integral into $|t|\le N^{-2/5}$ and its complement. On the first part the pre-exponential factor in Eq.~\eqref{MainJPDadditivedef_preexp} is its value at $t=0$ plus $o(1)$ uniformly, while the quadratic expansion of $\log A_R(t)$ gives the local central limit theorem. On the complement, non-lattice smoothness gives $|A_R(t)|^N\le e^{-cN^{1/5}}$ up to a fixed cutoff and $|A_R(t)|\le1-c$ beyond it. Therefore the multiplier may be frozen at $(R_*,0)$ at leading order.

Combining Eqs.~\eqref{MainJPDadditivedef_proof5}, \eqref{convolution_identity} and \eqref{R_integral_localclt} gives
\begin{equation}\label{fin_saddle-pointA}
{\cal P}_{\infty}(w,p)=
\left\langle e^{-p(R_*+|d|^2)}(R_*+|d|^2)\right\rangle
\frac{1}{\pi G_*}
\left[R_*G_*^2+
\left|\left\langle\frac{d}{(R_*+|d|^2)^2}\right\rangle\right|^2\right].
\end{equation}
Thus
\begin{equation}\label{fin_denA}
\rho^{NHRP}_{\infty}(w)=\frac{G_*}{\pi}
\left[R_*+
\left|\frac1{G_*}\left\langle\frac{d}{(R_*+|d|^2)^2}\right\rangle\right|^2\right],
\end{equation}
and, conditionally on an eigenvalue at $w$, the limiting density of $p=N|v_1|^2$ is
\begin{equation}\label{distr_pA}
{\cal P}^{NHRP}_{\infty}(p\mid w)=
\left\langle e^{-p(R_*+|d|^2)}(R_*+|d|^2)\right\rangle
=-\frac{d}{dp}\left\langle e^{-p(R_*+|d|^2)}\right\rangle.
\end{equation}
Accordingly, for $q>-1$,
\begin{equation}\label{mom_distr}
\mathbb E_{\infty}[p^q\mid w]=\Gamma(q+1)
\left\langle\frac1{(R_*+|d|^2)^q}\right\rangle.
\end{equation}
So far this establishes the limit of $\widehat{\mathcal P}^{NHRP}_N(w,p)$ at each fixed $p$, which by itself yields neither the convergence of
$\rho_N^{NHRP}(w)=\int_0^{\infty}\widehat{\mathcal P}^{NHRP}_N(w,p)\,dp$ nor the convergence of the conditional moments. Both follow, however, on applying
the same argument to the exact marked-moment formula Eq.~\eqref{MainJPDadditivedef_proof3_moments}, which after the substitutions preceding
Eq.~\eqref{MainJPDadditivedef_proof5} has exactly the structure of Eq.~\eqref{MainJPDadditivedef_proof5}, with $\mathcal F_N(R,t;p)$ replaced by a multiplier
$\mathcal F_N^{(q)}(R,t)$ obtained from Eq.~\eqref{MainJPDadditivedef_preexp} by the substitutions
\[
\frac{p^2}{N^2}\langle e^{-pY}\rangle\ \mapsto\ \frac{\Gamma(q+3)}{N^{q+3}}\left\langle Y^{-(q+3)}\right\rangle,\qquad
\frac{p}{N}\langle\cdot\, e^{-pY}\rangle\ \mapsto\ \frac{\Gamma(q+2)}{N^{q+2}}\left\langle \cdot\, Y^{-(q+2)}\right\rangle,
\]
and similarly in the last line, all other factors being unchanged. On the region $|R-R_*|=O(N^{-1/2})$, $|t|\le N^{-2/5}$ one has $|Y|\ge R\ge R_->0$, so these
multipliers are bounded uniformly in $N$ and, for $q$ ranging over a compact subset of $(-1,\infty)$, uniformly in $q$ as well; the freezing of the multiplier at
$(R_*,0)$ and the tail estimates therefore go through verbatim. This proves convergence of the finite-$N$ conditional moments to
Eq.~\eqref{mom_distr} for every $q>-1$, which is stronger than convergence in distribution alone; the case $q=0$ is the convergence of
$\rho_N^{NHRP}(w)$, and weak convergence of $\pi_N(\cdot\mid w)$ then follows from convergence of the moments of every positive integer order together with
the fact that the limiting law, having an exponential tail, is determined by its moments. Since $R_*>0$, that tail is indeed exponential and the moment integral is finite precisely for $q>-1$ at the origin.

The same convolution argument applies to other diagonal laws whenever $F(R_*)=1$ has an interior solution and the family $g_R$ satisfies the corresponding uniform local-CLT hypotheses.  We do not claim that $F(0)=\infty$ follows from finite variance alone.

Specifying now to the isotropic Gaussian complex random variable distributed according to
\begin{equation}\label{Gau_isotrop_comp}
d\nu:=\nu(a)d(\Re a) d(\Im a), \quad \nu(a)=\frac{1}{2\pi\sigma^2}e^{-\frac{(\Re a)^2+(\Im a)^2}{2\sigma^2}}
\end{equation}
one can easily show those equations to be exactly equivalent to Eqs.(\ref{NHRosenzweigPorter_R})-(\ref{NHRosenzweigPorter2}) featuring in the Theorem \ref{NHRP}.
We just demonstrate such equivalence on the example of bringing Eq.(\ref{saddlepoint}) to the form Eq.(\ref{NHRosenzweigPorter_R}) . Using $w=u+iv$ in this particular case
Eqs.(\ref{NHRosenzweigPorter_R}) can be written as
\begin{equation}\label{saddlepoint_gau1}
1=F(R), \quad F(R)=\frac{1}{2\pi\sigma^2}\int_{\mathbb{R}}\int_{\mathbb{R}} \frac{e^{-\frac{(\Re a)^2+(\Im a)^2}{2\sigma^2}}}{R+(\Re a-u)^2+(\Im a-v)^2}\,d(\Re a) d(\Im a).
\end{equation}
Introducing $t_1=\Re a-u, t_2=\Im a-v$ as new integration variables, and going to polar coordinates in $(t_1,t_2)$ plane: $t_1=\sqrt{t}\cos{\theta}, t_2=\sqrt{t}\sin{\theta}$,
the expression for $F(R)$ takes the form
\begin{equation}\label{saddlepoint_gau2}
 F(R)=\frac{e^{-\frac{u^2+v^2}{2\sigma^2}}}{2\sigma^2}  \int_{0}^{\infty} \frac{e^{-\frac{t}{2\sigma^2}}}{R+t}\,dt \int_0^{2\pi}
 e^{-\frac{\sqrt{t}}{\sigma^2}(u\cos{\theta}+v\sin{\theta})}\,\frac{d\theta}{2\pi}.
\end{equation}
Finally by noticing that the $\theta-$ integral yields the modified Bessel function $I_0\left(\frac{1}{\sigma^2}\sqrt{t(u^2+v^2)}\right)$ and finally changing $t\to Rt$ brings
this expression to the form of Eq.(\ref{NHRosenzweigPorter_R}).

Note that averaging over $a$ in Eq.(\ref{distr_pA}) amounts to evaluating a simple Gaussian integral which leads directly to Eq.(\ref{NHRosenzweigPorter1}).

%%%%%%%%%%%%%%%%%

\section{Appendices}
\subsection{Appendix A: The fundamental sphere--delta identity: analytic proofs}
\label{main_integral_identity_appendix}

This appendix proves the basic identity underlying the Kac--Rice representation used throughout the paper.    The first proof, based on Gaussian regularization and the rank-one Harish--Chandra--Itzykson--Zuber (HCIZ) integral, follows a suggestion of Martin Zirnbauer.  We then record a Fourier/Gaussian derivation of the same finite-$\epsilon$ formula which does not use HCIZ, and finally give a repaired proof by induction in the matrix size.  An independent geometric proof by coarea and transversality is given in Appendix~B.

For later reference set
\[
\mathfrak U_N:=\{B\in\mathcal M_N(\mathbb C):\mbox{rank}B\geq N-1\}.
\]
On the singular part of $\mathfrak U_N$ one has $\mbox{rank}B=N-1$, hence the differential of $B\mapsto\det B$ is nonzero.  Thus $\delta^{(2)}(\det B)$ is a well-defined pull-back distribution on $\mathfrak U_N$ in the standard submersion sense; see, e.g., H\"ormander~\cite[Thm.~6.1.2]{HormanderI}.  The matrix-space formulation needed for the inductive proof is
\begin{equation}\label{main_integral_identity_matrix_regular}
\frac1\pi\int_{\mathbb C^N}
\delta^{(2N)}(B{\bf v})\,\delta({\bf v}^*{\bf v}-1)\,d^{2N}{\bf v}
=\delta^{(2)}(\det B),
\qquad B\in\mathfrak U_N,
\end{equation}
as an identity in $\mathcal D'(\mathfrak U_N)$.  In particular, no statement is made here across the higher-corank locus, where $\det B=0$ is not a regular hypersurface.

For the applications in the main text we shall use the following one-parameter version.

\begin{proposition}[Holomorphic matrix pencil]\label{prop_holomorphic_pencil_analytic}
Let $\Omega\subset\mathbb C$ be open and let $B:\Omega\to\mathcal M_N(\mathbb C)$ be holomorphic.  Put $b(z)=\det B(z)$ and assume that every zero of $b$ in $\Omega$ is simple.  Then
\begin{equation}\label{main_integral_identity_AppA}
\delta^{(2)}(b(z))
=\frac1\pi\int_{\mathbf S_N^c}\delta^{(2N)}(B(z){\bf v})\,d\mu_N({\bf v})
\end{equation}
in $\mathcal D'(\Omega)$.  Coincidences among the nonzero singular values of $B(z)$ are allowed.
\end{proposition}

\subsubsection{Heat-kernel and HCIZ proof}
\label{Zirn}

We use the Gaussian regularization
\begin{equation}\label{main_integral_identity_AppA_reg1}
L_{\epsilon}^{(N)}(B^*B)
:=\frac{1}{(2\pi\epsilon)^N}
\int_{\mathbb C^N}
\exp{-\frac{{\bf v}^*B^*B{\bf v}}{2\epsilon}}
\,d\mu_N({\bf v}),
\qquad \epsilon>0.
\end{equation}
Write the squared singular values of $B$ as $x_1,\ldots,x_N\geq0$.  Since $d\mu_N$ is invariant under unitary rotations, $L_\epsilon^{(N)}$ depends only on these numbers.  For completeness, when $A$ has rank one with eigenvalues $a_1,0,\ldots,0$ and $B_0$ has pairwise distinct eigenvalues $b_1,\ldots,b_N$, the rank-one specialization of the HCIZ formula~\cite{HC1956,IZ} reads
\begin{equation}\label{IZHC_rank1}
\int_{U(N)}e^{t\,\mbox{Tr}(U^*B_0UA)}\,d\mu_H(U)
=\frac{\Gamma(N)}{(ta_1)^{N-1}}
\sum_{i=1}^N\frac{e^{tb_i a_1}}{\prod_{j\ne i}(b_i-b_j)}.
\end{equation}
Applying Eq.~\eqref{IZHC_rank1} with $a_1=1$, $b_i=x_i$, and $t=-(2\epsilon)^{-1}$ gives, for pairwise distinct $x_i$,
\begin{equation}\label{dent3new}
L_{\epsilon}^{(N)}(x_1,\ldots,x_N)
=\frac{(-1)^{N-1}}{2\epsilon}
\sum_{i=1}^N
\frac{e^{-x_i/(2\epsilon)}}{\prod_{j\ne i}(x_i-x_j)}.
\end{equation}
In particular, for $N=2$ 
\begin{equation}\label{main_integral_identity_AppA_reg4}
L_{\epsilon}^{(2)}(x_1,x_2)
=-\frac{1}{2\epsilon}
\left(
\frac{e^{-x_1/(2\epsilon)}}{x_1-x_2}
+\frac{e^{-x_2/(2\epsilon)}}{x_2-x_1}
\right).
\end{equation}
The apparent poles at coincident $x_i$ are removable.  Indeed, with
\[
f_\epsilon(x):=e^{-x/(2\epsilon)},
\]
Eq.~\eqref{dent3new} is
\begin{equation}\label{dent3new_divdiff}
L_{\epsilon}^{(N)}(x_1,\ldots,x_N)
=\frac{(-1)^{N-1}}{2\epsilon}
[x_1,\ldots,x_N]f_\epsilon,
\end{equation}
where $[x_1,\ldots,x_N]$ denotes the divided difference.  Divided differences have a unique confluent, or Hermite, continuation when some nodes coincide; see de Boor~\cite{deBoorDivided}.  Equivalently, the Hermite--Genocchi formula \cite{deBoorDivided} gives the representation
\begin{equation}\label{simplex_representation_AppA}
L_{\epsilon}^{(N)}(x_1,\ldots,x_N)
=\frac{1}{(2\epsilon)^N}
\int_{\Delta_{N-1}}
\exp{-\frac{1}{2\epsilon}\sum_{j=1}^N t_jx_j}\,dt,
\end{equation}
where 
\[
\Delta_{N-1}:=\{t_j\geq0:\ \sum_{j=1}^N t_j=1\}
\]
and $dt=dt_2\cdots dt_N$ with $t_1=1-\sum_{j=2}^Nt_j$.  Formula~\eqref{simplex_representation_AppA} is manifestly continuous and symmetric in all $x_j$ and therefore proves that Eq.~\eqref{dent3new} extends canonically to repeated singular values.  For example, the confluent limit of Eq.~\eqref{main_integral_identity_AppA_reg4} at $x_1=x_2=x$ equals $(2\epsilon)^{-2}e^{-x/(2\epsilon)}$.

We now take the singular limit in the variable in which the identity is actually used.  Let $z_a$ be a simple zero of $b(z)=\det B(z)$.  Then $B(z_a)$ has rank $N-1$.  In a neighbourhood of $z_a$ label the unique small squared singular value by $x_1(z)$ and put
\[
D_a:=\prod_{j=2}^N x_j(z_a)>0.
\]
No distinctness is assumed among $x_2(z_a),\ldots,x_N(z_a)$.  Since
\[
|b(z)|^2=\prod_{j=1}^N x_j(z),
\]
and $b(z)=b'(z_a)(z-z_a)+O(|z-z_a|^2)$, continuity of the remaining singular values gives
\begin{equation}\label{small_singular_value_expansion}
x_1(z)
=\frac{|b'(z_a)|^2}{D_a}|z-z_a|^2+o(|z-z_a|^2).
\end{equation}

To extract the local limit from Eq.~\eqref{simplex_representation_AppA}, write $t_j=2\epsilon s_j$ for $j\geq2$ and $t_1=1-2\epsilon\sum_{j=2}^Ns_j$.  For every fixed $R<\infty$, uniformly for $z=z_a+\sqrt\epsilon\,\zeta$ with $|\zeta|\leq R$, dominated convergence gives
\begin{equation}\label{local_L_epsilon_asymptotic}
L_{\epsilon}^{(N)}(z)
=\frac{1+o(1)}{2\epsilon D_a}
\exp{-\frac{x_1(z)}{2\epsilon}}.
\end{equation}
Together with Eq.~\eqref{small_singular_value_expansion}, this becomes
\begin{equation}\label{local_L_scaled}
L_{\epsilon}^{(N)}(z_a+\sqrt\epsilon\,\zeta)
=\frac{1+o(1)}{2\epsilon D_a}
\exp{-\frac{|b'(z_a)|^2}{2D_a}|\zeta|^2}.
\end{equation}
On compact subsets separated from the zero set of $b$, all singular values are bounded away from zero and Eq.~\eqref{simplex_representation_AppA} is exponentially small as $\epsilon\downarrow0$.  It remains to control the region $|\zeta|>R$ inside the small discs.  Since $z_a$ is a simple zero of $b$ and the remaining squared singular values $x_2(z),\ldots,x_N(z)$ stay bounded near $z_a$, Eq.~\eqref{small_singular_value_expansion} gives a uniform lower bound $x_1(z)\geq c\,|z-z_a|^2$ with $c>0$ on a sufficiently small punctured disc about $z_a$; by Eq.~\eqref{simplex_representation_AppA} and $t_1\le 1$ one then has $L_{\epsilon}^{(N)}(z)\le(2\epsilon)^{-N}\exp{-c|z-z_a|^2/(2\epsilon)}$ there, so that the contribution of $|\zeta|>R$ is $O(e^{-cR^2/2})$ uniformly in $\epsilon$ and disappears on letting $R\to\infty$ after $\epsilon\downarrow0$.  Standard Gaussian tail estimates applied in disjoint neighbourhoods of the simple zeros therefore imply that for every $\psi\in C_c^\infty(\Omega)$,
\begin{equation}\label{pencil_scalar_limit}
\lim_{\epsilon\downarrow0}
\frac1\pi\int_\Omega \psi(z)L_{\epsilon}^{(N)}(B(z)^*B(z))\,d^2z
=
\sum_{b(z_a)=0}\frac{\psi(z_a)}{|b'(z_a)|^2}.
\end{equation}
The right-hand side is exactly the action of $\delta^{(2)}(b(z))$.  This proves Proposition~\ref{prop_holomorphic_pencil_analytic}.

\begin{remark}
Repeated nonzero singular values are handled by the confluent divided
difference and cause no singularity in $L_\epsilon^{(N)}$. At a simple
determinant zero exactly one singular value vanishes.
\end{remark}

\subsubsection{The same finite-$\epsilon$ formula without HCIZ}

One can arrive at Eq.~\eqref{dent3new} without employing the HCIZ integral.  Instead, regularize the factor $\delta({\bf v}^*{\bf v}-1)$ in Eq.~\eqref{main_integral_identity_AppA_reg1} by
\begin{equation}\label{sphere_fourier_regularization}
\delta_{\gamma,g}({\bf v}^*{\bf v}-1)
=\int_{-\infty}^{\infty}
 e^{-\frac{\gamma}{2}k^2+ik({\bf v}^*{\bf v}-1)}\,\frac{dk}{2\pi},
\qquad \gamma>0.
\end{equation}
At fixed $\epsilon,\gamma>0$ the ${\bf v}$-integral is absolutely convergent and Gaussian.  In the singular-value basis one obtains
\begin{equation}\label{L_epsilon_gamma}
L_{\epsilon,\gamma}^{(N)}(x_1,\ldots,x_N)
=\frac{1}{(2\epsilon)^N}\int_{-\infty}^{\infty}
 e^{-\frac{\gamma}{2}k^2-ik}
 \frac{1}{\prod_{i=1}^N\left(\frac{x_i}{2\epsilon}-ik\right)}\,\frac{dk}{2\pi}.
\end{equation}
Assume first that $x_i>0$ are pairwise distinct.  The elementary partial-fraction identity
\begin{equation}\label{partial_fraction_fourier}
\frac{1}{\prod_{i=1}^N\left(\frac{x_i}{2\epsilon}-ik\right)}
=(2\epsilon)^{N-1}\sum_{i=1}^N
\frac{1}{\frac{x_i}{2\epsilon}-ik}
\frac{1}{\prod_{j\ne i}(x_j-x_i)}
\end{equation}
reduces Eq.~\eqref{L_epsilon_gamma} to one-dimensional integrals.  For every $a>0$,
\begin{align}\label{fourier_one_pole_limit}
\int_{-\infty}^{\infty}
\frac{e^{-\frac{\gamma}{2}k^2-ik}}{a-ik}\,\frac{dk}{2\pi}
&=\int_0^\infty e^{-at}
\frac{e^{-(t-1)^2/(2\gamma)}}{\sqrt{2\pi\gamma}}\,dt
\notag\\
&\longrightarrow e^{-a},\qquad \gamma\downarrow0.
\end{align}
Here we used $(a-ik)^{-1}=\int_0^\infty e^{-(a-ik)t}\,dt$ and then performed the Gaussian Fourier integral.  Therefore, taking
\[
\gamma\downarrow0\quad\hbox{first, at fixed }\epsilon>0,
\]
reproduces Eq.~\eqref{dent3new}.  Only afterwards is the limit $\epsilon\downarrow0$ taken in Proposition~\ref{prop_holomorphic_pencil_analytic}.

The Gaussian integral in Eq.~\eqref{main_integral_identity_AppA_reg1} is a
continuous symmetric function of the nonnegative squared singular values.
Moreover, Eq.~\eqref{dent3new} is equivalently represented by the divided
difference in Eq.~\eqref{dent3new_divdiff}. Hence the expression extends
uniquely to coincident $x_i$ and, by continuity, to the boundary where one
singular value vanishes.

\begin{remark}
A closely related Fourier/Gaussian regularization can also be used in the left--right construction discussed in Section~\ref{Towards}.  Because the latter contains an additional gauge constraint, its regularization and singular limit are treated separately rather than being inferred formally from Eq.~\eqref{L_epsilon_gamma}.
\end{remark}

\subsubsection{Alternative proof by induction}
\label{main_integral_identity_induction}

We carry out the Schur--complement induction on the class
$\mathfrak U_N$ of matrices of corank at most one. This class is preserved
by the reduction below and is therefore the natural setting for the
inductive argument. In particular, it contains the matrices arising at
simple determinant zeros.

Let
\[
J_N(B):=\int_{\mathbb C^N}
\delta^{(2N)}(B{\bf v})\,\delta({\bf v}^*{\bf v}-1)\,d^{2N}{\bf v}.
\]
We prove
\begin{equation}\label{induction_target_complex}
J_N(B)=\pi\,\delta^{(2)}(\det B)
\qquad\hbox{in }\mathcal D'(\mathfrak U_N).
\end{equation}
For $N=1$, test the left-hand side against a compactly supported smooth function of $b\in\mathbb C$.  On the support of $\delta(|v|^2-1)$ one has $|v|=1$, hence $\delta^{(2)}(bv)=\delta^{(2)}(b)$, while
\[
\int_{\mathbb C}\delta(|v|^2-1)\,d^2v=\pi.
\]
Thus Eq.~\eqref{induction_target_complex} holds for $N=1$.

Assume it for size $N-1$ and work first on the open chart $B_{11}\ne0$.  Write, exactly as in the original induction,
\begin{equation}\label{decomp}
B=\begin{pmatrix}B_{11}&{\bf b}_1^*\\ {\bf b}_2&B_{N-1}\end{pmatrix},
\qquad
{\bf v}=\begin{pmatrix}v_1\\ {\bf v}_{N-1}\end{pmatrix},
\end{equation}
and introduce
\begin{equation}\label{Schur_repaired}
S:=B_{N-1}-\frac{{\bf b}_2{\bf b}_1^*}{B_{11}},
\qquad
M:={\bf 1}_{N-1}+\frac{{\bf b}_1{\bf b}_1^*}{|B_{11}|^2}.
\end{equation}
The Schur determinant identity is
\begin{equation}\label{Schur}
\det B=B_{11}\det S.
\end{equation}
Integrating first over $v_1$ by means of
$\delta^{(2)}(B_{11}v_1+{\bf b}_1^*{\bf v}_{N-1})$ gives
\begin{equation}\label{main_integral_identity_variant_2}
J_N(B)=|B_{11}|^{-2}
\int_{\mathbb C^{N-1}}
\delta^{(2N-2)}(S{\bf v}_{N-1})
\delta({\bf v}_{N-1}^*M{\bf v}_{N-1}-1)\,d^{2N-2}{\bf v}_{N-1}.
\end{equation}
The change of variables ${\bf v}_{N-1}=M^{-1/2}{\bf w}_{N-1}$ therefore yields
\begin{equation}\label{main_integral_identity_variant_3}
J_N(B)=|B_{11}|^{-2}(\det M)^{-1}J_{N-1}(C),
\qquad C:=SM^{-1/2}.
\end{equation}
All transformations here are ordinary complex-linear changes of variables on the chart $B_{11}\ne0$ and hence are legitimate after testing against compactly supported smooth functions.  Moreover, applying the induction hypothesis to $J_{N-1}(C)$ amounts to pulling the identity in $\mathcal D'(\mathfrak U_{N-1})$ back along the map $B\mapsto C=SM^{-1/2}$; this is admissible because that map is a submersion of $\{B_{11}\ne0\}$ onto $\mathcal M_{N-1}(\mathbb C)$, as is seen by holding $B_{11},{\bf b}_1,{\bf b}_2$ fixed, when $B_{N-1}\mapsto S$ is a translation and $M^{-1/2}$ is a fixed invertible matrix.

The Schur--complement reduction preserves the corank-at-most-one class.
Indeed,
\[
\begin{pmatrix}
1&0\\
-B_{11}^{-1}{\bf b}_2&{\bf 1}_{N-1}
\end{pmatrix}B
=
\begin{pmatrix}
B_{11}&{\bf b}_1^*\\
0&S
\end{pmatrix},
\]
and therefore
\begin{equation}\label{rank_S_identity}
\mbox{rank}B=1+\mbox{rank}S.
\end{equation}
Since $M^{-1/2}$ is invertible, one also has
$\mbox{rank}C=\mbox{rank}S$. Consequently,
\[
B\in\mathfrak U_N
\quad\Longrightarrow\quad
C\in\mathfrak U_{N-1},
\]
so the induction closes on the class $\mathfrak U_N$.
Using Eq.~\eqref{induction_target_complex} for $C$ in Eq.~\eqref{main_integral_identity_variant_3}, together with
\[
\det C=\det S\,(\det M)^{-1/2},
\]
and the complex scaling rule $\delta^{(2)}(cw)=|c|^{-2}\delta^{(2)}(w)$, gives
\begin{align*}
J_N(B)
&=\pi |B_{11}|^{-2}(\det M)^{-1}
\delta^{(2)}\!\left((\det M)^{-1/2}\det S\right)\\
&=\pi |B_{11}|^{-2}\delta^{(2)}(\det S)
=\pi\delta^{(2)}(\det B),
\end{align*}
where the last equality follows from Eq.~\eqref{Schur}.  This proves the identity on the chart $B_{11}\ne0$.

It remains only to cover the part of the determinant hypersurface on which $B_{11}=0$.  There is no restriction to $\det B\ne0$: both sides of Eq.~\eqref{induction_target_complex} are supported on $\det B=0$.  If $B\in\mathfrak U_N$ and $\det B=0$, then $\mbox{rank}B=N-1$, so at least one entry $B_{ij}$ is nonzero.  The open sets
\[
\mathcal V_{ij}:=\{B\in\mathfrak U_N:B_{ij}\ne0\}
\]
cover the regular determinant hypersurface.  On each $\mathcal V_{ij}$, row and column permutation matrices move $B_{ij}$ to the $(1,1)$ position.  The left-hand side of Eq.~\eqref{induction_target_complex} is invariant under these unitary permutations after the corresponding change of variables on the sphere, and the right-hand side is unchanged because the determinant changes only by a unit-modulus factor.  A partition of unity subordinate to the finite family $\{\mathcal V_{ij}\}$ therefore completes the proof of Eq.~\eqref{main_integral_identity_matrix_regular}.

\begin{remark}[Real analogue]
The same induction, with real rather than complex Dirac distributions and orthogonal instead of unitary changes of variables, yields on the regular determinantal locus $\{B\in\mathcal M_N(\mathbb R):\mbox{rank}B\geq N-1\}$ the real-sphere identity
\[
\int_{\mathbb R^N}\delta^{(N)}(B{\bf v})\,\delta({\bf v}^T{\bf v}-1)\,d^N{\bf v}
=\delta(\det B).
\]
The corresponding left--right gauge-fixed identities require a separate treatment and are not inferred by multiplying singular Dirac factors.
\end{remark}

\subsubsection{From the scalar identity to the phase-averaged eigenpair measure}

For completeness we finish the analytic proof of Theorem~\ref{MainMetaTheorem} without invoking coarea.  Put $B(z)=X-z{\bf 1}_N$ and define, for $\epsilon>0$,
\begin{equation}\label{regularized_marked_measure_AppA}
d\nu_{X,\epsilon}(z,{\bf v})
:=\frac1\pi |p_X'(z)|^2
\delta_{g,\epsilon}^{(2N)}((X-z{\bf 1}_N){\bf v})
\,d^2z\,d\mu_N({\bf v}).
\end{equation}
Its $z$-marginal is
\[
\frac1\pi |p_X'(z)|^2
L_\epsilon^{(N)}((X-z{\bf 1}_N)^*(X-z{\bf 1}_N))\,d^2z.
\]
By Eq.~\eqref{pencil_scalar_limit}, this converges weakly to $\sum_a\delta_{z_a}$.  On every compact $z$-set the measures in Eq.~\eqref{regularized_marked_measure_AppA} have uniformly bounded mass; away from the zero set of $(X-z{\bf 1}_N){\bf v}$ the Gaussian factor is exponentially small.  Hence every weak subsequential limit is supported on
\[
\bigcup_{a=1}^N\{(z_a,e^{i\phi}{\bf v}_a):0\leq\phi<2\pi\}.
\]
For every $\epsilon>0$ the measure $\nu_{X,\epsilon}$ is invariant under ${\bf v}\mapsto e^{i\alpha}{\bf v}$.  Its restriction in the limit to each of the above circles must therefore be a constant multiple of Haar phase measure.  The already determined $z$-marginal shows that this constant is one for every $a$.  Consequently
\[
\nu_{X,\epsilon}\Longrightarrow
\sum_{a=1}^N\delta_{z_a}(d^2z)
\int_0^{2\pi}\delta_{e^{i\phi}{\bf v}_a}(d{\bf v})\,\frac{d\phi}{2\pi},
\]
which is exactly Theorem~\ref{MainMetaTheorem}.

\subsection{Appendix B: Coarea and transversality: an alternative geometric proof}
\label{coarea_alternative_appendix}

We give here an independent geometric proof of Theorem~\ref{MainMetaTheorem} and Eq.~\eqref{main_integral_identity}.  It identifies the Gaussian limits used in Appendix~A with the standard coarea pull-backs and makes the underlying transversality explicit.  We use the coarea formula in its standard form; see Federer~\cite[Sec.~3.2.12]{FedererGMT} or Evans--Gariepy~\cite[Chap.~3]{EvansGariepy}.

\begin{lemma}[Gaussian regularization and coarea]\label{lemma_coarea_delta}
Let $M$ be a smooth Riemannian manifold, let $F:M\to\mathbb R^k$ be smooth, and let $\varphi\in C_c^\infty(M)$.  Assume that $0$ is a regular value of $F$ on a neighbourhood of $\mbox{supp}\varphi\cap F^{-1}(0)$.  Put
\[
J_F(x)=\sqrt{\det\!\left(DF_xDF_x^*\right)}
\]
on the normal space to the level set.  Then
\begin{equation}\label{coarea_delta_limit}
\lim_{\epsilon\downarrow0}\int_M\varphi(x)\,
\delta_{g,\epsilon}^{(k)}(F(x))\,d\mathrm{vol}_M(x)
=
\int_{F^{-1}(0)}\frac{\varphi(x)}{J_F(x)}\,d\mathcal H^{\dim M-k}(x).
\end{equation}
In particular the left-hand side defines a positive Radon measure, denoted $\delta_0(F)\,d\mathrm{vol}_M$.
\end{lemma}

\begin{proof}
By the coarea formula the left-hand side before taking the limit equals
\[
\int_{\mathbb R^k}\delta_{g,\epsilon}^{(k)}(y)
\left[\int_{F^{-1}(y)}
\frac{\varphi(x)}{J_F(x)}\,d\mathcal H^{\dim M-k}(x)\right]dy.
\]
By the submersion theorem and compactness of the relevant part of $\mbox{supp}\varphi$, the expression in square brackets is continuous for $y$ in a neighbourhood of the origin.  As $\epsilon\downarrow 0$, the Gaussian kernel converges to the Dirac
distribution in the sense of distributions, which gives
Eq.~\eqref{coarea_delta_limit}.
\end{proof}

We next compute the normal Jacobian in the eigenpair problem.  Let
\[
\mathcal F_X:\mathbb C\times\mathbf S_N^c\longrightarrow\mathbb C^N,
\qquad
\mathcal F_X(z,{\bf v})=(X-z{\bf 1}_N){\bf v}.
\]
Suppose $z_a$ is a simple eigenvalue, let ${\bf v}_a$ be a unit right eigenvector, and let ${\boldsymbol\ell}_a$ be a unit left eigenvector:
\[
(X-z_a{\bf 1}_N){\bf v}_a=0,\qquad
(X-z_a{\bf 1}_N)^*{\boldsymbol\ell}_a=0.
\]
Set $B_a=X-z_a{\bf 1}_N$.  The nonzero singular values of $B_a$ will be denoted by $s_1,\ldots,s_{N-1}$.  Since $z_a$ is simple,
\begin{equation}\label{adjugate_simple}
\mbox{adj}(B_a)=c_a\,{\bf v}_a{\boldsymbol\ell}_a^*,
\qquad
|c_a|=\prod_{j=1}^{N-1}s_j,
\end{equation}
and therefore
\begin{equation}\label{pprime_singular}
|p_X'(z_a)|
=
\left(\prod_{j=1}^{N-1}s_j\right)
|{\boldsymbol\ell}_a^*{\bf v}_a|>0.
\end{equation}

At $(z_a,e^{i\phi}{\bf v}_a)$,
\[
D\mathcal F_X(\dot z,\dot{\bf v})
=
B_a\dot{\bf v}-\dot z\,e^{i\phi}{\bf v}_a.
\]
The kernel is the one-dimensional real tangent direction generated by the phase, $\dot{\bf v}=i\,e^{i\phi}{\bf v}_a$.  On the orthogonal complement of this direction, the restriction of $B_a$ to ${\bf v}_a^\perp$ contributes the real Jacobian
$\prod_{j=1}^{N-1}s_j^2$, while the complex variable $\dot z$ contributes, modulo $\mbox{Ran}B_a={\boldsymbol\ell}_a^\perp$, the factor
$|{\boldsymbol\ell}_a^*{\bf v}_a|^2$.  Consequently
\begin{equation}\label{normal_jacobian_eigenpair}
J_{\mathcal F_X}(z_a,e^{i\phi}{\bf v}_a)
=
\left(\prod_{j=1}^{N-1}s_j^2\right)
|{\boldsymbol\ell}_a^*{\bf v}_a|^2
=
|p_X'(z_a)|^2.
\end{equation}
This also proves directly that $0$ is a regular value of $\mathcal F_X$ along the zero set.

Since $d\mu_N=\frac12dS$ on $\mathbf S_N^c$, Lemma~\ref{lemma_coarea_delta} and Eq.~\eqref{normal_jacobian_eigenpair} give, for every
$F\in C_c^\infty(\mathbb C\times\mathbf S_N^c)$,
\[
\frac1\pi\int F(z,{\bf v})\,
\delta^{(2N)}(\mathcal F_X(z,{\bf v}))|p_X'(z)|^2
\,d^2z\,d\mu_N({\bf v})
=
\frac1{2\pi}\sum_{a=1}^N\int_0^{2\pi}
F(z_a,e^{i\phi}{\bf v}_a)\,d\phi,
\]
which is Theorem~\ref{MainMetaTheorem}.

For completeness we record the slightly more general one-parameter identity used in Section~2.

\begin{lemma}[Holomorphic matrix pencil]\label{lemma_holomorphic_pencil}
Let $\Omega\subset\mathbb C$ be open and let $B:\Omega\to\mathcal M_N(\mathbb C)$ be holomorphic.  If every zero of $b(z)=\det B(z)$ is simple, then
\[
\delta^{(2)}(b(z))
=
\frac1\pi\int_{\mathbf S_N^c}
\delta^{(2N)}(B(z){\bf v})\,d\mu_N({\bf v})
\]
in $\mathcal D'(\Omega)$.
\end{lemma}

\begin{proof}
At a zero $z_0$, $\mbox{rank}B(z_0)=N-1$.  Let ${\bf v}$ and ${\boldsymbol\ell}$ be unit right and left null vectors and let $s_1,\ldots,s_{N-1}$ be the nonzero singular values.  The normal Jacobian of
$(z,{\bf v})\mapsto B(z){\bf v}$ is
\[
\left(\prod_{j=1}^{N-1}s_j^2\right)
|{\boldsymbol\ell}^*B'(z_0){\bf v}|^2
=
|b'(z_0)|^2,
\]
Indeed, at a rank-$N-1$ matrix one has
$\operatorname{adj}B(z_0)=c\,{\bf v}{\boldsymbol\ell}^*$ with
$|c|=\prod_{j=1}^{N-1}s_j$, while Jacobi's formula gives
\[
b'(z_0)
=
\operatorname{Tr}\!\left(\operatorname{adj}B(z_0)B'(z_0)\right)
=
c\,{\boldsymbol\ell}^*B'(z_0){\bf v}.
\]  The coarea calculation therefore gives
\[
\frac1\pi\int_{\mathbf S_N^c}\delta^{(2N)}(B(z){\bf v})\,d\mu_N({\bf v})
=
\sum_{b(z_0)=0}\frac{\delta^{(2)}(z-z_0)}{|b'(z_0)|^2},
\]
which is exactly the standard scalar identity for $\delta^{(2)}(b(z))$.
\end{proof}

The crucial point is that repeated \emph{nonzero} singular values cause no difficulty: only the rank-$N-1$ transversality at a simple zero is used.

\subsection{Appendix C: A transverse left--right eigenvector identity}
\label{Appendix C left-right}

We record a companion identity needed only for the left--right extension discussed at the end of Section~2.  For ${\bf v}\in\mathbf S_N^c$ let
$P_{\bf v}={\bf 1}_N-{\bf v}{\bf v}^*$ and let
$\delta^{(2N-2)}_{{\bf v}^\perp}$ denote the Euclidean Dirac distribution on the complex space ${\bf v}^\perp$.  Equivalently, if
$V:\mathbb C^{N-1}\to{\bf v}^\perp$ is any unitary isometry, then
\[
\delta^{(2N-2)}_{{\bf v}^\perp}({\bf y})
:=
\delta^{(2N-2)}(V^*{\bf y});
\]
this is independent of the choice of $V$.

Let $z_a$ be simple, let ${\bf v}_a$ be a unit right eigenvector, and choose the left eigenvector ${\bf u}_a$ by
\[
(X-z_a{\bf 1}_N)^*{\bf u}_a=0,\qquad {\bf u}_a^*{\bf v}_a=1.
\]
For fixed $(z_a,{\bf v}_a)$ the real map
\[
{\bf u}\longmapsto
\left(
V^*(X-z_a{\bf 1}_N)^*{\bf u},\,
{\bf u}^*{\bf v}_a-1
\right)\in\mathbb C^{N-1}\times\mathbb C
\]
has the unique zero ${\bf u}_a$.  Using the same singular-value decomposition as in Appendix~B, its real Jacobian at that zero is
\begin{equation}\label{left_transverse_jacobian}
\left(\prod_{j=1}^{N-1}s_j^2\right)
|{\boldsymbol\ell}_a^*{\bf v}_a|^2
=
|p_X'(z_a)|^2.
\end{equation}
Consequently, as a measure in ${\bf u}$,
\begin{equation}\label{left_transverse_delta}
|p_X'(z_a)|^2
\delta^{(2N-2)}_{{\bf v}_a^\perp}
\!\left((X-z_a{\bf 1}_N)^*{\bf u}\right)
\delta^{(2)}({\bf u}^*{\bf v}_a-1)\,d^{2N}{\bf u}
=
\delta_{{\bf u}_a}(d{\bf u}).
\end{equation}
Combining Eq.~\eqref{left_transverse_delta} with Theorem~\ref{MainMetaTheorem} gives the rigorous left--right marked empirical measure stated in Section~\ref{Towards}.  Notice that the adjoint equation contributes only $2N-2$ independent real constraints.  This is precisely what is lost if one writes a full
$\delta^{(2N)}((X-z{\bf 1})^*{\bf u})$ on the singular support.

\subsection{Appendix D: Proof of the Lemma \ref{IntLemma}}
For convenience of the reader we repeat the statement of the Lemma below:\\[0.5ex]
{\it
 The identity
\begin{equation}\label{sphereintegralApp}
\int_{\mathbf{S}^{(c)}_N}  f(|{\bf v}^T{\bf v}|^2)\,d\mu_H({\bf v})=\frac{N-1}{2}\int_0^1(1-u)^{(N-3)/2}f(u)\,du,
\end{equation}
with  $\mathbf{S}^{(c)}_N$ being  the unit complex sphere, holds true for any function $f(u), \, u\in \mathbb{R}$ such that the above integrals exist, assuming $N\ge 2$.
}

\begin{proof}
It is convenient to parametrize ${\bf v}\in \mathbb{C}^N$ as ${\bf v}={\bf p}_1+i{\bf p}_2$ with ${\bf p}_{1,2}\in \mathbb{R}^N$
which implies
\[
|{\bf v}^T{\bf v}|^2=({\bf p}^T_1{\bf p}_1-{\bf p}_2^T{\bf p}_2)^2+4\left({\bf p}^T_1{\bf p}_2\right)^2
\]
and further express the Haar's measure on the unit complex sphere $\mathbf{S}^{(c)}_N$ using the Dirac $\delta-$function as
\begin{equation}\label{Haar}
d\mu_H({\bf v})\propto \delta\left({\bf p}^T_1{\bf p}_1+{\bf p}_2^T{\bf p}_2-1\right)d{\bf p}_1d{\bf p}_2\, ,
\end{equation}
where here and henceforth we omit arising multiplicative factors as their product can be easily fixed in the very end.
Hence in the new coordinates the integral in question takes the form
\begin{equation}\label{sphereintegral_A}
\int_{\mathbf{S}^{(c)}_N}  f(|{\bf v}^T{\bf v}|^2)\,d\mu_H({\bf v})\propto \int_{\mathbb{R}^N} d{\bf p}_1\int_{\mathbb{R}^N} d{\bf p}_2 \delta\left({\bf p}^T_1{\bf p}_1+{\bf p}_2^T{\bf p}_2-1\right)f\left[({\bf p}^T_1{\bf p}_1-{\bf p}_2^T{\bf p}_2)^2+4\left({\bf p}^T_1{\bf p}_2\right)^2\right].
\end{equation}
To evaluate such an integral we may use a trick suggested in \cite{YF2002} of introducing the (non-negative definite) real symmetric matrix
$Q=\left(\begin{array}{cc}{\bf p}^T_1{\bf p}_1 & {\bf p}^T_1{\bf p}_2\\
{\bf p}^T_2{\bf p}_1 & {\bf p}^T_2{\bf p}_2\end{array}\right)\ge 0$  and using it as the new integration variable in Eq.(\ref{sphereintegral_A}).
Such a change incurs a certain Jacobian and the corresponding formulas can be found, e.g. in Eqs. (10)--(12) in \cite{FyoSom2007}. Here $dQ=dQ_{11}\,dQ_{22}\,dQ_{12}$ denotes Lebesgue measure on the three-dimensional space of real symmetric $2\times2$ matrices. Following this route one finds after straightforward manipulations
\begin{equation}\label{sphereintegral_B}
\int_{\mathbf{S}^{(c)}_N}  f(|{\bf v}^T{\bf v}|^2)\,d\mu_H({\bf v})\propto \frac{\pi^{N-1/2}}{\Gamma\left(\frac{N}{2}\right)\Gamma\left(\frac{N-1}{2}\right)} \int_{Q\ge 0} dQ  \det{Q}^{\frac{N-3}{2}} \delta\left(\mbox{\small Tr}Q-1\right)
f\left[\left(\mbox{\small Tr}Q\right)^2-4\det{Q}\right].
\end{equation}
Writing $Q$ in terms of the eigenvalues $q_1$ and $q_2$ as $Q=O\,\mbox{diag}(q_1,q_2)\,O^*$, with $O^TO={\bf 1}_2$ implying $dQ\propto |q_1-q_2|dq_1dq_2\,d\mu(O)$
we further rewrite the above as
\begin{equation}\label{sphereintegral_C}
\int_{\mathbf{S}^{(c)}_N}  f(|{\bf v}^T{\bf v}|^2)\,d\mu_H({\bf v})\propto  \int_{0}^{\infty}dq_1\int_{0}^{\infty}dq_2\,|q_1-q_2|   (q_1q_2)^{\frac{N-3}{2}} \delta(q_1+q_2-1)
f[1-4q_1q_2].
\end{equation}
and after performing the integration over $q_2$ one arrives at
\begin{equation}\label{sphereintegral_D}
\int_{\mathbf{S}^{(c)}_N}  f(|{\bf v}^T{\bf v}|^2)\,d\mu_H({\bf v}) \propto \int_{0}^{1}dq_1|2q_1-1|   \left[q_1(1-q_1)\right]^{\frac{N-3}{2}}
f[\left(2q_1-1\right)^2].
\end{equation}
Finally, one may notice the invariance of the integrand with respect to the change $q_1\to 1-q_1$, so it is equal to twice the same integral with the upper limit $q_1=1/2$.
Introducing the variable $u=(2q-1)^2$ so that $q_1(1-q_1)=(1-u)/4$ and restoring the accumulated multiplicative constant by setting $f=1$ one arrives at the statement of the lemma.

 \end{proof}

\subsection{Appendix E: Exact equivalence with the Hikami--Pnini formula}
\label{AppendixE_HP_equivalence}

In this appendix we prove, for every finite $N$, the equivalence between the
mean eigenvalue density obtained from the Kac--Rice formula for a normal
additive deformation and the earlier contour representation of Hikami and
Pnini~\cite{HP1996}.  The proof is most transparent on the open set where
all numbers $g_i=|z-a_i|^2$ are positive and pairwise distinct.  The final
identity extends to coinciding $g_i$ by the unique confluent continuation of
the divided differences which occur below.

We use throughout
\[
d_i=z-a_i,\qquad g_i=|d_i|^2,
\]
and the two symmetric functions already introduced in the main text,
\begin{equation}\label{HP_FG_def}
F(g_1,\ldots,g_N)
 =\int_0^\infty e^{-R}\prod_{j=1}^N(R+g_j)\,dR,
\qquad
G(g_1,\ldots,g_N)
 =\sum_{i=1}^N\frac{e^{-g_i}}{\prod_{j\ne i}(g_j-g_i)}.
\end{equation}
We abbreviate $F_i=\partial_{g_i}F$, $F_{ij}=\partial_{g_i}\partial_{g_j}F$
and similarly for $G$.  Equation~\eqref{mean_dens_deformed_norm} obtained by
the Kac--Rice route reads
\begin{equation}\label{HP_KR_density_repeated}
p_N^{(A,\mathrm{norm})}(z)
=-\frac1\pi\left[
\sum_{i=1}^N F_iG_i
+\sum_{i<j}|d_i-d_j|^2F_{ij}G_{ij}
\right].
\end{equation}
We shall show that the Hikami--Pnini representation gives exactly the same
expression.

\subsubsection{The Hikami--Pnini potential}

In the variance-one Ginibre convention used in the present paper the
Hikami--Pnini representation takes the form
\begin{equation}\label{HikamiPnini0}
p_N^{(HP)}(z)=\frac1\pi\,\partial_z\partial_{\bar z}
\int_0^1\frac{dp}{p^2}\oint_\gamma\frac{dt}{2\pi i}\,
\frac{e^{-pt}}{t}
\prod_{i=1}^N\frac{t(1-p)-g_i}{t-g_i},
\end{equation}
where $\gamma$ encloses $0,g_1,\ldots,g_N$.  Because the integral at
$p=0$ is not absolutely convergent, Eq.~\eqref{HikamiPnini0} is understood
with a cutoff $p\geq\eta>0$, the derivatives are taken first, and only then
$\eta\downarrow0$.

For pairwise distinct $g_i$, the contour integral is evaluated by the
residues at $t=0$ and $t=g_i$:
\begin{equation}\label{HP_contour_residue_general}
\oint_\gamma\frac{dt}{2\pi i}\,\frac{e^{-pt}}{t}
\prod_{i=1}^N\frac{t(1-p)-g_i}{t-g_i}
=1-p\sum_{i=1}^N e^{-pg_i}
\prod_{k\ne i}\frac{g_i(1-p)-g_k}{g_i-g_k}.
\end{equation}
The $p^{-2}$ contribution of the first term is independent of $z$.  The
remaining logarithmic divergence is likewise $z$-independent.  A convenient
finite representative is therefore
\begin{equation}\label{HP_Phi_ren_general}
\Phi_{\rm ren}(g)
:=\sum_{i=1}^N\int_0^1
\left[
e^{-pg_i}\prod_{k\ne i}\frac{g_i(1-p)-g_k}{g_i-g_k}-1
\right]\frac{dp}{p},
\end{equation}
for which
\begin{equation}\label{HP_density_Phi}
p_N^{(HP)}(z)
=-\frac1\pi\,\partial_z\partial_{\bar z}\Phi_{\rm ren}(g).
\end{equation}
Since only derivatives will be used, we may equivalently omit the common
subtracted constant and, after $\tau=pg_i$ in the $i$th term, write
\begin{equation}\label{HP_Phi_formal_general}
\Phi(g)
=\sum_{i=1}^N\int_0^{g_i}e^{-\tau}
\prod_{k\ne i}\left(1-\frac{\tau}{g_i-g_k}\right)
\frac{d\tau}{\tau}.
\end{equation}
All derivatives of Eq.~\eqref{HP_Phi_formal_general} below are understood
through the finite expression~\eqref{HP_Phi_ren_general}.

Put
\[
S:=\sum_{i=1}^N\Phi_i,\qquad
E:=\sum_{i=1}^Ng_i\partial_{g_i}.
\]
Since $\partial_zg_i=\bar d_i$ and $\partial_{\bar z}g_i=d_i$,
\[
\partial_z\partial_{\bar z}\Phi
=\sum_i\Phi_i+\sum_{i,j}\bar d_i d_j\Phi_{ij}.
\]
Using
\[
\bar d_i d_j+\bar d_j d_i
=g_i+g_j-|d_i-d_j|^2
\]
and collecting the diagonal and off-diagonal terms gives
\begin{equation}\label{HP_Laplacian_reduction}
\partial_z\partial_{\bar z}\Phi
=(1+E)S-\sum_{i<j}|d_i-d_j|^2\Phi_{ij}.
\end{equation}
Consequently the equivalence of Eqs.~\eqref{HP_KR_density_repeated} and
\eqref{HP_density_Phi} follows from the two identities
\begin{equation}\label{HP_two_key_identities}
\Phi_{ij}=-F_{ij}G_{ij}\quad(i\ne j),
\qquad
(1+E)S=\sum_{i=1}^NF_iG_i.
\end{equation}
We now prove them separately.

\subsubsection{Mixed derivatives}

Let
\[
P(t):=\prod_{k=1}^N(t-g_k).
\]
The function $G$ has the contour representation
\begin{equation}\label{HP_G_contour}
G=(-1)^{N-1}\oint_\Gamma\frac{e^{-t}}{P(t)}\frac{dt}{2\pi i},
\end{equation}
where $\Gamma$ surrounds $g_1,\ldots,g_N$ but no other singularity.  Fix
$i\ne j$ and put
\[
Q_{ij}(t):=\prod_{k\ne i,j}(t-g_k),
\qquad
P(t)=(t-g_i)(t-g_j)Q_{ij}(t).
\]
Differentiating Eq.~\eqref{HP_G_contour} twice gives
\begin{equation}\label{HP_Gij_contour}
G_{ij}=(-1)^{N-1}\oint_\Gamma
\frac{e^{-t}}{(t-g_i)^2(t-g_j)^2Q_{ij}(t)}\frac{dt}{2\pi i}.
\end{equation}

It is convenient to compute the same mixed derivative on the original
Hikami--Pnini side before residues.  Denote by $\mathcal J(g)$ the inner
$p,t$ integral in Eq.~\eqref{HikamiPnini0}, with the above cutoff
prescription.  For $i\ne j$ the two derivatives remove the singular
$p^{-2}$ factor, because
\[
\partial_{g_i}\left(1-\frac{pt}{t-g_i}\right)
=-\frac{pt}{(t-g_i)^2}.
\]
Hence
\begin{equation}\label{HP_Jij_before_p}
\partial_{g_i}\partial_{g_j}\mathcal J
=\oint_\Gamma\frac{dt}{2\pi i}
\frac{1}{(t-g_i)^2(t-g_j)^2Q_{ij}(t)}
\int_0^1 t e^{-pt}Q_{ij}((1-p)t)\,dp.
\end{equation}
With $u=(1-p)t$,
\begin{equation}\label{HP_p_integral}
\int_0^1 t e^{-pt}Q_{ij}((1-p)t)\,dp
=e^{-t}\int_0^t e^uQ_{ij}(u)\,du.
\end{equation}
Let $H_{ij}$ be the unique polynomial satisfying
\begin{equation}\label{HP_Hij_def}
H_{ij}'+H_{ij}=Q_{ij}.
\end{equation}
Equivalently,
$H_{ij}=Q_{ij}-Q_{ij}'+Q_{ij}''-\cdots$.  Then
\[
e^{-t}\int_0^t e^uQ_{ij}(u)\,du
=H_{ij}(t)-e^{-t}H_{ij}(0).
\]
The contour integral of the term containing $H_{ij}(t)$ vanishes: the
corresponding rational function decays as $t^{-4}$ at infinity.  Therefore
\begin{equation}\label{HP_Jij_H0}
\partial_{g_i}\partial_{g_j}\mathcal J
=-H_{ij}(0)\oint_\Gamma
\frac{e^{-t}}{(t-g_i)^2(t-g_j)^2Q_{ij}(t)}\frac{dt}{2\pi i}.
\end{equation}
On the other hand, integrating Eq.~\eqref{HP_Hij_def} from $-\infty$ to $0$
gives
\begin{align}\label{HP_H0_Fij}
H_{ij}(0)
&=\int_{-\infty}^0e^uQ_{ij}(u)\,du \notag\\
&=(-1)^{N-2}\int_0^\infty e^{-R}
\prod_{k\ne i,j}(R+g_k)\,dR
=(-1)^{N-2}F_{ij}.
\end{align}
Combining Eqs.~\eqref{HP_Gij_contour}, \eqref{HP_Jij_H0} and
\eqref{HP_H0_Fij} yields
\begin{equation}\label{HP_Jij_equals}
\partial_{g_i}\partial_{g_j}\mathcal J=F_{ij}G_{ij}.
\end{equation}
But Eq.~\eqref{HP_contour_residue_general} shows that
$\mathcal J=-\Phi+$ a $g$-independent term.  Hence
\begin{equation}\label{HP_mixed_identity_proved}
\boxed{\Phi_{ij}=-F_{ij}G_{ij}},\qquad i\ne j.
\end{equation}

\subsubsection{The first-order identity}

Introduce the Lagrange interpolation polynomials
\begin{equation}\label{HP_Lagrange}
L_i(x):=\prod_{j\ne i}\frac{x-g_j}{g_i-g_j},
\end{equation}
so that $L_i(g_j)=\delta_{ij}$.
Then Eq.~\eqref{HP_Phi_formal_general} becomes
\begin{equation}\label{HP_Phi_Lagrange}
\Phi(g)=\sum_{i=1}^N\int_0^{g_i}e^{-\tau}
L_i(g_i-\tau)\frac{d\tau}{\tau}.
\end{equation}
Scale all nodes simultaneously, $g_i\mapsto\lambda g_i$.  Since
$L_i^{(\lambda g)}(\lambda x)=L_i^{(g)}(x)$, the substitution
$\tau=\lambda s$ gives
\[
\Phi(\lambda g)=\sum_i\int_0^{g_i}e^{-\lambda s}
L_i(g_i-s)\frac{ds}{s}.
\]
Differentiating at $\lambda=1$ therefore yields
\begin{equation}\label{HP_EPhi}
E\Phi=-K(g),\qquad
K(g):=\sum_{i=1}^N\int_0^{g_i}e^{-s}L_i(g_i-s)\,ds.
\end{equation}
Let $D:=\sum_i\partial_{g_i}$.  Since $[E,D]=-D$, one has
$D E=(1+E)D$, and hence
\begin{equation}\label{HP_ES_DK}
(1+E)S=D(E\Phi)=-DK.
\end{equation}

We now evaluate $DK$ by translating all nodes.  After $x=g_i-s$,
\[
K(g)=\sum_i e^{-g_i}\int_0^{g_i}e^xL_i(x)\,dx.
\]
Under $g_i\mapsto g_i+c$ one has
$L_i^{(g+c)}(x+c)=L_i^{(g)}(x)$, so
\[
K(g+c)=\sum_i e^{-g_i}\int_{-c}^{g_i}e^xL_i(x)\,dx.
\]
Differentiation at $c=0$ gives
\begin{equation}\label{HP_DK_L0}
DK=\sum_{i=1}^Ne^{-g_i}L_i(0),
\end{equation}
and therefore
\begin{equation}\label{HP_ES_L0}
(1+E)S=-\sum_{i=1}^Ne^{-g_i}L_i(0).
\end{equation}

It remains to identify the right-hand side with $\sum_iF_iG_i$.  From
Eq.~\eqref{HP_FG_def}, after putting $x=-R$,
\begin{equation}\label{HP_Fi_x}
F_i=(-1)^{N-1}\int_{-\infty}^0e^x\frac{P(x)}{x-g_i}\,dx,
\end{equation}
whereas differentiation of Eq.~\eqref{HP_G_contour} gives
\begin{equation}\label{HP_Gi_contour}
G_i=(-1)^{N-1}\oint_\Gamma
\frac{e^{-t}}{(t-g_i)P(t)}\frac{dt}{2\pi i}.
\end{equation}
Summing over $i$ and using
\begin{equation}\label{HP_logder_identity}
\sum_i\frac{P(x)}{(x-g_i)P(t)(t-g_i)}
=\frac{P'(x)P(t)-P(x)P'(t)}{(t-x)P(t)^2}
=(\partial_x+\partial_t)\frac{P(x)}{(t-x)P(t)},
\end{equation}
we obtain
\begin{equation}\label{HP_sum_FiGi_before_parts}
\sum_iF_iG_i
=\int_{-\infty}^0e^x\,dx\oint_\Gamma e^{-t}
(\partial_x+\partial_t)\frac{P(x)}{(t-x)P(t)}\frac{dt}{2\pi i}.
\end{equation}
Integration by parts in $x$ and around the closed $t$-contour cancels the
two bulk terms and leaves only the boundary $x=0$:
\begin{equation}\label{HP_sum_FiGi_boundary}
\sum_iF_iG_i
=\oint_\Gamma e^{-t}\frac{P(0)}{tP(t)}\frac{dt}{2\pi i}.
\end{equation}
Choose $\Gamma$ to surround the $g_i$ but not the origin.  Taking residues,
\begin{equation}\label{HP_sum_FiGi_residues}
\sum_iF_iG_i
=\sum_i e^{-g_i}\frac{P(0)}{g_iP'(g_i)}
=-\sum_i e^{-g_i}L_i(0).
\end{equation}
Together with Eq.~\eqref{HP_ES_L0}, this proves
\begin{equation}\label{HP_first_identity_proved}
\boxed{(1+E)S=\sum_{i=1}^NF_iG_i}.
\end{equation}

\subsubsection{Conclusion and confluent continuation}

Substituting Eqs.~\eqref{HP_mixed_identity_proved} and
\eqref{HP_first_identity_proved} into Eq.~\eqref{HP_Laplacian_reduction}
gives
\begin{align}\label{HP_final_equivalence}
p_N^{(HP)}(z)
&=-\frac1\pi\left[
\sum_iF_iG_i
+\sum_{i<j}|d_i-d_j|^2F_{ij}G_{ij}
\right]\notag\\
&=p_N^{(A,\mathrm{norm})}(z).
\end{align}
Thus the Hikami--Pnini and Kac--Rice formulas are exactly equivalent for
every finite $N$.

The derivation was carried out for positive pairwise distinct $g_i$ only to
avoid carrying confluent notation through the residue calculations.  The
original Hikami--Pnini contour integral is regular when some $g_i$ coincide,
and the function $G$ and all expressions obtained from it are divided
differences with their unique Hermite continuation.  Therefore
Eq.~\eqref{HP_final_equivalence} extends by continuity to coinciding $g_i$
and, subsequently, to the boundary where one or more $g_i$ vanish.  The
explicit $N=2$ calculation displayed in the main text is thus a special case
of the general identity.

\subsection{Appendix F: Proof of the Proposition \ref{cor2.10}}
We aim at analysing implications of Theorem~\ref{Theorem2} for a general rank-one {\it non-normal} deformation matrix $A$ parametrized  by a single complex eigenvalue $a\in \mathbb{C}$ and two complex vectors: the left ${\bf l}$ and the right ${\bf r}$ so that
\begin{equation}\label{rank_one_gen_proof}
 A=a\,{\bf r}\otimes {\bf l}^*, \quad \mbox{with}\quad a\in \mathbb{C}\ \quad \mbox{and s.t.}\,\, {\bf l}^*{\bf r}=1.
 \end{equation}
 In fact in this case we find it is slightly easier to start with an equivalent form  Eq.(\ref{MainJPDadditivedef}) repeated for convenience below:
 \begin{equation}\label{JPDadditive_fin_proof}
{\cal P}^{(A)}_N(z,{\bf v})=\frac{1}{\pi\Gamma(N)}e^{-\mathbf{v}^*A^*_zA_z{\bf v}} \int_0^\infty e^{-R}\, \det{\left[R\,
{\bf 1}_{N-1}+T({\bf v})\right]}\,dR,
\end{equation}
where the entries of $(N-1)\times (N-1)$ matrix are given by $T_{ij}={\bf v}^*_iA^*_zP({\bf v})A_z {\bf v}_j$, with $P({\bf v})={\bf 1}_N-{\bf v}\otimes {\bf v}^*$ and the vectors ${\bf v},{\bf v}_1, \ldots, {\bf v}_{N-1}$ forming an orthonormal basis of $\mathbb{C}^N$.  It is also convenient to introduce the notations:
\begin{equation}\label{lr_notations}
{\bf p}:=P({\bf v}){\bf r}, \quad {\tilde{\bf p}}:=\left(\begin{array}{c}{\bf v}_1^*{\bf p}\\{\bf v}_2^*{\bf p} \\ \ldots \\ \ldots \\ {\bf v}_{N-1}^*{\bf p}  \end{array}\right),
\quad {\tilde{\bf l}}:=\left(\begin{array}{c}{\bf v}_1^*{\bf l}\\{\bf v}_2^*{\bf l} \\ \ldots \\ \ldots \\ {\bf v}_{N-1}^*{\bf l}  \end{array}\right),
\end{equation}
in terms of which the matrix $T({\bf v})$ can be readily expressed as
\begin{equation}\label{Tmat_proof}
T({\bf v})=|z|^2{\bf 1}_{N-1}+\tilde{T}({\bf v}), \quad \tilde{T}({\bf v})=|a|^2\left({\bf p}^*{\bf p}\right)\tilde{{\bf l}}\otimes \tilde{{\bf l}}^*-a\overline{z}\tilde{{\bf p}}\otimes \tilde{{\bf l}}^*
-\overline{a}z\tilde{{\bf l}}\otimes \tilde{{\bf p}}^*.
\end{equation}
The matrix $\tilde{T}({\bf v})$ has rank at most $2$.  Away from degenerate cases where the rank drops, it has $(N-3)$ zero eigenvalues and two possibly nonzero eigenvalues $\lambda_1,\lambda_2$
corresponding to eigenvectors ${\bf u}_{1,2}$ belonging to the space of linear combinations   ${\bf u}=\alpha\,\tilde{{\bf l}}+\beta  \tilde{{\bf p}}, \, (\alpha,\beta)\ne (0,0)$. This implies the system of 2 linear equations satisfied by  $(\alpha,\beta)$ and $\lambda$:
\begin{equation}
\begin{array}{c}
\alpha\left\{\left[|a|^2\left({\bf p}^*{\bf p}\right)\left(\tilde{{\bf l}}^*\tilde{{\bf l}}\right)-\overline{a}z\left(\tilde{{\bf p}}^*\tilde{{\bf l}}\right)\right]-\lambda\right\}+\beta\left[|a|^2\left({\bf p}^*{\bf p}\right)\left(\tilde{{\bf l}}^*\tilde{{\bf p}}\right)-\overline{a}z\left(\tilde{{\bf p}}^*\tilde{{\bf p}}\right)\right]=0\\ \alpha\,\overline{z}a\left(\tilde{{\bf l}}^*\tilde{{\bf l}}\right)+\beta\left[a\overline{z}\left(\tilde{{\bf l}}^*\tilde{{\bf p}}\right)+\lambda \right]=0
\end{array},
\end{equation}
with the corresponding characteristic equation
\[
\lambda^2+\lambda\left[z\overline{a}\left(\tilde{{\bf p}}^*\tilde{{\bf l}}\right)+a\overline{z}\left(\tilde{{\bf l}}^*\tilde{{\bf p}}\right)
-|a|^2\left({\bf p}^*{\bf p}\right)\left(\tilde{{\bf l}}^*\tilde{{\bf l}}\right)\right]+|a|^2|z|^2
\left[\left({\bf p}^*{\bf l}\right)\left(\tilde{{\bf l}}^*\tilde{{\bf p}}\right)-\left(\tilde{{\bf l}}^*\tilde{{\bf l}}\right)\left(\tilde{{\bf p}}^*\tilde{{\bf p}}\right)\right]=0.
\]
Taking into account the relations
\[
\tilde{{\bf p}}^*\tilde{{\bf p}}={\bf p}^*{\bf p}={\bf r}^*P({\bf v}){\bf r}, \quad \tilde{{\bf l}}^*\tilde{{\bf p}}={\bf l}^*P({\bf v}){\bf p}={\bf l}^*P({\bf v}){\bf r}
\]
one finds that the characteristic equation implies the following relations for the nontrivial eigenvalues $\lambda_{12}$:
\begin{equation}\label{Vieta}
\begin{array}{c}
\lambda_1+\lambda_2=-\overline{a}z\,{\bf r}^*P({\bf v}){\bf l}-\overline{z}a\,{\bf l}^*P({\bf v}){\bf r}+|a|^2\,\left({\bf r}^*P({\bf v}){\bf r}\right)\,
\left({\bf l}^*P({\bf v}){\bf l}\right)\\ \lambda_1\lambda_2=-|a|^2|z|^2\left[\left({\bf l}^*P({\bf v}){\bf l}\right)\,\left({\bf r}^*P({\bf v}){\bf r}\right)-
\left({\bf l}^*P({\bf v}){\bf r}\right)\,\left({\bf r}^*P({\bf v}){\bf l}\right)\right]
\end{array},
\end{equation}
Now the determinant entering Eq.(\ref{JPDadditive_fin_proof}) can be obviously evaluated as
\[\det{\left[R\,
{\bf 1}_{N-1}+T({\bf v})\right]}=(R+|z|^2)^{N-3}(R+|z|^2+\lambda_1)(R+|z|^2+\lambda_2).
\]
Combining this with relations Eq.(\ref{Vieta}) and using that in the present case
\[
{\bf v}^*A_z^*A_z{\bf v}=|z|^2+|a|^2({\bf r}^*{\bf r})|{\bf v}^*{\bf l}|^2-\overline{z}a({\bf v}^*{\bf r}) ({\bf l}^*{\bf v})-\overline{a}z({\bf v}^*{\bf l}) ({\bf r}^*{\bf v})
\]
and employing the definition Eq.(\ref{gamma_incom}) of the incomplete $\Gamma$-function one immediately arrives at the statement of {\bf Proposition \ref{cor2.10}}.

%%%%%%%%%%%%%%%%%%%%%%

\subsection{Appendix G: proof of Proposition \ref{asydenprop}}
We rescale $z=\sqrt N\,w$ and $a=\sqrt N\,\alpha$ and keep $w,\alpha$ fixed.  Then $\varkappa=N\Delta$ with $\Delta=|\alpha|^2-(\alpha\bar w+\bar\alpha w)$.  For fixed integer $k$,
\[
I^{(k)}(N,N\Delta)=\int_0^1e^{-Nq\Delta}(1-q)^{N+k-2}\,dq
\]
can be written as
\[
I^{(k)}(N,N\Delta)=\int_0^1\exp\{Nf_\Delta(q)\}(1-q)^{k-2}\,dq,
\qquad f_\Delta(q)=-\Delta q+\log(1-q).
\]
If $\Delta>-1$, then $f_\Delta'(0)=-(1+\Delta)<0$ and the maximum is the endpoint $q=0$. Setting $q=x/N$ and using dominated convergence on every fixed $x$-interval, followed by the elementary exponential tail bound coming from $f_\Delta(q)\le-cq$, gives
\begin{equation}\label{asyIplus}
I^{(k)}(N,N\Delta)=\frac{1+O(N^{-1})}{N(1+\Delta)}.
\end{equation}
If $\Delta<-1$, there is a unique interior maximizer $q_*=1-|\Delta|^{-1}$ with $f_\Delta''(q_*)=-|\Delta|^2$. Since $f_\Delta(q_*)=-1-\Delta-\log|\Delta|$ and $(1-q_*)^{k-2}=|\Delta|^{2-k}$, the standard one-dimensional Laplace expansion yields
\begin{equation}\label{asyIminus}
I^{(k)}(N,N\Delta)\sim
\sqrt{\frac{2\pi}{N}}\frac{e^{-N(1+\Delta)}}{|\Delta|^{N+k-1}}.
\end{equation}
Both estimates are uniform when $\Delta$ stays in a compact subset of the indicated open regime. The borderline $\Delta=-1$ is a coalescing endpoint/interior-saddle regime and requires a separate scaling.

For fixed integer $l$,
\begin{equation}\label{asyGamma}
\frac{\Gamma(N-l,N|w|^2)}{\Gamma(N)}\sim
\begin{cases}
N^{-l},& |w|<1,\\[1ex]
\displaystyle
\frac{e^{N(1-|w|^2)}}{N^l\sqrt{2\pi N}}
\frac{|w|^{2(N-l)}}{|w|^2-1},& |w|>1.
\end{cases}
\end{equation}
Inside the unit disk one has $\Delta>-1$ for all $\alpha$: for $\alpha\ne0$ this follows from
\[
|w|\cos(\theta-\theta_\alpha)<\frac12\left(|\alpha|+|\alpha|^{-1}\right),
\]
while $\alpha=0$ is immediate.  Combining Eq.~\eqref{asyIplus} with the first line of Eq.~\eqref{asyGamma} in Eq.~\eqref{mean den rank_one_norm} gives Eq.~\eqref{inside}.

Let now $|w|>1$.  For $\Delta<-1$, Eqs.~\eqref{asyIminus} and \eqref{asyGamma} give the rate ${\cal F}(w)$ and the prefactor in Eq.~\eqref{asydenout}.  For the complementary exterior sector $\Delta>-1$, using Eq.~\eqref{asyIplus} for $k=0,1,2$ and collecting the first nonvanishing terms gives
\[
p_N^{(r1,\,norm)}(\sqrt N\,w)
\sim
\frac{e^{-N(|w|^2-1-\log|w|^2)}}{\pi\sqrt{2\pi N}}
\frac{|w-\alpha|^2}{|w|^2(|w|^2-1)(1+\Delta)},
\]
which is Eq.~\eqref{asydenout_gin_branch}.  In particular, for $\alpha=0$,
\[
p_N^{(0)}(\sqrt N\,w)
\sim\frac{e^{-NI_{\rm Gin}(|w|^2)}}{\pi\sqrt{2\pi N}(|w|^2-1)},
\]
the standard Ginibre exterior tail.

\subsection{Appendix H: Source regularization and gauge fixing for the left--right problem}
\label{Appendix H left-right regularization}
The transverse identity of Appendix~C is the rigorous distributional formula
used in the main text. It is also useful to have a complementary construction
based on source regularization. This provides a concrete gauge-fixing
mechanism for the complex scaling freedom of a left--right eigenpair:
the source resolves the gauge orbit into isolated zeros to which the ordinary
multivariate Kac--Rice formula applies, and the phase-averaged marked
empirical measure is recovered when the source is removed. Besides giving an
alternative derivation, this construction makes the regularized geometry
particularly transparent: each eigenvalue gives rise to two nearby branches,
whose combined contribution acquires the corresponding factor $1/2$ in the
singular limit.

Let $X$ have simple eigenvalues $z_1,\ldots,z_N$.  Choose right and left eigenvectors ${\bf v}_a,{\bf u}_a$ bi-orthogonally,
\begin{equation}\label{LR_biorthogonal_basis}
X{\bf v}_a=z_a{\bf v}_a,\qquad
X^*{\bf u}_a=\overline z_a{\bf u}_a,\qquad
\|{\bf v}_a\|=1,\qquad
{\bf u}_a^*{\bf v}_b=\delta_{ab}.
\end{equation}
No unit-norm condition on ${\bf u}_a$ is imposed.  Let $\alpha_a\ne0$ and set
\begin{equation}\label{LR_sources}
{\bf h}_1:=\sum_{a=1}^N\alpha_a{\bf v}_a,
\qquad
{\bf h}_2:=\sum_{a=1}^N\alpha_a{\bf u}_a.
\end{equation}
For $\sigma>0$ and an auxiliary phase $\theta\in[0,2\pi)$ consider
\begin{equation}\label{LR_sigma_system}
(X-z{\bf 1}_N){\bf v}=\sigma e^{i\theta}{\bf h}_1,
\qquad
(X-z{\bf 1}_N)^*{\bf u}=\sigma e^{i\theta}{\bf h}_2,
\qquad
{\bf u}^*{\bf v}=1.
\end{equation}
At $\sigma=0$ the equations are invariant under
\[
({\bf v},{\bf u})\longmapsto(c{\bf v},\overline c^{-1}{\bf u}),
\qquad c\in\mathbb C\setminus\{0\},
\]
which is the gauge degeneracy that prevents a direct application of the ordinary Kac--Rice formula to the unreduced equations.  The source terms in Eq.~\eqref{LR_sigma_system} remove this degeneracy.

\subsubsection{The scalar secular equation and its two branches}

Put $B(z):=X-z{\bf 1}_N$.  For $z\notin\{z_1,\ldots,z_N\}$ the first two equations in Eq.~\eqref{LR_sigma_system} determine
\begin{equation}\label{LR_sigma_vectors}
{\bf v}=\sigma e^{i\theta}B(z)^{-1}{\bf h}_1,
\qquad
{\bf u}=\sigma e^{i\theta}B(z)^{-*}{\bf h}_2.
\end{equation}
Substituting this into the normalization condition gives the scalar equation
\begin{equation}\label{LR_sigma_secular}
1=q_\sigma(z),\qquad
q_\sigma(z):=\sigma^2{\bf h}_2^*B(z)^{-2}{\bf h}_1
=\sigma^2\sum_{a=1}^N\frac{|\alpha_a|^2}{(z_a-z)^2}.
\end{equation}
Multiplication by $p_X(z)^2$, where $p_X(z)=\det B(z)$, turns Eq.~\eqref{LR_sigma_secular} into a polynomial equation of degree $2N$.

Fix $a$ and write $z=z_a+\delta$.  Then
\[
1=\sigma^2\left(\frac{|\alpha_a|^2}{\delta^2}+r_a(z_a+\delta)\right),
\]
where $r_a$ is holomorphic near $z_a$.  With $\delta=\sigma y$ this becomes
\[
1=\frac{|\alpha_a|^2}{y^2}+\sigma^2r_a(z_a+\sigma y).
\]
At $\sigma=0$ it has the two simple roots $y=\pm|\alpha_a|$.  The implicit-function theorem therefore gives, for sufficiently small $\sigma$, two simple branches
\begin{equation}\label{LR_sigma_roots}
\zeta_a^{\pm}(\sigma)=z_a\pm\sigma|\alpha_a|+O(\sigma^3).
\end{equation}
These $2N$ local roots exhaust the polynomial equation.  Equivalently, outside disjoint small neighbourhoods of the $z_a$ one has $q_\sigma(z)=O(\sigma^2)$ uniformly and hence no additional solutions occur for small $\sigma$.

Let
\[
c_a:=\frac{\alpha_a}{|\alpha_a|}.
\]
For $s=\pm1$, Eqs.~\eqref{LR_sigma_vectors} and \eqref{LR_sigma_roots} imply
\begin{equation}\label{LR_sigma_vector_limits}
{\bf v}_a^{(s)}(\sigma)
=-s\,e^{i\theta}c_a{\bf v}_a+O(\sigma),
\qquad
{\bf u}_a^{(s)}(\sigma)
=-s\,e^{i\theta}c_a{\bf u}_a+O(\sigma).
\end{equation}
The same phase multiplies the left and right vectors.  This is required by ${\bf u}^*{\bf v}=1$: under a gauge transformation ${\bf v}\mapsto c{\bf v}$ one has ${\bf u}\mapsto\overline c^{-1}{\bf u}$, which for $|c|=1$ is the same phase rotation of both vectors.

\subsubsection{Exact finite-$\sigma$ Kac--Rice Jacobian}

The regularized equations form an ordinary holomorphic zero problem if one uses $(z,{\bf v},\overline{\bf u})$ as complex coordinates.  More explicitly, define
\begin{equation}\label{LR_holomorphic_map}
\mathcal F_{\sigma,\theta}(z,{\bf v},\overline{\bf u})
:=
\begin{pmatrix}
B(z){\bf v}-\sigma e^{i\theta}{\bf h}_1\\
B(z)^T\overline{\bf u}-\sigma e^{-i\theta}\overline{\bf h}_2\\
\overline{\bf u}^{\,T}{\bf v}-1
\end{pmatrix}
\in\mathbb C^{2N+1}.
\end{equation}
At a zero of Eq.~\eqref{LR_sigma_system}, $B(z)$ is invertible, and the complex Jacobian matrix is
\begin{equation}\label{LR_complex_Jacobian_matrix}
D_{\mathbb C}\mathcal F_{\sigma,\theta}
=
\begin{pmatrix}
B&0&-{\bf v}\\
0&B^T&-\overline{\bf u}\\
\overline{\bf u}^{\,T}&{\bf v}^T&0
\end{pmatrix}.
\end{equation}
Taking the Schur complement of $\mbox{diag}(B,B^T)$ gives
\begin{equation}\label{LR_complex_Jacobian_det1}
\det D_{\mathbb C}\mathcal F_{\sigma,\theta}
=2\,p_X(z)^2\,{\bf u}^*B(z)^{-1}{\bf v}.
\end{equation}
On the other hand, differentiating Eq.~\eqref{LR_sigma_secular} yields
\begin{equation}\label{LR_qprime}
q_\sigma'(z)
=2\sigma^2{\bf h}_2^*B(z)^{-3}{\bf h}_1
=2{\bf u}^*B(z)^{-1}{\bf v}.
\end{equation}
Consequently
\begin{equation}\label{LR_complex_Jacobian_det2}
\boxed{\det D_{\mathbb C}\mathcal F_{\sigma,\theta}
=p_X(z)^2q_\sigma'(z).}
\end{equation}
Since the real Jacobian of a holomorphic map is the squared modulus of its complex Jacobian, the exact Kac--Rice factor is therefore
\begin{equation}\label{LR_real_Jacobian}
J_\sigma(z,{\bf v},{\bf u})
=|p_X(z)|^4|q_\sigma'(z)|^2.
\end{equation}
In particular all zeros corresponding to the branches in Eq.~\eqref{LR_sigma_roots} are nondegenerate for sufficiently small $\sigma$.

For fixed $\theta$ the ordinary multivariate Kac--Rice formula now gives the positive atomic measure
\begin{align}\label{LR_sigma_KR_measure}
d\Pi^{LR}_{\sigma,\theta}
={}&J_\sigma(z,{\bf v},{\bf u})
\,\delta^{(2N)}\!\left(B(z){\bf v}-\sigma e^{i\theta}{\bf h}_1\right)
\,\delta^{(2N)}\!\left(B(z)^*{\bf u}-\sigma e^{i\theta}{\bf h}_2\right)\notag\\
&\times\delta^{(2)}({\bf u}^*{\bf v}-1)
\,d^2z\,d^{2N}{\bf v}\,d^{2N}{\bf u}.
\end{align}
Equivalently,
\begin{equation}\label{LR_sigma_atomic}
\Pi^{LR}_{\sigma,\theta}
=\sum_{a=1}^N\sum_{s=\pm1}
\delta_{(\zeta_a^{(s)}(\sigma),\,{\bf v}_a^{(s)}(\sigma),\,{\bf u}_a^{(s)}(\sigma))}.
\end{equation}
Thus the source regularization produces two isolated representatives of every gauge orbit and the total mass is $2N$.

\subsubsection{Removal of the source}

The doubling in Eq.~\eqref{LR_sigma_atomic} must be compensated before the source is removed.  Define the phase-averaged, de-doubled regularized marked measure
\begin{equation}\label{LR_sigma_phase_average}
\widehat\Pi_\sigma^{LR}
:=\frac12\int_0^{2\pi}\Pi^{LR}_{\sigma,\theta}\,\frac{d\theta}{2\pi}.
\end{equation}
Then for every compactly supported continuous test function $F$,
\begin{align}\label{LR_sigma_weak_limit_test}
\lim_{\sigma\downarrow0}\langle\widehat\Pi_\sigma^{LR},F\rangle
&=\frac12\sum_{a=1}^N\sum_{s=\pm1}
\int_0^{2\pi}
F\!\left(z_a,-s e^{i\theta}c_a{\bf v}_a,-s e^{i\theta}c_a{\bf u}_a\right)
\frac{d\theta}{2\pi}\notag\\
&=\sum_{a=1}^N\int_0^{2\pi}
F(z_a,e^{i\phi}{\bf v}_a,e^{i\phi}{\bf u}_a)
\frac{d\phi}{2\pi}.
\end{align}
Hence
\begin{equation}\label{LR_sigma_weak_limit}
\boxed{\widehat\Pi_\sigma^{LR}\Longrightarrow\Pi_X^{LR},\qquad \sigma\downarrow0,}
\end{equation}
where $\Pi_X^{LR}$ is precisely the phase-averaged left--right marked empirical measure in Eq.~\eqref{empir_right_left_eigenv}.  The limiting measure is independent of the auxiliary coefficients $\alpha_a$: their moduli determine only the $O(\sigma)$ splitting of the two branches, while their phases are absorbed by the $\theta$-average.

The source-regularization construction makes the gauge fixing and the
two-branch structure explicit. Since the source vectors in
Eq.~\eqref{LR_sources} are defined through the spectral basis, this
construction is most naturally viewed as a proof device rather than as a
formula for direct ensemble averaging. In the zero-source limit the relevant
distributional statement is the transverse identity
Eq.~\eqref{left_right_KacRice}, proved in Appendix~C.

\end{document}